%% file: main.tex
\documentclass[fleqn,usenatbib]{mnras}
\usepackage[utf8]{inputenc}
\usepackage{geometry}
\include{preamble.tex}

\title{The source of electrons at comet 67P}
\author[Stephenson et al.]{
P. Stephenson$^{1}$\thanks{pps18@ic.ac.uk},
A. Beth$^{1,2}$, 
J. Deca$^{3}$,
M. Galand$^{1}$,
C. Goetz$^{4}$, 
P. Henri$^{5,6}$, 
K. Heritier$^{1}$,
\newauthor
Z. Lewis,$^{1}$
A. Moeslinger,$^{2}$
H. Nilsson$^{7}$ 
and M. Rubin$^{8}$
\\
$^{1}$Department of Physics, Imperial College London, London, UK \\
$^{2}$Department of Physics, Ume\r{a} University, Ume\r{a}, Sweden\\
$^{3}$Laboratory for Atmospheric and Space Physics, University of Colorado, Boulder, Colorado \\
$^{4}$Department of Mathematics, Physics and Electrical Engineering, Northumbria University, Newcastle Upon Tyne, UK\\
$^{5}$Lagrange, OCA, CNRS, UCA, Nice, France\\
$^{6}$LPC2E, CNRS, Orleans, France\\
$^{7}$ Swedish Institute of Space Physics, Kiruna, Sweden\\
$^{8}$Physikalisches Institut, University of Bern, Bern, Switzerland 
}
\date{Jun 2023}

\begin{document}

\label{firstpage}
\maketitle
\begin{abstract}
We examine the origin of electrons in a weakly outgassing comet, using Rosetta mission data and a 3D collisional model of electrons at a comet. We have calculated a new dataset of electron-impact ionization (EII) frequency throughout the Rosetta escort phase, with measurements of the Rosetta Plasma Consortium’s Ion and Electron Sensor (RPC/IES). The EII frequency is evaluated in 15-minute intervals and compared to other Rosetta datasets.

Electron-impact ionization is the dominant source of electrons at 67P away from perihelion and is highly variable (by up to three orders of magnitude). Around perihelion, EII is much less variable and less efficient than photoionization at Rosetta. Several drivers of the EII frequency are identified, including magnetic field strength and the outgassing rate. Energetic electrons are correlated to the Rosetta-upstream solar wind potential difference, confirming that the ionizing electrons are solar wind electrons accelerated by an ambipolar field.

The collisional test particle model incorporates a spherically symmetric, pure water coma and all the relevant electron-neutral collision processes. Electric and magnetic fields are stationary model inputs, and are computed using a fully-kinetic, collisionless Particle-in-Cell simulation. Collisional electrons are modelled at outgassing rates of $Q=10^{26}$~s$^{-1}$ and $Q=1.5\times10^{27}$~s$^{-1}$. Secondary electrons are the dominant population within a weakly outgassing comet. These are produced by collisions of solar wind electrons with the neutral coma.

The implications of large ion flow speed estimates at Rosetta, away from perihelion, are discussed in relation to multi-instrument studies and the new results of the EII frequency obtained in the present study.

\end{abstract}
\begin{keywords}
Comets:general -- Comets: individual: 67P/CG
\end{keywords}

\section{Introduction}
As a comet approaches the Sun, near-surface ices are heated and sublimate, producing an envelope of expanding neutral gas known as the coma. The coma can be ionized by several processes, generating a population of ions and electrons within it. These form a  cometary ionosphere, which interacts with the solar wind. Ionization can be driven by absorption of high energy, extreme ultraviolet (EUV) photons or by collisions of energetic electrons. Solar wind ions can collide with cometary neutrals, causing ionization or a charge exchange process to produce a different ion species. 
The Rosetta mission escorted comet 67P/Churyumov-Gerasimenko (67P) for two years along its orbit from \edits{early} Aug 2014 to 30 Sep 2016. During the escort phase, Rosetta gathered an extensive and unique dataset of an evolving cometary ionosphere \citep[e.g.,][]{Goetz2016CavStruct, Heritier2017IonComp}. The cometary plasma environment at Rosetta was probed by instruments of the Rosetta Plasma Consortium \citep[RPC;][]{Carr2007} and the Rosetta Orbiter Spectrometer for Ion and Neutral {Analysis} \citep[ROSINA;][]{Balsiger2007}. At the end of mission, Rosetta made a final, controlled descent to the surface of 67P, measuring the vertical profile of a cometary ionosphere for the first time \citep{Heritier2017Vert}. 

The electrons within the coma of 67P cannot be described as a single population. \comments{Cold} ($<1$~eV), warm ($\sim 10$~eV) and hot ($>15$~eV) \comments{populations} have been identified within the coma \citep{Clark2015, Broiles2016Supratherm, Broiles2016Kappa, Eriksson2017, Engelhardt2018, Myllys2019, Gilet2020, Wattieaux2020, Myllys2021}. The warm population is made up by newly born cometary electrons, produced by ionization within the coma. The cold electrons are produced by cooling the warm population through collisions with the neutral coma \citep{Stephenson2022TestPl}. The hot population is likely made up of solar wind electrons that have been \edits{energised} by an ambipolar field that forms around the comet \citep{Myllys2019, Stephenson2022TestPl}. 
The cold electron population was observed frequently throughout the Rosetta mission, when \edits{the neutral gas} should not have been dense enough to substantially cool the warm electrons \citep{Engelhardt2018, Gilet2020, Wattieaux2020}. These estimates assumed that cometary electrons travelled approximately radially away from the nucleus, like the neutral gas in which they are embedded. 
\comments{To evaluate the energy degradation of electrons at a comet, \cite{Stephenson2022TestPl} applied a 3D collisional electron model.} With trapping in the ambipolar potential well and the gyration around the magnetic field, electron cooling becomes substantially more efficient and a weakly outgassing comet can generate and sustain a cold electron population.

\comments{At a comet with a low outgassing rate, $Q<2\times10^{27}$\,s$^{-1}$}, the ambipolar field forms a potential well around the nucleus \citep{Deca2017, Deca2019, Divin2020}. The potential well, which exceeds 60~V in depth at $Q=10^{25}$~s$^{-1}$, traps cometary electrons in the inner coma \citep{Sishtla2019trapping, Stephenson2022TestPl}.
As well as causing electron cooling, the ambipolar field accelerates and funnels solar wind electrons towards the nucleus \citep{Deca2017, Stephenson2022TestPl}. The acceleration of solar wind electrons by the ambipolar field is a likely source of ionizing electrons within the coma. Measurements of the solar wind ions by the Ion Composition Analyser \citep[RPC/ICA;][]{Nilsson2007} have been used to estimate the potential difference between Rosetta and the upstream solar wind. This potential fluctuates rapidly, but often exceeds several hundred electron volts \citep{Nilsson2022}. 

The electron density at Rosetta was measured by the Langmuir Probes \citep[LAP;][]{Eriksson2007} and Mutual Impedance Probes \citep[MIP;][]{Trotignon2007} of the Rosetta Plasma Consortium \citep[RPC; ][]{Carr2007}. Multi-instrument models have been used to explain the electron density during periods at large heliocentric distances \citep{Galand2016, Heritier2017Vert, Heritier2018source}. These models include photoionization and electron-impact ionization (EII) as a source of electrons, with EII frequencies derived from measurements of energetic electrons by the Ion and Electron Sensor \citep[RPC/IES;][]{Burch2007}. EII was found to be a substantial but variable source of cometary electrons over the considered intervals, sometimes dominating over photoionization \citep{Galand2016, Heritier2017Vert, Heritier2018source}. The EII frequency has not been assessed across the whole escort phase. Additionally, the source and drivers of the ionizing electrons are not well known. 

The variation of the energetic electron population throughout the coma is also not well established. \comments{At large heliocentric distances,} multi-instrument {analysis} of far ultraviolet (FUV) emissions is consistent with a constant energetic electron population with cometocentric distance \citep{Chaufray2017}. FUV emissions from the coma of 67P were largely driven by dissociative excitation by electron impact \citep{Galand2020, Stephenson2021FUV}. 
\comments{The models, which assume a constant energetic electron flux along the line-of-sight and an ion bulk speed, $u_i$, equal to the neutral bulk speed, $u_{gas}$, accurately reproduce the electron density and emission brightness through a multi-instrument approach \citep{Galand2016, Galand2020, Heritier2017Vert, Heritier2018source, Stephenson2022TestPl}. }
The strong agreement between modelled and observed FUV emissions during limb observation demonstrates that emissions are driven by electrons accelerated on large scales \citep{Galand2020, Stephenson2021FUV}.

RPC/LAP and RPC/ICA measurements have been used to derive ion flow speeds in excess of 10~km\,s$^{-1}$ within 30~km of the nucleus \edits{at large heliocentric distances} \citep{Nilsson2020, Johansson2021plasDens}. The cometary ions are born at the neutral outflow speed \citep[0.4-1~km\,s$^{-1}$;][]{Marshall2017, Biver2019}. For such high speeds to be achieved at Rosetta, the cometary ions must undergo substantial acceleration in the inner coma, which conflicts with the findings of the  \comments{studies combining multiple instruments and models, where the ions are tied to the neutral gas \citep[$u_i \approx u_n$;][]{beth2022cometary}}. 

In addition to the cases at large heliocentric distances, there have been a number of estimates of the ion velocity close to perihelion, particularly near the diamagnetic cavity \citep{Vigren2017, Odelstad2018, Bergman2021cavSpeeds,Bergman2021CavityFlow}. These show accelerated ions at speeds of 7\,km\,s$^{-1}$ inside the cavity, as well as ions streaming back towards the cavity. The multi-instrument models of plasma density and FUV emissions are only applicable to cases at large heliocentric distances and the weakly collisional regime \citep{Galand2016, Heritier2018source}, and cannot be compared with the perihelion ion speed measurements. \comments{To extend the multi-instrument models to higher outgassing rates, additional processes would need to be included, particularly consideration of ion-neutral chemistry and electron-ion recombination \citep{beth2022cometary}.}

In this paper, we examine the source of the cometary electrons through two methods (see Section~\ref{sec: Methods}). In Section~\ref{sec: data methods}, we calculate the ionization frequency throughout the escort phase of Rosetta, through photoionization and electron-impact ionization, based on in situ measurements. Section~\ref{sec: test pl methods} outlines the collisional test particle model used to model electrons in the cometary environment and the simulation parameters.

The new dataset of electron-impact ionization frequency throughout the Rosetta mission is compared to other measurements of the cometary environment, from RPC and ROSINA, and properties of the spacecraft trajectory in Section~\ref{sec: EII Rosetta}. In Section~\ref{sec: test pl results}, the collisional test particle model is used to examine the origin of electrons in the cometary environment, distinguishing between photoelectrons, solar wind electrons and secondary electrons. Section~\ref{sec: discussion} discusses the measurements of large ion flow speeds (>5~km\,s$^{-1}$) by the RPC/LAP and ICA instruments \citep{Johansson2021plasDens, Nilsson2020} and the difficulty of reconciling these measurements with multi-instrument studies of the electron density and FUV emissions. The results are summarized in Section~\ref{sec: conclusion}.
\section{Methods}\label{sec: Methods}
We address the source of electrons at comet 67P through two methods: {analysis} of Rosetta data (see Section~\ref{sec: data methods}) and the application of a collisional test particle model \citep[Section~\ref{sec: test pl methods};][]{Stephenson2022TestPl}. 
\subsection{Data from Rosetta mission}\label{sec: data methods}

We calculate the ionization frequency from photoionization and electron-impact ionization (EII) throughout the Rosetta mission. Section~\ref{sec: EII freq methods} outlines the calculation of the EII frequency from in-situ measurements.
\subsubsection{Ionization Frequencies}\label{sec: ioni freqs general}

The photoionization frequency of water is given by
\begin{equation}\label{eq: photoionisation freq}
    \nu_{h\nu, \text{H}_2\text{O}}^{ioni} = \int\limits_{\lambda_{min}}^{\lambda_{th}} \sigma_{h\nu, \text{H}_2\text{O}}^{ioni}(\lambda) I(\lambda) \mathrm{d}\lambda.
\end{equation}
where $I(\lambda)$ is the photon flux at 67P.  $\sigma_{h\nu, \text{H}_2\text{O}}^{ioni}(\lambda)$ is the photoionization cross section of water \citep{Vigren2013}. 

The photon flux, $I(\lambda)$, is measured at 1~au by TIMED/SEE \citep[1 day resolution][]{Woods2005}, and then extrapolated to the position of comet 67P. The extrapolation assumes the flux follows $1/r_h^2$, \comments{with heliocentric distance $r_h$,} and includes a time-shift to correct for the solar phase angle \citep{Galand2016, Heritier2018source}. We assume there is no attenuation of the solar flux in the coma, which is valid for $Q< 2 \times10^{27}$~s$^{-1}$. We do not have measurements of the neutral gas column between Rosetta and the Sun, but this is not significant at Rosetta during the mission, \edits{as the cometocentric distance of the spacecraft varied with outgassing} \citep{Heritier2018source, Beth2019}.

Solar wind ion impact ionization and solar wind charge exchange frequencies in the coma were derived from in situ RPC/ICA measurements by \comments{\cite{SimonWedlund2019b}}. These have a resolution up to 192\,s. Close to perihelion, Rosetta was inside the solar wind ion cavity \comments{\citep{Behar2017, Nilsson2017}}, so there are no estimates available for the solar wind ion impact or charge exchange frequencies. 
\begin{table*}
    \centering
    \caption{Time periods considered during the mission \edits{(see Figure~\ref{fig: CPR ioni freq fig}b)}.}
    \label{tab: ioni freq time periods}
    \begin{tabular}{c c c c}
         Period & Start Date & End Date & Heliocentric Distance [au]  \\
         \hline
         Early Mission & 08-Sep-2014 & 05-Feb-2015 & 3.4 - 2.4 \\
         Around perihelion & 10-May-2015 & 02-Dec-2015 & 1.24 - 1.8 \\
         Late Mission & 01-Apr-2016 & 30-Sep-2016 & 2.7 - 3.8 \\
         Pre-perihelion & 08-Sep-2014 & 11-Aug-2015 & 3.4 - 1.24 \\
         Post-perihelion & 11-Aug-2015 & 30-Sep-2016 & 1.24 - 3.8\\
         Pre-excursion & 08-Sep-2014 & 21-Sep-2015 & 3.4 - 1.24 \\
         Post-excursion & 23-Oct-2015 & 30-Sep-2016 & 1.51 - 3.8 \\
    \end{tabular}
    
\end{table*}

\begin{table*}
    \centering
    \caption{Datasets from the Rosetta mission, retrieved from the \comments{European Space Agency's Planetary Science Archive} (PSA), which are compared to the electron-impact ionization frequency. The potential difference between Rosetta and the solar wind, $\Delta U$, is  calculated by \citet{Nilsson2022}. $^{1}$~\citet{Balsiger2007}. $^{2}$~\citet{Trotignon2007}. $^{3}$~\citet{Eriksson2007}. $^{4}$~\citet{Glassmeier2007MAG}. $^{5}$~\citet{Nilsson2007}.}
    \label{tab: ioni freq comp datasets}
    \begin{tabular}{c c c c}
         Name & Equation/Symbol & Units & Instrument  \\
         \hline
         Heliocentric distance & $r_{h}$ & au & - \\
         Cometocentric Distance & $r_{Ros}$ & km & - \\
         Spacecraft Latitude & - & $^\circ$& -\\
         Spacecraft Longitude & - & $^\circ$& -\\
         Neutral Density & $n_{\ce{H2O}}$ & cm$^{-3}$ & ROSINA $^{1}$\\
         Local Outgassing Proxy & $n_{\ce{H2O}} \times r_{Ros}^2 $ & cm$^{-1}$ & ROSINA $^{1}$\\
         Electron Density & $n_e$ & cm$^{-3}$ & RPC/MIP $^{2}$ \& LAP $^{3}$\\
         Spacecraft Potential & $V_{S/C}$ & V & RPC/LAP $^{3}$\\
         Magnetic Field Strength & $B$ & nT & RPC/MAG $^{4}$\\
         Clock Angle & $\arctan(B_y/B_z)$ & ${}^\circ$ & RPC/MAG $^{4}$\\
         Cone Angle & $\arctan( \sqrt{B_y^2 +B_z^2}/B_x )$ & $^\circ$ & RPC/MAG $^{4}$\\
         \ce{H2O^+} Density ($<60$~eV) & $n_{\ce{H2O+}}$ & cm$^{-3}$ & RPC/ICA $^{5}$\\
         SW-Rosetta Potential Diff. & $\Delta U$ & V & RPC/ICA $^{5}$\\

    \end{tabular}

\end{table*}
\subsubsection{Electron impact ionization frequency}\label{sec: EII freq methods}
We calculate the electron-impact ionization frequency throughout the Rosetta mission, from measurements of the energetic electron population RPC/IES \citep{Burch2007}.
RPC/IES measured the electron \comments{differential} particle flux, $J(E)$, from 4.32\,eV up to 17\,keV at the detector, throughout the mission with scans typically taking $\sim2$ mins. In April 2015, half of the anodes degraded, so these have been excluded from the {analysis} throughout the mission. The electron \comments{differential} particle flux, $J(E)$ [cm$^{-2}$s$^{-1}$eV$^{-1}$], is computed as described in \cite{Stephenson2021FUV}, assuming that the electron flux is isotropic. 
\comments{The electron fluxes are calculated using half the anodes of RPC/IES as the other half degraded in April 2015 \citep{Broiles2016Kappa}. While there could be up to a maximum factor of 2 uncertainty arising from the assumption of isotropy, the thermal speed of the energetic electron population is large compared to the bulk solar wind speed throughout the cometary environment, so no significant anisotropy in the electron distribution is expected.}
The flux is also corrected for the persistently negative substantial spacecraft potential ($V_{S/C}<-10$~V) measured by the Langmuir probe, \edits{using Liouville's theorem \comments{(in which the phase space density is conserved along the particle trajectories)}:} 
\begin{equation}
    \frac{J(E)}{E} = \frac{J_{IES}}{E_{IES}} \text{, with } E \, \text{[\si{\electronvolt}]}= E_{IES} \, \text{[\si{\electronvolt}]}- qV_{S/C}\, \text{[\si{\volt}]}.
\end{equation}

\edits{The $IES$ subscript indicates the quantities measured in situ by the instrument, while those unlabelled are the quantities in the local cometary environment. The spacecraft potential} repelled low energy electrons away from the detector and limited the threshold electron energy for detection \comments{(electrons in the coma must have at least $E_{\text{min}} \text{[eV]} = 4.32\,\text{[eV]} - qV_{S/C}$ [V] to be detected for $V_{S/C}<4.32$\,V)}.

The electron particle flux is calculated in 15-minute intervals throughout the mission, \comments{typically containing three or four scans of RPC/IES. After July 2016, 38\% of the 15-minute intervals contain two RPC/IES scans, 12\% contain one scan, with the other 50\% containing three or more.} The electron flux in each interval is extended to energies below the detection limit, assuming a constant electron flux. The extrapolation to low energies does not substantially affect the ionization frequency, as the electron-impact ionization cross sections are small near the threshold ($E_{Th} \approx12~eV$) \citep{Itikawa2005}. 

The electron-impact ionization frequency for collisions with water ($\nu^{\text{ioni}}_{e, \ce{H2O}}$ [s$^{-1}$]) is given by 
\begin{equation}\label{eq: e ioni freq}
\nu^{\text{ioni}}_{e, \ce{H2O}} = \sum\limits_X \int\limits_{E_{Th,X}}^{E_{\text{Max}}} \sigma^{\text{ioni},X}_{e, \ce{H2O}}(E) \: J(E) \mathrm{d}E,
\end{equation}
where the summation is over the different ion products, $X$ (e.g., \ce{H2O+}, \ce{OH+}, \ce{O+}, \comments{\ce{H+}}). $\sigma^{\text{ioni},X}_{e, \ce{H2O}}(E)$ is the electron-impact ionization cross section for electron collisions with water.
\comments{The electron-impact ionization cross-sections for water are based on measurements by \cite{Straub1994}, with a minor update to the instrument calibration  applied by \cite{LindsayMangan2003}, as recommended by \cite{Itikawa2005} and \cite{Song2021}.}

The upper limit, $E_{\text{Max}}=300$~eV, is where the energetic electron flux measured by RPC/IES becomes small \comments{(orders of magnitude less than those measured at $E<100$\,eV), reaching the 1 count per scan level. The retrieved ionization frequencies are not dependent on the upper energy limit of the integration. An increased upper energy limit from 300\,eV to 450\,eV only results in an increase in $\nu_e^{\text{ioni}}$ of $<0.1$\%.}

\subsection{Comparison with other Rosetta datasets}

The ionization frequencies (particularly from EII) are compared to properties of the coma (e.g. electron density) and spacecraft trajectory (e.g. spacecraft latitude) to confirm the source of electrons and identify drivers of EII.
These are listed in Table~\ref{tab: ioni freq comp datasets}, with the relevant instruments for each dataset. The comparison datasets are time-averaged over the same 15-minute intervals as the EII frequency. 

The EII frequency is generally compared to other datasets using 2D histogram. The counts in each column along the $x$-axis are normalised to 1, to attempt to minimize sampling biases in the spacecraft trajectory. Mean, median and quartiles are given for each column.

\subsection{Test particle modelling}\label{sec: test pl methods}
\begin{table}
\centering
\caption{Parameters of the test particle simulations. \comments{$Q$ is the cometary outgassing rate, while $u_{gas}$ is the neutral outflow velocity. The photoionization frequency ($\nu_{h\nu}^{ioni}$) and photoelectron temperature($T_{h\nu, e}$) are also given. $n_{SW}, T_{SW,e}, u_{SW, x}, B_{SW}$ are the density, electron temperature, $x$ component of the bulk velocity and magnetic field strength of the undisturbed upstream solar wind, respectively. $\mathrm{d}x_{EB}$ gives the resolution of the electric and magnetic fields used as an input to the simulation (see Section~\ref{sec: test pl methods}).} \comments{$\mathrm{d}t$ is the timestep used to generate the test particle trajectories. $N_{h\nu, pls}$ and $N_{SW, pls}$ are the number of macroparticles used in the photoelectron and solar wind electron simulations, respectively.}}
\label{tab: test pl parameters}
\begin{tabular}{l l c c}
\hline
Parameter & & Simulation 1  & Simulation 2 \\
\hline
\\
$Q$ &[s$^{-1}$] &  $10^{26}$ & $1.5 \times 10^{27}$ \\
$u_{gas}$ & [km\,s$^{-1}$]  & 1 & 1 \\
$\nu_{h\nu}^{ioni}$ &[s$^{-1}$] & $1.32\times 10^{-7}$ & $1.32\times 10^{-7}$ \\
$T_{h\nu, e}$ &[eV] & 10 & 10 \\
$n_{SW}$ &[cm$^{-3}$] & 1 & 1 \\
$T_{e,SW}$ &[eV] & 10 & 10 \\
$u_{SW, x}$ &[km\,s$^{-1}$] & 400 & 400 \\
$B_{SW,y}$ &[nT] & 6 & 6 \\
$\mathrm{d}x_{EB}$ & [km] & 7.66 & 10 \\
$\mathrm{d}t$ & [s] & $10^{-6}$ & $10^{-6}$ \\
$N_{SW, pls}$ & - & $15\times10^6$ & $3.48\times10^6$\\
$N_{h\nu, pls}$ & - & $4.41\times10^6$ & $1.43\times10^6$ \\
\end{tabular}

\end{table}

In addition to {analysis} of the Rosetta dataset, we have modelled electron impact ionization throughout the coma using a 3D collisional electron test particle model \citep{Stephenson2022TestPl}. The model includes photoelectrons, solar wind electrons and secondary electrons, generated within the coma and at the simulation boundaries. \edits{The $x$-axis of the domain points in the Sun-comet direction, the $y$-axis is aligned with the interplanetary magnetic field, and the $z$-axis completes the set. The extent of the domain is given by $-770\text{~km}< x <1430\text{~km}$ and $-1100\text{~km}< y,z <1100\text{~km}$, with the comet nucleus at the origin.}

The electrons move through the domain, pushed by stationary electric and magnetic fields. The input electric and magnetic fields are taken from a fully-kinetic but collisionless iPiC3D simulation \citep{Deca2017, Deca2019}, run using the same simulation parameters \comments{as these authors} (see Table~\ref{tab: test pl parameters}). The fields in the PiC model are evaluated on a $\mathrm{d}x_{EB} = 7.66$~km grid for the lower outgassing rate and 10~km grid for the higher outgassing rate. The PiC model uses a reduced ion-electron mass ratio ($m_e = m_p/100$) to reduce computation time, whereas the physical electron mass is used in the test particle model \comments{\citep[see discussion of the impact of the electron mass in][]{Stephenson2022TestPl}.}

In our model, the electrons can undergo collisions with molecules in the neutral coma. The neutral coma is pure water and spherically symmetric, with a constant outflow velocity for the neutral gas \citep{Haser1957}. The relevant electron-water collisions across the energy ranges of interest \citep{Itikawa2005} are included in the model. This includes elastic scattering \citep{cho2004measurements, faure2004low} \comments{and inelastic collisions such as} excitations and electron-impact ionization. The collision processes are treated as stochastic processes.

The model outputs are the energy distribution function (edf) of electrons, $f(\bm{x}, E)$ \comments{[cm$^{-3}$eV$^{-1}$]}, for each electron population (photoelectrons, solar wind and secondaries of each). The density of each electron population is given by: 
\begin{equation}\label{eq: test pl el dens}
    n_e(\bm{x}) = \int f(\bm{x}, E)\; \mathrm{d}E
\end{equation}
The EII frequency of each population is given by
\begin{equation}\label{eq: eimpact ioni freq edf}
\nu^{\text{ioni}}_{e}(\bm{x}) = \sum\limits_X \int_{E_{Th, X}}^{E_{\text{Max}}}  \sigma_{e}^{\text{ioni},X}(E) \times f(\bm{x}, E) \times \sqrt{\frac{2E}{m_e}} \: \mathrm{d}E
\end{equation}
The \comments{speed} factor $\sqrt{\frac{2E}{m_e}}$ converts the edf to an electron flux. \comments{As for the measured electron fluxes, there are few electrons with energies exceeding the integration limit and the derived ionization frequencies are not sensitive to $E_{\text{Max}}$ ($<0.1$\% change beyond 300\,eV).}

The secondary electrons produced within the simulation are generated through stochastic collisions, rather than using Eq.~\ref{eq: eimpact ioni freq edf} \citep{Stephenson2022TestPl}.
However, far from the nucleus the frequency of ionization collisions is quite low, so the statistics are better when directly calculating the EII frequency from the edf. Closer to the nucleus, the EII frequency calculated from the edf is the same as produced through stochastic collisions \comments{within 10\%. Up to seven generations of electrons are simulated, with each impact ionization collision produced 10 macroparticles with their total weight equal to that of the ionizing electron \citep{Stephenson2022TestPl}. Beyond seven generations, the weight of each macroparticle is reduced by $10^7$ compared to the particles of the first generation. The total number of macroparticles also falls with generation, so any subsequent generations have negligible impact on the simulation outputs.}

\section{Ionization frequency throughout the Rosetta mission}\label{sec: EII Rosetta}
We first analyse the new dataset of electron-impact ionization frequency throughout the Rosetta mission to determine the source of electrons within the coma and some key drivers of the e-impact ionization frequency. Section~\ref{sec: elec source data} examines the source of the electrons within the coma, while Section~\ref{sec: ionizing elecs data} identifies the origin of the ionizing electrons.

\begin{figure*}
    \centering
    \begin{tikzpicture}
    \node (CPR) at (0,0) {\includegraphics[width=\textwidth]{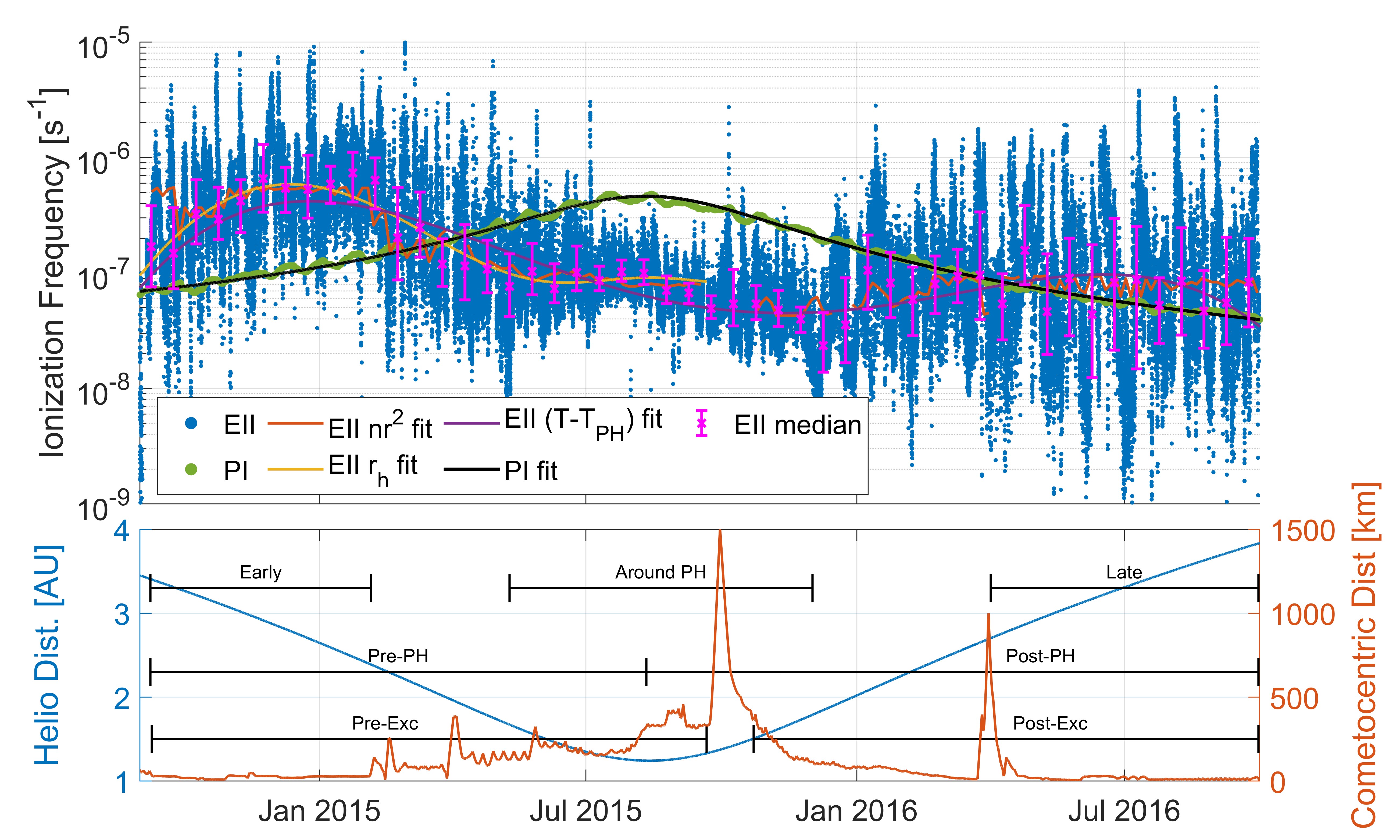}};
    \node [above left = 4.3 and 8 of CPR.center, font = \bf] (a) {(a)};
    \node [below left = 1 and 8 of CPR.center, font = \bf] (b) {(b)};
    \end{tikzpicture} 
    \caption{(a) Ionization processes throughout the Rosetta escort phase. The electron-impact ionization frequency (blue) is calculated in 15-minute intervals, from RPC/IES measurements \citep{Burch2007}. \comments{Median, 25th and 75th percentiles for the EII freqeuncy are given for 15 day periods across the escort phase (pink).} Photoionization frequency (green) is calculated from daily measurements of solar flux at 1au by TIMED/SEE \citep{Woods2005} and then extrapolated to 67P. The electron-impact ionization frequency has been fitted with respect to (orange) local outgassing rate, (yellow) heliocentric distance and (purple) \comments{$(T-T_{PH})$}. The coefficients and goodness of the fits are listed in Table~\ref{tab: ioni freq fits} and the time periods for the fits are outlined in Table~\ref{tab: ioni freq time periods}. The photoionization frequency is fitted using Eq.~\ref{eq: photoioni fit} (black). (b) Heliocentric distance (blue) and cometocentric distance (orange) throughout the Rosetta mission. The time intervals used in the {analysis} and listed in Table~\ref{tab: ioni freq time periods} are shown.}
    \label{fig: CPR ioni freq fig}
    
\end{figure*}

\subsection{Source of electrons in the coma}\label{sec: elec source data}
\subsubsection{Comparison of ionization sources}
Figure~\ref{fig: CPR ioni freq fig} shows the frequency of the two major ionization processes at comet 67P throughout the Rosetta mission, photoionization and electron-impact ionization.

Photoionization (green, Figure~\ref{fig: CPR ioni freq fig}a) is the steadiest process throughout the mission. The photoionization frequency is evaluated at Rosetta, assuming there is no absorption of the flux between the Sun and the spacecraft. The frequency varies roughly with $1/r_h^2$ from $10^{-7}$~s$^{-1}$ at the start of mission to $5.2\times10^{-7}$~s$^{-1}$ on 17 Jul 2015, due to the variation in the photon flux. Shorter timescale oscillations are driven by the solar rotation with a period of $\sim 28$ days. Additionally, the solar activity decreased over the Rosetta mission leading to smaller photon fluxes towards the end of mission. 

The electron impact ionization frequency (blue) fluctuated much more than photoionization. Far from perihelion, the EII frequency varied by three orders of magnitude and was often much higher than the photoionization frequency. Near the end of mission, the EII frequency reached up to $1.4\times10^{-6}$~s$^{-1}$, whereas photoionization frequency was only  $5\times10^{-8}$~s$^{-1}$.
\comments{In the late mission period (see Table~\ref{tab: ioni freq time periods}), the photoionization frequency was $(4-8)\times10^{-8}$\,s$^{-1}$ while the EII frequency varied between $10^{-9}$\,s$^{-1}$ and $4\times10^{-6}$\,s$^{-1}$ with a median of $1.68\times10^{-7}$\,s$^{-1}$}. 
At times, photoionization was still the main source of cometary electrons, as the EII frequency was so \comments{variable}. This is consistent with previous studies of the electron-impact ionization frequency over shorter time intervals \citep{Galand2016, Heritier2017Vert, Heritier2018source}. Towards the start of the escort phase, the minima in EII frequency were comparable to the photoionization frequency at $1.2\times10^{-7}$~s$^{-1}$.

Close to perihelion, the variability in EII frequency was greatly diminished to only a factor of 10 over short intervals. As a result, it was smaller than the photoionization frequency throughout this period (up to $5\times10^{-7}$~s$^{-1}$). This may not be true throughout the coma, as photoabsorption is significant in the inner coma at large outgassing rates \citep[\edits{$r<70$\,km for $Q=3\times 10^{28}$\,s$^{-1}$ at the terminator;}][]{Heritier2018source, Beth2019}.

In addition to photoionization and electron-impact, charge exchange (CX) and solar wind ion impact (SWII) are sources of cometary ions. Throughout the Rosetta mission, the frequencies of CX and SWII were typically both very small compared to the electron impact ionization frequency \citep[$10^{-8}$\,s$^{-1}$ for CX and $10^{-10}$\,s$^{-1}$ for SWII;][]{SimonWedlund2019b}. On short timescales, solar wind charge exchange was a substantial source of cometary ions, but this was infrequent.
Charge exchange does not have a net contribution to the plasma production, as no free electrons are generated. However, it can modify the plasma density through changes in transport and the fields due to mass loading of the solar wind. 

\comments{At large heliocentric distances, EII is often the major source of electrons within the coma, although it is highly variable. When the EII is at a minimum, photoionization is also an important source of electrons away from perihelion. Around perihelion, the EII frequency is much less variable than at large heliocentric distances and was consistently lower than the photoionization frequency at Rosetta. The median EII frequencies (pink, Figure~\ref{fig: CPR ioni freq fig}a) are similar between perihelion and towards the end of mission, but the peaks in the EII frequency were much larger towards the end of mission.}
At large heliocentric distances, EII is the major source of electrons within the coma, although it is highly variable. When the EII is at a minimum, photoionization is also an important source of electrons away from perihelion. Around perihelion, the EII frequency is smaller than found at large heliocentric distances and is weaker than photoionization. 

Fits to the median values of the photoionization frequency and electron-impact ionization frequency are shown in Figure~\ref{fig: CPR ioni freq fig}a. The photoionization frequency at 1au decreased approximately linearly over the Rosetta mission, due to the decrease in solar activity \citep{Heritier2018source}. The approximate photoionization frequency (in s$^{-1}$) at Rosetta (black, Figure~\ref{fig: CPR ioni freq fig}a) is given by:
\begin{equation}\label{eq: photoioni fit}
    \nu_{h\nu}^{\text{ioni}} =  \frac{1}{r_h^2}(p_0 + p_1 (T-T_{PH}))
\end{equation}
where $T-T_{PH}$ is the number of days since perihelion, $r_h$ is in au, $p_0 =7.18\times10^{-7}$\,s$^{-1}$\,au$^2$ and $p_1=-3.23\times10^{-10}$\,s$^{-1}$au$^2$\,days$^{-1}$.

The electron impact ionization frequency has been fitted using several variables: heliocentric distance ($r_h$ [au]), local outgassing rate ($nr^2$ [cm$^{-1}$]) and days since perihelion. The fits with heliocentric distance and outgassing are split into two periods around the excursion in September 2015 (see Table~\ref{tab: ioni freq time periods} \comments{and red, Fig.~\ref{fig: CPR ioni freq fig}a}), as the EII frequency behaved differently through these periods (see Sections~\ref{sec: eii vs ogr}). After the excursion, heliocentric distance was not a good predictor of the EII frequency, so the fit over this period is not shown in Figure~\ref{fig: CPR ioni freq fig}a. The median EII frequency is much more \comments{variable} after the excursion, which may partially be driven by the arrival of corotating interaction regions at 67P from Jun-Sep 2016 \citep{Edberg2016CIR, Hajra2018CIRs, Stephenson2021FUV}.

The \comments{dependence} between the variables $x = r_h$ \comments{(yellow, Fig.~\ref{fig: CPR ioni freq fig}a)}, $nr^2$ \comments{(orange)}, $(T-T_{PH})$ \comments{(purple)} and the EII frequency are given by:
\begin{equation}\label{eq: EII fit func}
    \log_{10}(\nu_e^{\text{ioni}}) = \sum\limits_n p_n x^n,
\end{equation}
where the polynomial coefficients and adjusted-$R^2$ values are listed in Table~\ref{tab: ioni freq fits}. \comments{The plotted fit for $x=nr^2$ has been calculated using the median local outgassing rates throughout the mission, with the pre and post-excursion periods (see Table~\ref{tab: ioni freq time periods}) split into 100 equal parts of 90 and 82 hours respectively.}
 
The \comments{dependence} of the EII frequency on the local outgassing rate (orange, Figure~\ref{fig: CPR ioni freq fig}) is discussed in Section~\ref{sec: eii vs ogr} and seen in Figure~\ref{fig: ioni freq vs. outgassing inb vs outb}.
\begin{table*}
    \centering
    \caption{Coefficients, $p_n$, of fits to the median electron-impact ionization frequencies throughout the Rosetta mission (see Eq.~\ref{eq: EII fit func}). The independent variable, $x$, for each fit is given in addition to the adjusted-R-squared value for each fit. The time periods for each fit are outlined in Table~\ref{tab: ioni freq time periods}. Each of the fits is shown in Figure\,\ref{fig: CPR ioni freq fig}a. The fit to the local outgassing rate is shown in Figure~\ref{fig: ioni freq vs. outgassing inb vs outb} and the pre-excursion fit to the heliocentirc distance is shown in Figure\,\ref{fig: EII vs rh}a.}
    \label{tab: ioni freq fits}
    \begin{tabular}{c c c c c c c c }
         Variable & Period & $p_0$ & $p_1$ & $p_2$ & $p_3$ & $p_4$ & Adj-R-sq \\
         \hline
         $r_h$ [au]& Pre-Exc &-2.4918 &  -7.8105	&4.16 & -0.6599 & -	& 0.65 \\
         $\log_{10}$($nr^2$ [cm$^{-1}$]) & Pre-Exc & -2127 & 744.6 & -97.62 & 5.665 & -0.1228 & 0.922 \\
         $\log_{10}$($nr^2$ [cm$^{-1}$]) & Post-Exc & 540.3 & -225.3 & 34.28 & -2.29 & 0.05674 & 0.6399 \\
         $T-T_{PH}$ [days] & All & -7.1690 & $-3.34\times10^{-3}$ & $1.36\times10^{-5}$ &  $3.14\times10^{-8}$ &  $-1.14\times10^{-10}$ & 0.56
         
    \end{tabular}
    
\end{table*}

\subsubsection{EII vs electron density}\label{sec: eii vs el dens}
In order to confirm that EII is the major source of electrons away from perihelion, we compare the EII frequency with the electron density at Rosetta, derived from the cross-calibration between RPC/MIP and LAP \citep{Johansson2021plasDens}. The electron and ion density are dependent on cometocentric distance, with $n_e\propto\frac{r-r_{67P}}{r^2}$ when transport is significant \citep{Galand2016, Heritier2017Vert, Heritier2018source, beth2022cometary}. \edits{For the sake of simplicity, we assume a spherical comet with an equivalent radius of} $r_{67P}=1.7$~km \citep{Jorda2016}. \edits{In order to remove the dependence of Rosetta's distance to the nucleus, where $n_e$ is measured,} the electron density is corrected. \comments{For an ionization frequency (and gas outflow speed, $u_{gas}$) independent of cometocentric distance the corrected electron density is given by:}
\begin{equation}
    n_{e,corr} = n_e \times\frac{r^2}{r-r_{67P}} = \frac{Q\nu^{ioni}}{4\pi u_{gas} u_i},
\end{equation}
\comments{where $u_i$ is the bulk radial ion speed \citep{Galand2016, Heritier2018source, beth2022cometary}. This has the same unit as a column density.} The correction to the electron density does not significantly change the correlation early and late in the mission as the bulk of the measurements is taken over a narrow range of cometocentric distances (see Figure~\ref{fig: CPR ioni freq fig}b).

\begin{figure*}
    \centering
\begin{tikzpicture}
       \node (eIoni) at (-2,0) {\includegraphics[width =\textwidth]{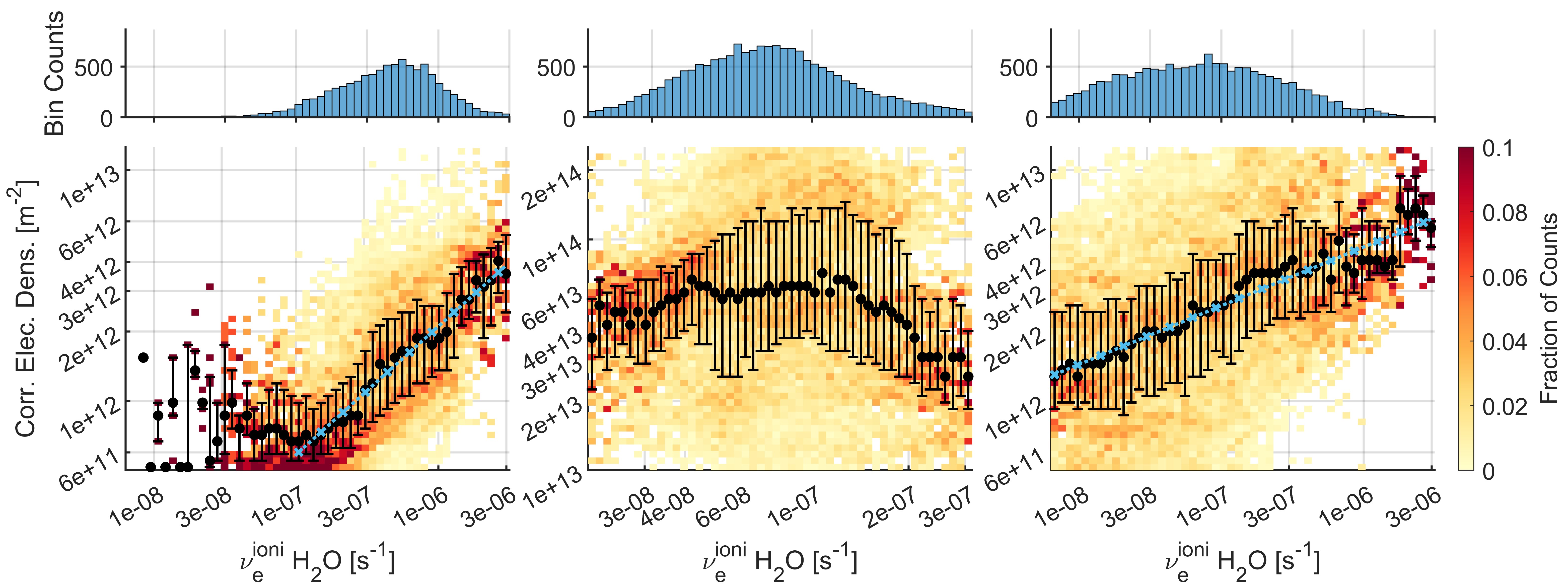}};
       
        \node [above left = 2.8 and 4.1 of eIoni.center, font = \bf] (EarlyLab) {Early Mission};
        \node [above = 2.8 of eIoni.center, font = \bf] (PHLab) {Around PH};
        \node [above right = 2.8 and 4.1 of eIoni.center, font = \bf] (LateLab) {Late Mission};
        \node [above left = 6.3 and 7.3 of eIoni.base, font =\bf] (A) {(a)};
        \node [above left = 6.3 and 2.1 of eIoni.base, font =\bf] (B) {(b)};
        \node [above right =6.3 and 1.9 of eIoni.base, font =\bf] (C) {(c)};
        \node [above left = 3.6 and 3.0 of eIoni.base, align=left, font =\bf] (R2a) {$\begin{aligned}\log_{10}\big(\frac{y}{\mathrm{m}^{-2}}\big)& = \\0.555&\log_{10}\big(\frac{x}{\mathrm{s}^{-1}}\big) +15.65\end{aligned}$\\$R^2=0.978$};
        
        \node [above right = 1.3 and 3.9 of eIoni.base, align=right, font =\bf] (R2a) {$\begin{aligned}\log_{10}\big(\frac{y}{\mathrm{m}^{-2}}\big) =\\ 0.254\log_{10}\big(\frac{x}{\mathrm{s}^{-1}}\big)+14.19\end{aligned}$\\$R^2=0.936$};

\end{tikzpicture}

    \caption{(a-c) Histogram of the cross-calibrated electron density measured at Rosetta by RPC/MIP and LAP \citep{Trotignon2007, Eriksson2007, Johansson2021plasDens}, corrected for cometocentric distance \citep[assuming $n_e\propto1/r$,][]{Heritier2017Vert}, against electron-impact ionization frequency. Median (black circles) and quartile values (error bars) are plotted for each bin along $x$. \comments{Linear fits are given for the early and late periods (blue crosses). The corrected electron density has the same units as a column density.}}
    \label{fig: ioni freq vs el dens}
\end{figure*}

Figure~\ref{fig: ioni freq vs el dens} shows the relationship between the corrected electron density and electron impact ionization frequency. The corrected electron density is positively correlated with EII frequency in the early and late mission periods, when the EII frequency is not small \comments{compared to the photoionization frequency} (median values have $R^2=0.978$ for $\nu_e^{\text{ioni}}>10^{-7}$~s$^{-1}$ for the early period) When the EII frequency is \comments{relatively} small, the electron density does not vary significantly with EII frequency. This occurs when the photoionization frequency is comparable to or larger than the EII frequency, so most cometary electrons are photoelectrons. This occurs at $\nu_e^{\text{ioni}}<1.5 \times10^{-7}$~s$^{-1}$ in the early mission period, \comments{although the statistics are smaller over this range (only 550 intervals with $\nu_e^{\text{ioni}}<10^{-7}$\,s$^{-1}$ out of 8926 in Figure \ref{fig: ioni freq vs el dens}a).}

The corrected electron density is well correlated to the total ionization frequency (photoionization and EII) across the whole range of frequencies, with the lower limit being the photoionization frequency. 
Around perihelion, the \comments{median} corrected electron density shows no significant dependence on the EII frequency. Electron-impact ionization is \comments{significantly} weaker than photoionization around perihelion, so it is not the key source of cometary electrons over the period (see Figure~\ref{fig: CPR ioni freq fig}). The comparison to the total ionization frequency is not useful around perihelion, as it is dominated by photoionization and does not vary substantially. Other sources of variation in the coma are more important for the electron density around perihelion, such as the outgassing rate. 

The EII frequency is also strongly correlated to the spacecraft potential early and late in the mission, with more EII leading to more negative potentials (see Appendix~\ref{sec: EII vs VSC}). \comments{A hotter and denser electron population drives more negative spacecraft potentials \citep{Odelstad2015b, Johansson2021plasDens}, while containing more energetic electrons that can drive impact ionization \citep[e.g.][]{Cravens1987ImpactIoni}}. Around perihelion, the spacecraft potential was not strongly dependent on EII frequency, supporting the conclusion that EII was not the major source of electrons close to the Sun. 

\subsubsection{EII vs water ion density}\label{sec: eii vs ion dens}
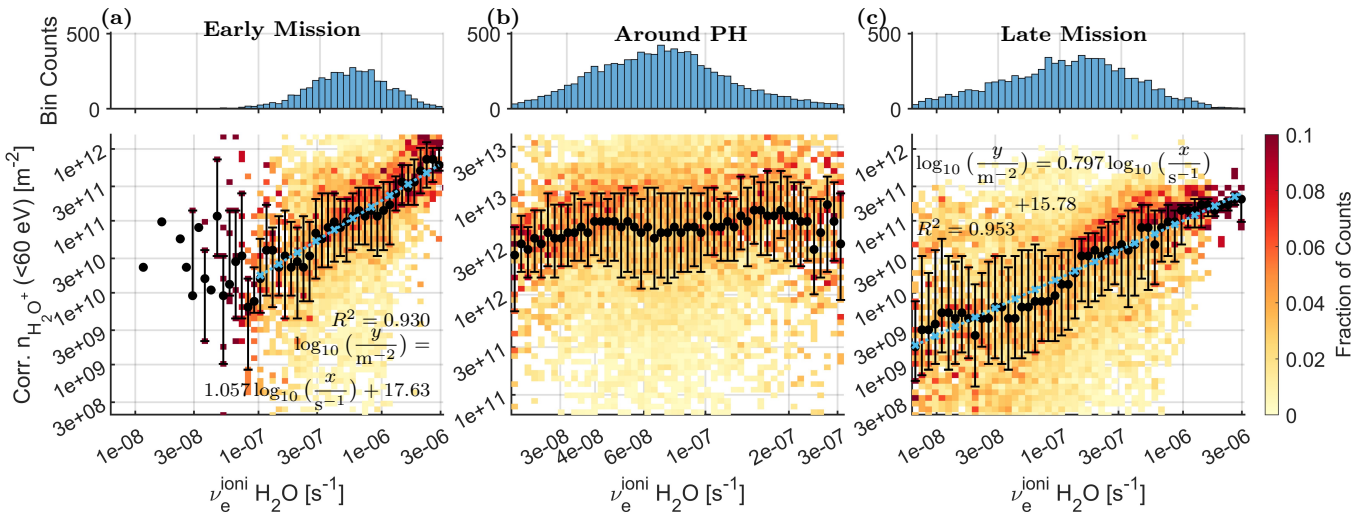
\begin{figure*}
    \centering
\begin{tikzpicture}
       
        \node (H2OIon) at (0,0) {\includegraphics[width = \textwidth]{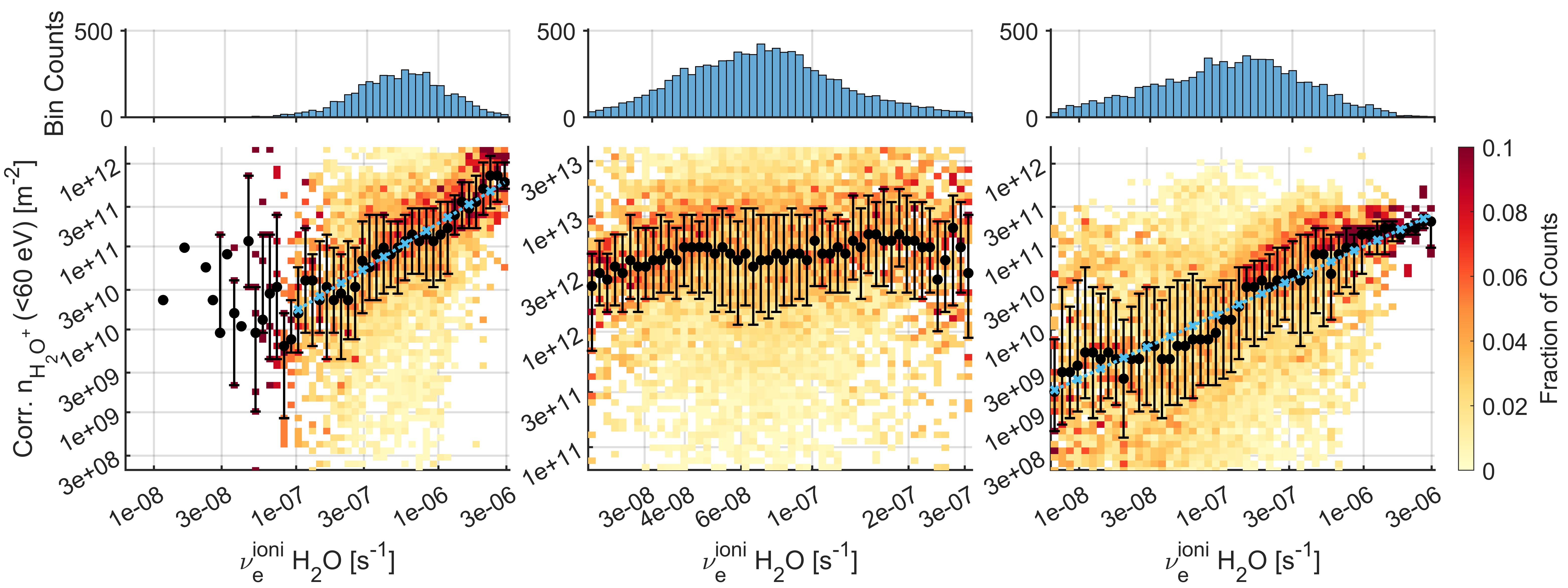}};

        \node [above left = 2.8 and 4.1 of H2OIon.center, font = \bf] (EarlyLab) {Early Mission};
        \node [above = 2.8 of H2OIon.center, font = \bf] (PHLab) {Around PH};
        \node [above right = 2.8 and 4.1 of H2OIon.center, font = \bf] (LateLab) {Late Mission};
        \node [above left = 6.3 and 7.1 of H2OIon.base, font =\bf] (A) {(a)};
        \node [above left = 6.3 and 2.0 of H2OIon.base, font =\bf] (B) {(b)};
        \node [above right =6.3 and 2.2 of H2OIon.base, font =\bf] (C) {(c)};
       
        \node [above left = 1.3 and 1.2 of eIoni.base, align=right, font =\bf, scale=0.9] (R2a) {$R^2=0.930$\\$\begin{aligned}\log_{10}\big(\frac{y}{\mathrm{m}^{-2}}\big) =\\ 1.057\log_{10}\big(\frac{x}{\mathrm{s}^{-1}}\big)+17.63\end{aligned}$};
        
        \node [above right = 3.6 and 5.0 of eIoni.base, align=left, font =\bf, scale =0.9] (R2a) {$\begin{aligned}\log_{10}\big(\frac{y}{\mathrm{m}^{-2}}\big) & = 0.797\log_{10}\big(\frac{x}{\mathrm{s}^{-1}}\big)\\+&15.78\end{aligned}$\\$R^2=0.953$};

\end{tikzpicture}

    \caption{Electron-impact ionization frequency against low-energy cometary water ion density (a) early in the mission, (b) around perihelion and (c) late in the mission. Median (black circles) and quartile values (error bars) are plotted for each bin along $x$. \comments{Linear fits are given for the early and late periods (blue crosses).}}
    \label{fig: ioni freq vs Low energy ion Pop}
\end{figure*}

Figure~\ref{fig: ioni freq vs Low energy ion Pop} shows the correlation between the EII frequency and the low energy water ion density, derived from moments of the RPC/ICA data \citep{Nilsson2007}. The density estimate includes contributions from all ions near 18 amu, such as \ce{H2O+}, \ce{H3O+} and \ce{NH4+}. \comments{The RPC/ICA measurements of the low energy cometary ions were less frequently available than the RPC/LAP-MIP electron density or RPC/IES electron flux measurements, with 4113, 9598 and 8065 intervals included in Figures\,\ref{fig: ioni freq vs Low energy ion Pop}a, b and c, respectively (compared to 8926, 17746 and 15000 intervals in the corresponding plots of Figure~\ref{fig: ioni freq vs el dens}).}The moments of the RPC/ICA data are 1 to 1.5 orders of magnitude smaller than the electron densities measured concurrently by the RPC/LAP and RPC/MIP instruments. Therefore, RPC/ICA must be missing a large part of the cometary ion population. \comments{This may be a result of the limited field of view of the instrument, or a consequence of the strong deflection of low energy cometary ions by the spacecraft potential \citep{Bergman2020VSC}.} However, we are interested in the variation rather than the magnitude of the ion population. The density calculation also neglects the deflection of the cometary ions near the detector \citep{Bergman2020VSC}. 

As with the electron density, the low energy cometary ion density, \comments{$n_i$}, is normalized for the cometocentric distance by using $n_{i,corr} = n_{i} \times \frac{r^2}{r-r_{67P}}$. The corrected low-energy cometary ion density is well correlated to the EII frequency, both early and late in the mission. Far from perihelion, an increasing EII frequency is associated with higher cometary ion densities. Late in the mission, the cometary water ion density is fairly constant with EII frequency when the EII frequency is small ($<3\times10^{-8}$~s$^{-1}$). At larger ionization frequencies, there is a clear positive correlation \comments{(median values have $R=0.9645$, and $R=0.9760$ early and late in the mission, respectively)}. Around perihelion, the density of cometary ions is not strongly dependent on the EII frequency \comments{($R=-0.41$)}.

This supports the conclusions in Section~\ref{sec: eii vs el dens}, that ionization within the coma is primarily driven by EII when 67P was far from perihelion. The bulk of ions and electrons in the coma are  produced by EII, \comments{as expected from Figure~\ref{fig: CPR ioni freq fig}}. Around perihelion, EII is a weaker process and less ionizing than photoionization, so the plasma density in the coma is not strongly dependent on the EII frequency. 
\subsection{Drivers of electron-impact ionization}\label{sec: ionizing elecs data}
Having confirmed that EII is the major source of cometary electrons at large heliocentric distances, we next identify some key drivers of the ionizing population of electrons.
\subsubsection{EII vs magnetic field strength}\label{sec: eii vs B field}
\begin{figure*}
    \centering
\begin{tikzpicture}
        \node (BMag) at (0,0) {\includegraphics[width = \textwidth]{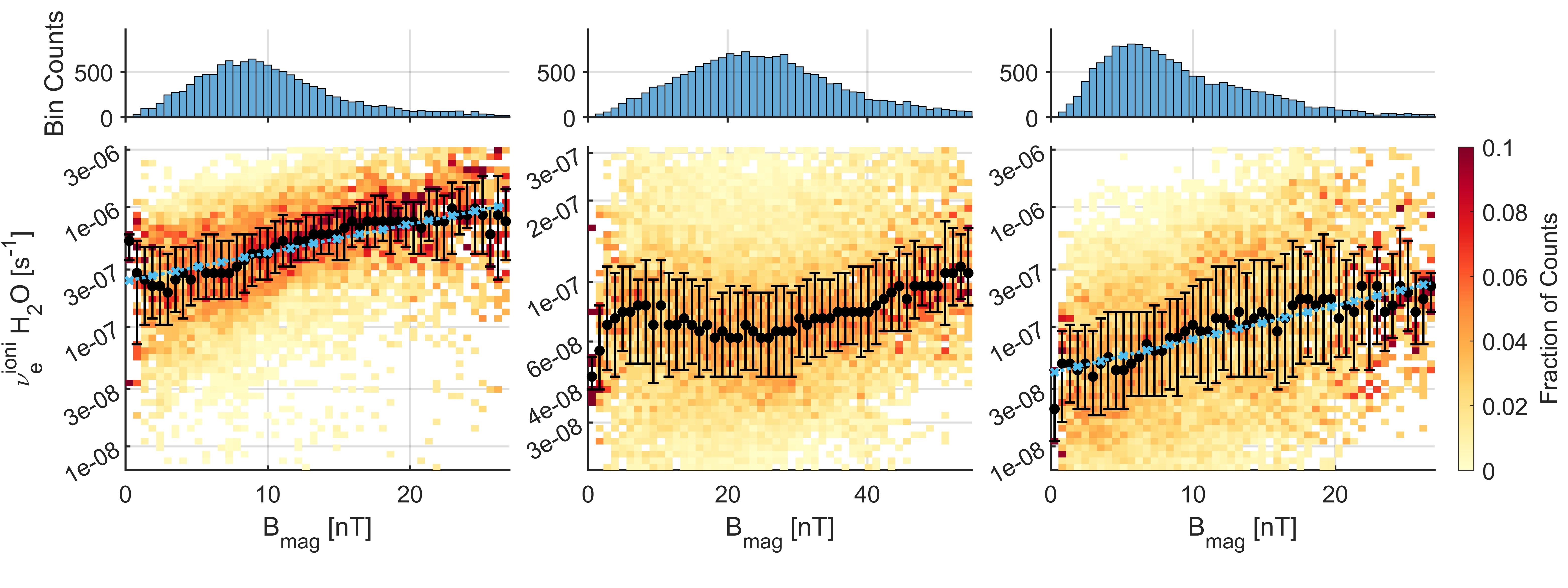}};
        
        \node [above left = 6.2 and 4.1 of BMag.base, font = \bf] (EarlyLab) {Early Mission};
        \node [above = 6.2 of BMag.base, font = \bf] (PHLab) {Around PH};
        \node [above right = 6.2 and 4.1  of BMag.base, font = \bf] (LateLab) {Late Mission};
        \node [above left = 6.2 and 7.1 of BMag.base, font =\bf] (A) {(a)};
        \node [above left = 6.2 and 2.0 of BMag.base, font =\bf] (B) {(b)};
        \node [above right =6.2 and 2.2 of BMag.base, font =\bf] (C) {(c)};
        \node [above left = 1.3 and 1.3 of eIoni.base, align=left, font =\bf] (R2a) {$\log_{10}(\frac{y}{\mathrm{s}^{-1}}) = 0.0238\frac{x}{\mathrm{nT}} -6.62$\\$R^2=0.825$};
        \node [above right = 4.2 and 5.0 of eIoni.base, align=left, font =\bf] (R2c) {$\log_{10}(\frac{y}{\mathrm{s}^{-1}}) = 0.0281\frac{x}{\mathrm{nT}} -7.39$\\$R^2=0.844$};

\end{tikzpicture}

    \caption{Electron-impact ionization frequency vs. magnetic field magnitude (a) early in the mission, (b) around perihelion and (c) late in the mission. Median (black circles) and quartile values (error bars) are plotted for each bin along the $x$-axis. \comments{Linear fits are given for the early and late periods (blue crosses).}}
    \label{fig: ioni freq vs B Mag}
\end{figure*}
Figure~\ref{fig: ioni freq vs B Mag} shows the relation between the EII frequency and magnetic field strength. Early and late in the mission, the ionization frequency increases with increasing magnetic field strength, \comments{with a steeper gradient late in the mission (0.0238 in the early period vs. 0.0281 in the late period).}
The correlation holds well for weak magnetic fields  (\comments{$<17$~nT in the early period, and $<12$\,nT for the late period}), but plateaus at $\nu^{\text{ioni}}_{e, \text{H}_2\text{O}} = 10^{-7}$~s$^{-1}$ in the late mission period. However, at the higher field strengths, \comments{there are far fewer data points with less than 100 per bin beyond  $B=20$~nT, where the median EII frequencies show more variability.} 

Around perihelion, there is no significant variation of the EII frequency with magnetic field strength (Figure~\ref{fig: ioni freq vs B Mag}b). \comments{The median EII frequency only varied within a factor of 2 between $B=0$\,nT and $B=50$\,nT, whereas the EII frequency increased by a factor of 5 between $B=0$\,nT and $B=25$\,nT at large heliocentric distances.} The EII frequency decreases slightly with increasing magnetic field strength up to 25\,nT, and then increases.  

The EII frequency varies non-linearly with the clock and cone angles of the magnetic field, both early and late in the mission. \comments{However, the non-linear relationship with the orientation of the field likely reflects the correlation between the angles and magnetic field strength (see Appendix\,\ref{sec: EII vs mag angles})}.

The relationship with magnetic field strength early and late in the mission may indicate that EII is strongest in regions where pile up is substantial. The pile-up of the solar wind magnetic field is also associated with an increase in the solar wind electron density. It is found deeper in the coma, where the ambipolar potential well is \comments{more} likely to be deep. This indicates that solar wind electrons, accelerated by the ambipolar field, are piling up with the magnetic field and driving ionization. Additionally, magnetic mirroring can lead to an increase in the perpendicular electron energy in regions of high magnetic field. A large perpendicular energy, compared to parallel energy,  leads to more efficient electron trapping within the potential well and therefore \comments{the trapped electrons spend} more time in a region where ionization can occur.

\subsubsection{EII vs. potential well}\label{sec: eii vs pot well}
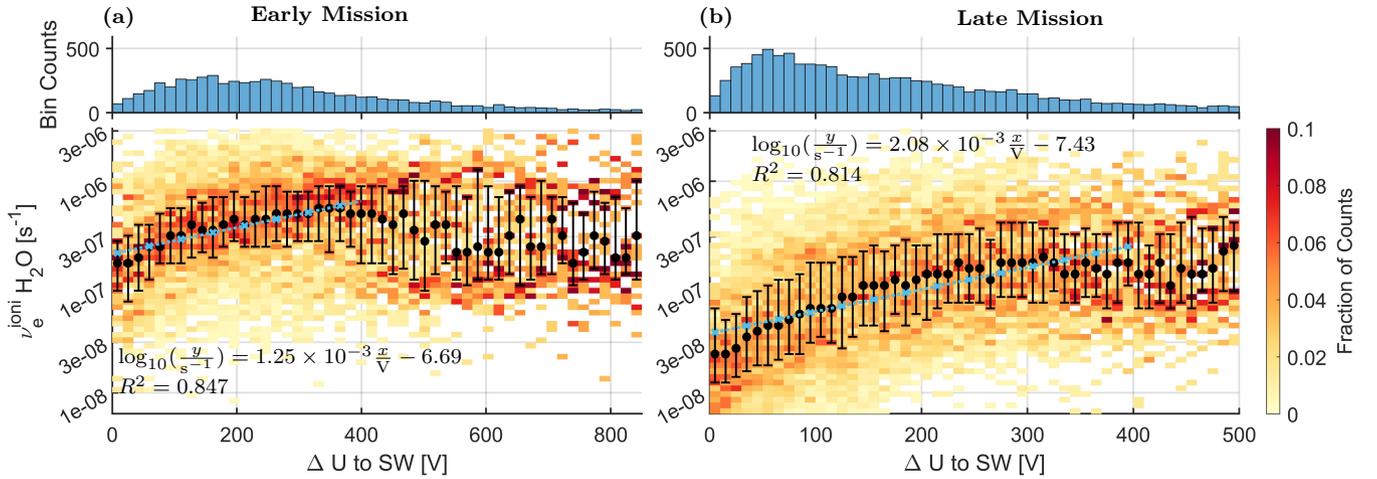
\begin{figure*}
\begin{tikzpicture}

\node (USW) at (0,0) {\includegraphics[width = \textwidth]{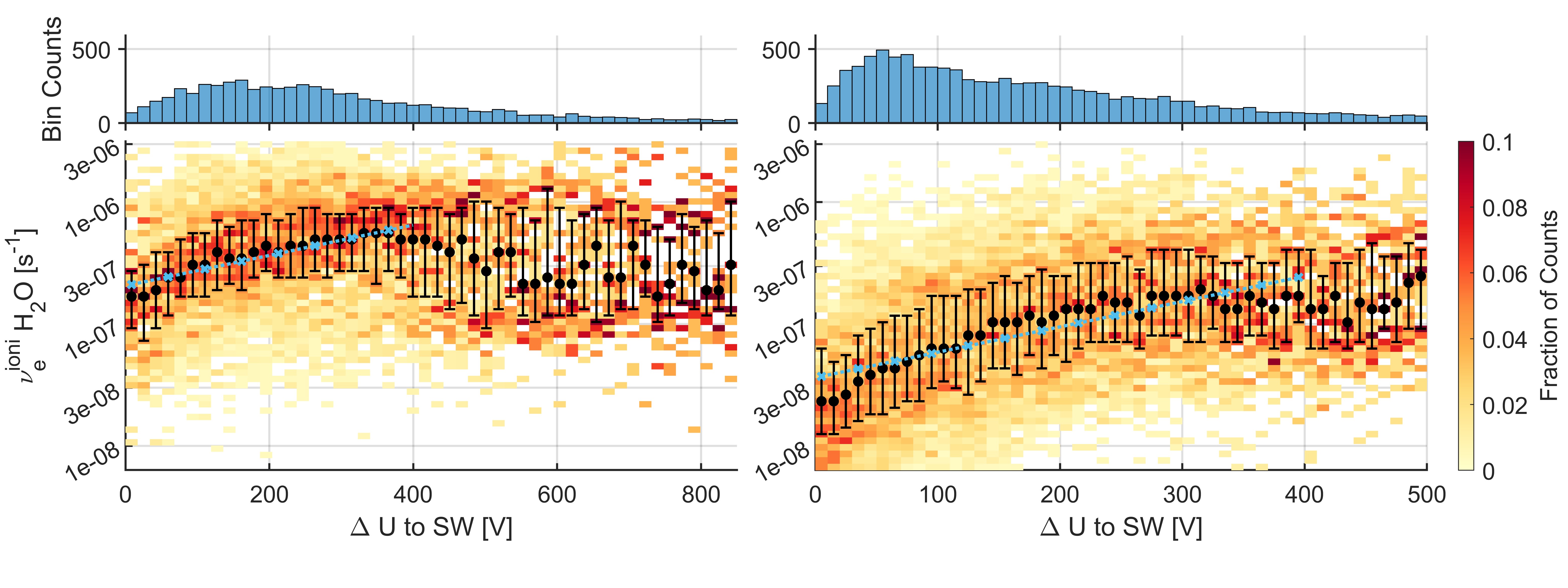}};
 \node [above left = 3.0 and 3.5 of USW.center, font = \bf] (USWLab) {Early Mission};
    \node [above right = 3.0 and 3.5 of USW.center, font = \bf] (OutBLab) {Late Mission};
    \node [above left = 6.3 and 7.1 of USW.base, font =\bf] (A) {(a)};
    \node [above right = 6.3 and 0.1 of USW.base, font =\bf] (B) {(b)};
    \node [above left = 1.5 and 0.8 of eIoni.base, align=left, font =\bf] (R2c) {$\log_{10}(\frac{y}{\mathrm{s}^{-1}}) = 1.25\times 10^{-3}\frac{x}{\mathrm{V}} -6.69$\\$R^2=0.847$};
    \node [above right = 4.3 and 2.8 of eIoni.base, align=left, font =\bf] (R2c) {$\log_{10}(\frac{y}{\mathrm{s}^{-1}}) = 2.08\times 10^{-3}\frac{x}{\mathrm{V}} -7.43$\\$R^2=0.814$};
\end{tikzpicture}
\caption{EII frequency against SW-Rosetta potential difference (a) early in the mission and (b) late in the mission. Median (black circles) and quartile values (error bars) are plotted for each bin along the $x$-axis. \comments{Linear fits are given for the early and late periods (blue crosses).}}
\label{fig: EII vs potwell}
\end{figure*}

Figure~\ref{fig: EII vs potwell} shows the correlation between the Rosetta-upstream solar wind potential difference ($\Delta U$) and the EII frequency. The electric potential difference between the observation point and the upstream solar wind is derived from in situ measurements of the solar wind ions (\ce{H+}, \ce{He^{++}}) by RPC/ICA \citep{Nilsson2007, Nilsson2022}.

Both early and late in the Rosetta mission, the EII is very well correlated to the solar wind potential difference \comments{at low outgassing rates}. Early in the mission, the median EII frequency increased from $2.5\times10^{-7}$~s$^{-1}$ at 50~V to $8\times10^{-7}$~s$^{-1}$ at 300 V. Late in the mission, the EII frequency increased from $3\times10^{-8}$~s$^{-1}$ to $2\times10^{-7}$~s$^{-1}$ between 40 and 250~V. \comments{At large values of $\Delta U$ ($>500$\,V in the early mission and $>400$\,V in the late period) the median values show much more variability and the positive correlation seems to break down, but the statistics here are weaker with fewer than 30 points per bin.}

It was not possible to estimate the potential difference to the solar wind around perihelion, as Rosetta was in the solar wind ion cavity \citep{Nilsson2017}.

The solar wind potential difference is a local measurement of the ambipolar potential well that formed around 67P. As the EII frequency is well correlated to the potential difference, it is very likely that the ionizing electron population are solar wind electrons that have been accelerated by the ambipolar field.



\subsubsection{EII vs outgassing rate}\label{sec: eii vs ogr}
\begin{figure*}
    \centering
    \begin{tikzpicture}
    \node (InB) at (0,0) {\includegraphics[width = \textwidth]{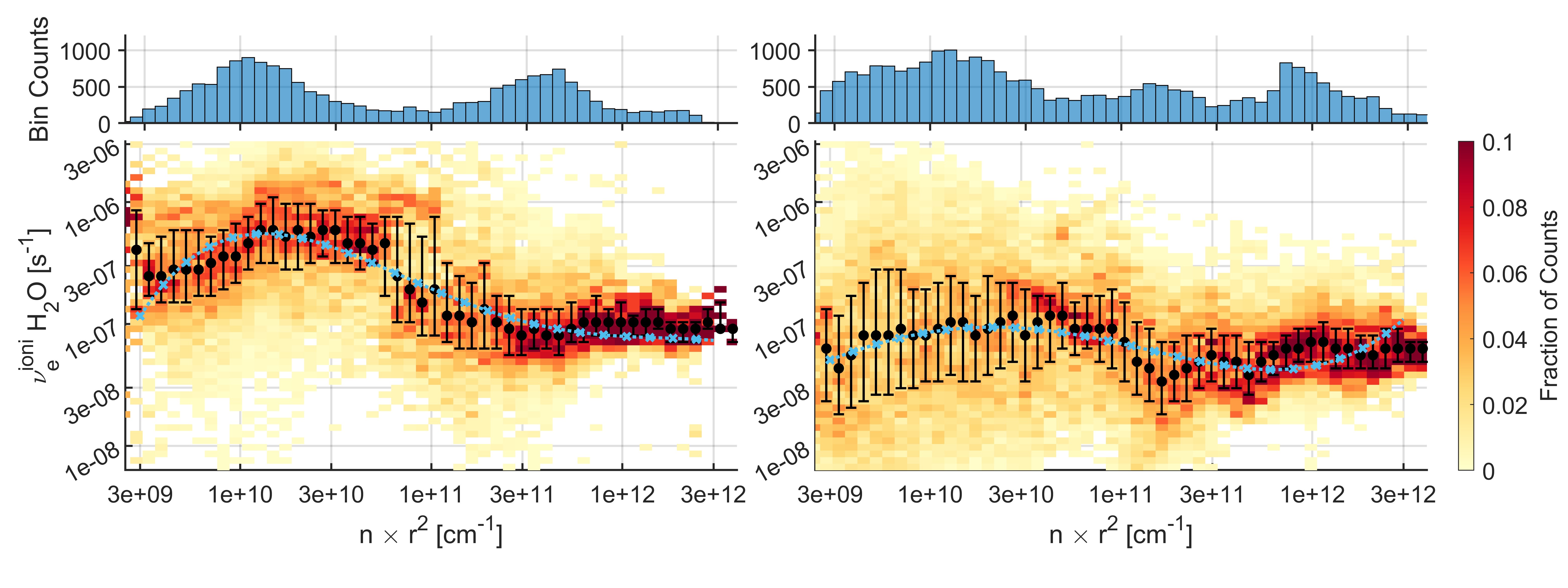}};

    \node [above left = 3.0 and 3.5 of InB.center, font = \bf] (InBLab) {Pre-PH};
    \node [above right = 3.0 and 3.5 of InB.center, font = \bf] (OutBLab) {Post-PH};
    \node [above left = 6.3 and 7.1 of InB.base, font =\bf] (A) {(a)};
    \node [above right = 6.3 and 0.1 of InB.base, font =\bf] (B) {(b)};
    
    \end{tikzpicture}
    \caption{Electron-impact ionization frequency against a proxy measurement of the outgassing rate \citep[$n\times r^2$, where $n$ is the neutral density measured by ROSINA][]{Balsiger2007} for the (a) pre-perihelion and (b) post-perihelion phases. Median (black circles) and quartile values (error bars) are plotted for each bin along the $x$-axis. The fit to the median EII frequency as a function of local outgassing rate is shown in each period in blue crosses (see Table~\ref{tab: ioni freq fits}).}
    \label{fig: ioni freq vs. outgassing inb vs outb}
\end{figure*}
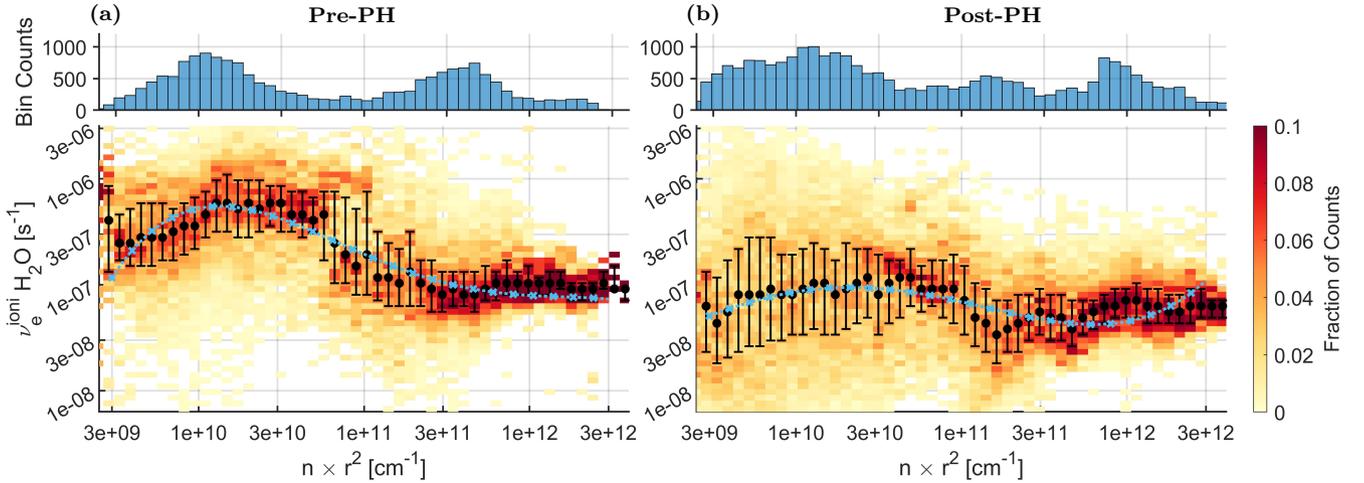
Figure~\ref{fig: ioni freq vs. outgassing inb vs outb} shows the variation of the EII frequency with a proxy of the \comments{local} outgassing rate, $n\times r^2$. The neutral density, $n$, is measured locally by ROSINA/COPS, with a correction for the composition using ROSINA/DFMS \citep{Balsiger2007, Gasc2017MNRAS}. This neglects variation of the neutral gas outflow velocity, which varied between 400 m\;s$^{-1}$ and 900 m\;s$^{-1}$ throughout the escort phase \citep{Biver2019}. \comments{Both pre- and post-perihelion, the EII frequency has a non-linear relationship (see Figure~\ref{fig: ioni freq vs. outgassing inb vs outb}) with the proxy for the local outgassing rate and shows little variability around the wider trend (especially in comparison to the relationship with heliocentric distance, Section~\ref{sec: EII vs rh}).} At low outgassing rates ($n\times r^2 < 3 \times 10^{10}$~cm$^{-1}$), the ionization frequency increases with outgassing rate. As the outgassing increases further, the ionization frequency remains fairly constant before beginning to decrease. The EII frequency then falls until $n\times r^2 = 3 \times 10^{11}$~cm$^{-1}$, after which the ionization frequency remains constant or shows a marginal increase.

Post perihelion, the magnitude of the ionization frequency is smaller than early in the mission, particularly at large heliocentric distances (see Figure~\ref{fig: CPR ioni freq fig}). This may be driven by the decreasing solar activity throughout the mission, \comments{which results in a reduction in the EUV flux for a given heliocentric distance \citep{Lean2003, Woods2002, Woods2012}. The EUV flux drives photoionization in the coma which is critical for generating the ambipolar potential well, into which solar wind electrons can be accelerated.}

The variation of the ionization frequency with outgassing may be indicative of the evolution of the ambipolar potential well throughout the mission. Far from perihelion, the coma is very weakly collisional and increased outgassing produces more electrons \comments{locally} and strengthens the electron pressure gradient in the coma \comments{($\bm{E}_{Ambi} = -\frac{1}{e n_e}\nabla(n_e k_B T_e)$)}. This increases the ambipolar field strength and causes increased acceleration of solar wind electrons into the coma. As the coma becomes denser, it also becomes increasingly collisional. The electron-neutral collisions, in conjunction with trapping in the ambipolar field, become strong enough to cool electrons within the inner coma \citep{Stephenson2022TestPl}. This weakens the pressure gradient in the inner coma and the ambipolar field strength. In this way, negative feedback from the cooling on the potential well is established \citep{Stephenson2022TestPl}. This also results in less acceleration and funnelling of the solar wind electrons into the inner coma. \edits{The feedback of collisions onto the ambipolar field is discussed further in Section~\ref{sec: discussion}.}

At higher outgassing rates, the electron cooling and feedback become stronger, further damping the ambipolar potential well. Close to perihelion, the ambipolar field is effectively quenched in the inner coma by the electron-neutral collisional cooling. 

A similar relationship is also observed between the EII frequency and heliocentric distance (see Section~\ref{sec: EII vs rh}), but with much more variability, especially post-perihelion. This may be caused by the variation with cometocentric distance which is unaccounted for or by the outgassing variation across the surface of the comet, especially between hemispheres. 

In the early and late mission periods (see Table~\ref{tab: ioni freq time periods}), the correlation with outgassing rate is also observed in the relation with spacecraft latitude. In the early period, the EII frequency increases with latitude from the southern to the northern hemisphere, where the outgassing was stronger \citep{Hansen2016, Fougere2016, Laeuter2018}. Late in the mission, the strongly outgassing southern latitudes were associated with the highest EII frequencies. Both the early and late periods have outgassing rates in the lower range of those seen in Figure~\ref{fig: ioni freq vs. outgassing inb vs outb} ($n\times r^2 < 3 \times 10^{10}$~cm$^{-1}$). In the two periods,  outgassing was positively correlated with EII frequency, so the more outgassing hemisphere typically saw larger ionization rates. Closer to perihelion (i.e., $n\times r^2 > 3 \times 10^{10}$~cm$^{-1}$), outgassing and EII frequency are negatively correlated, so the less outgassing latitudes may see higher levels of EII.
\subsection{Electron impact ionization vs. cometocentric distance}\label{sec: eii vs rcom}
\begin{figure*}
    \centering
\begin{tikzpicture}
        \node (rEarly) at (0,0) {\includegraphics[width = \textwidth]{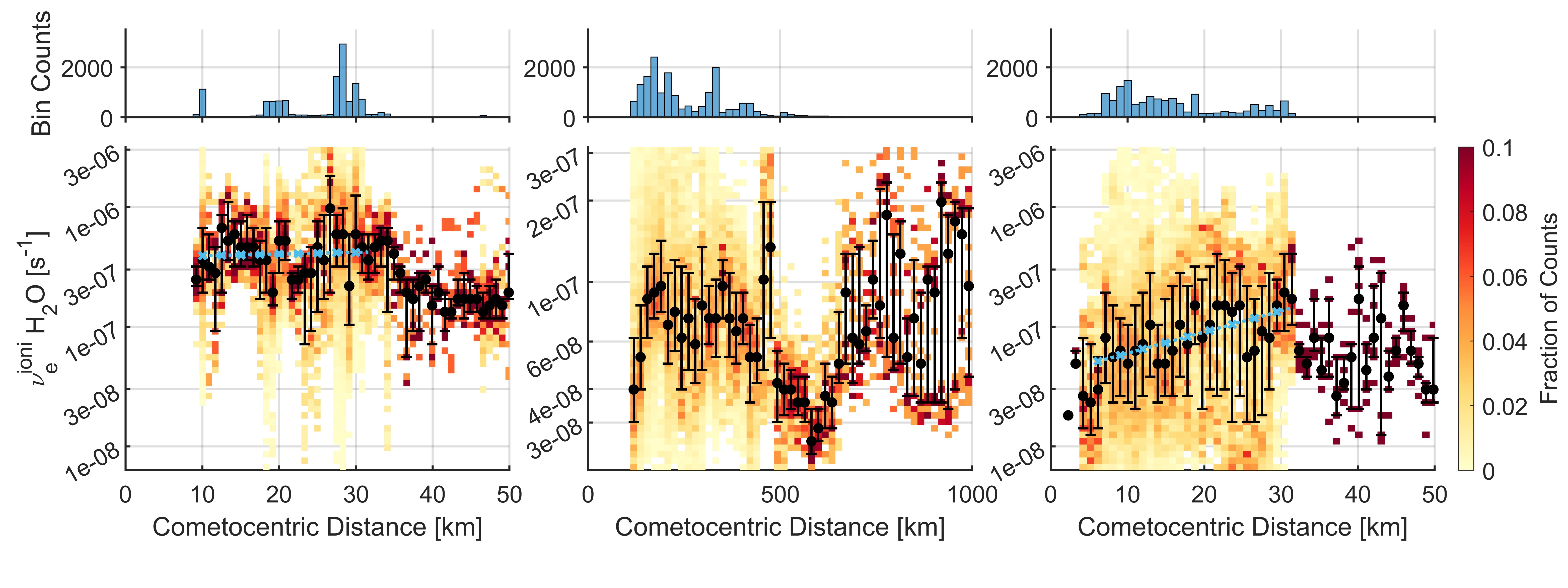}};
        
        \node [above left = 2.8 and 4.1  of rEarly.center, font = \bf] (EarlyLab) {Early Mission};
        \node [above = 2.8 of rEarly.center, font = \bf] (PHLab) {Around PH};
        \node [above right = 2.8 and 4.1  of rEarly.center, font = \bf] (LateLab) {Late Mission};
        \node [above left = 6.3 and 7.1 of rEarly.base, font =\bf] (A) {(a)};
        \node [above left = 6.3 and 2.1 of rEarly.base, font =\bf] (B) {(b)};
        \node [above right =6.3 and 2.3 of rEarly.base, font =\bf] (C) {(c)};
        \node [above left = 1.5 and 3.3 of rEarly.base, align=left, font =\bf] (R2a) {$\log_{10}(\frac{y}{\mathrm{s}^{-1}}) = 0.0014\frac{x}{\mathrm{km}} -6.42$\\$R^2=0.055$};
        \node [above right = 4.2 and 3.0 of rEarly.base, align=left, font =\bf] (R2c) {$\log_{10}(\frac{y}{\mathrm{s}^{-1}}) = 0.0177\frac{x}{\mathrm{km}} -7.40$\\$R^2=0.660$};
\end{tikzpicture}

    \caption{Electron-impact ionization frequency vs. cometocentric distance (a) early in the mission, (b) around perihelion and (c) late in the mission. Median (black circles) and quartile values (error bars) are plotted for each bin along $x$, in addition to the mean values (black crosses). \comments{Linear fits to the median values (blue crosses) are shown for the early and late mission periods below in the ranges 10 to 32\,km and 5 to 32\,km, respectively.}}
    \label{fig: ioni freq vs R com hist}
\end{figure*}
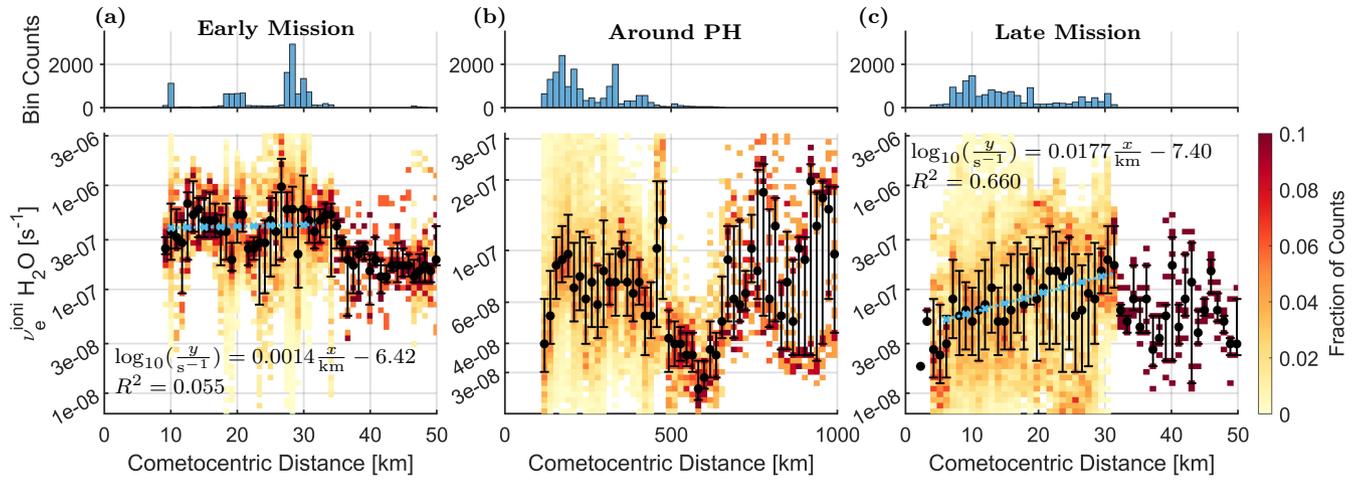

Figure~\ref{fig: ioni freq vs R com hist} shows the correlation between EII frequency and the cometocentric distance during three periods of the mission. Both early and late in the mission, there is no obvious correlation between the EII frequency and cometocentric distance. Early in the mission, the cometocentric distance did not vary substantially and the bulk of the measurements were taken at 20~km or 30~km from the nucleus. At other distances, there is little data over the early mission period and no clear correlation with cometocentric distance.

Late in the mission, Rosetta collected data more evenly between cometocentric distances of 5 and 30~km. This period shows a slight increase in the electron-impact ionization frequency with increasing cometocentric distance. There are measurements up to 500~km in the late period, but not enough datapoints \comments{beyond 32~km (only 131 points compared to 14,500 for $r<32$\,km)} to draw any significant conclusions.

\edits{The trajectory of Rosetta was tuned throughout the mission to approximately maintain the neutral density at the spacecraft. Therefore, larger cometocentric distances are linked to higher outgassing rates. Furthermore, in the late mission period, the coma was weakly collisional and higher outgassing rates led to increased EII frequencies. Hence, these two factors could be a driver of the increase in EII frequency with cometocentric distance late in the mission (see Figure~\ref{fig: ioni freq vs R com hist}c).}

Around perihelion, the EII frequency does exhibit a dependence on cometocentric distance. Between 200 and 600~km, the EII frequency decreases with increasing cometocentric distance from $\nu_e^{\text{ioni}} = 10^{-7}$~s$^{-1}$ to $4\times10^{-8}$~s$^{-1}$ at 600~km. This could be driven by a process accelerating electrons towards the nucleus. \edits{However, beyond 400~km there is limited data and this largely originates from the excursion around perihelion (see Figure~\ref{fig: CPR ioni freq fig}b), so could be driven by temporal changes. For example, a CME was observed during the inbound leg of the \comments{excursion} at $\sim800$~km \citep{Edberg2016CME}, which drives the peak in ionization \comments{frequency} at that cometocentric distance.} 

Closer to the nucleus ($<200$~km), the EII frequency decreases with decreasing cometocentric distances. This may indicate the region in which electron-neutral collisions can efficiently cool the suprathermal electrons. 

During the mission, there were a number of periods where Rosetta rapidly changed cometocentric distance, such as the excursion at the end of March 2016. The EII frequency did not behave in the same way between manoeuvres or between the outbound and inbound motion of the spacecraft. At times, the EII frequency would decrease as the spacecraft moved away from the nucleus, such as the increase from 28~km to 260~km from 15-17 Feb 2015. However, as the spacecraft returned to 100~km on 25 Feb 2015, the EII frequency fluctuates but shows no overall change. The fluctuations of the EII frequency throughout the manoeuvres greatly exceeded the overall change in magnitude with cometocentric distance.

\subsection{Summary of ionization frequency throughout the Rosetta mission}
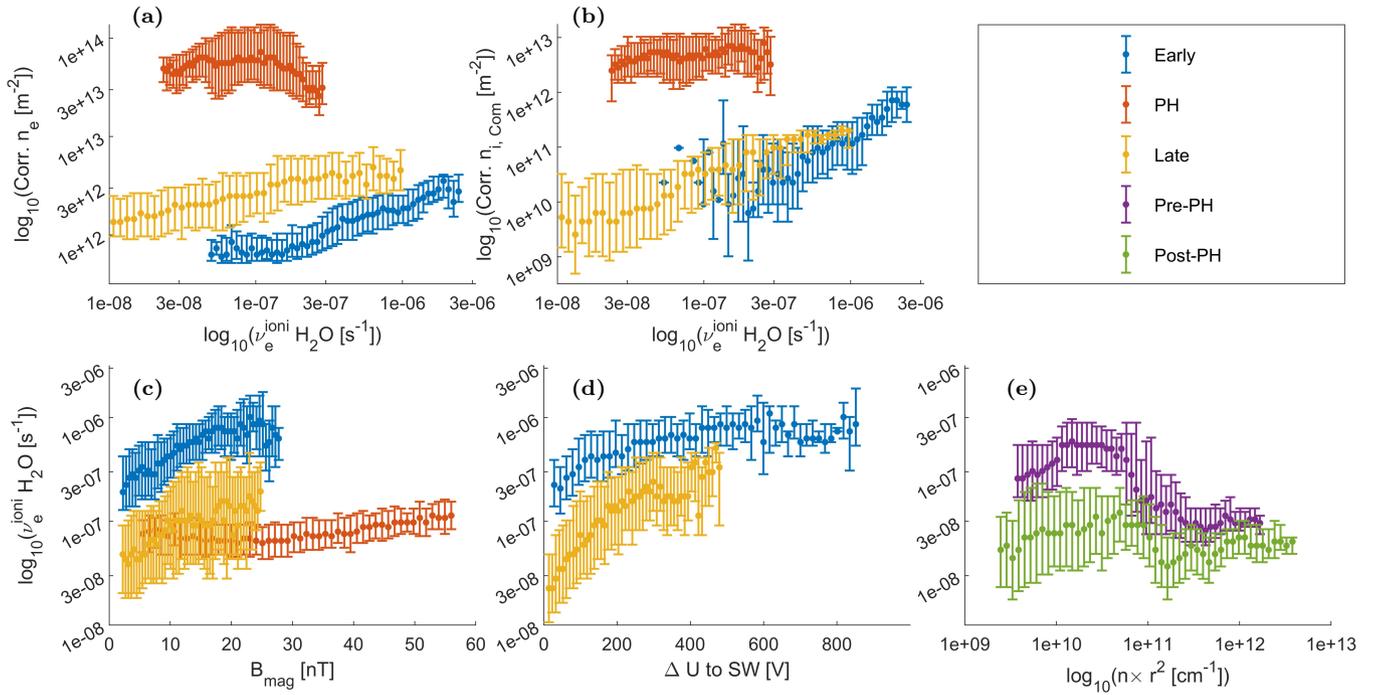
\begin{figure*}
    \centering
    \begin{tikzpicture}
    \node (EII) at (0,0) {\includegraphics[width = \textwidth]{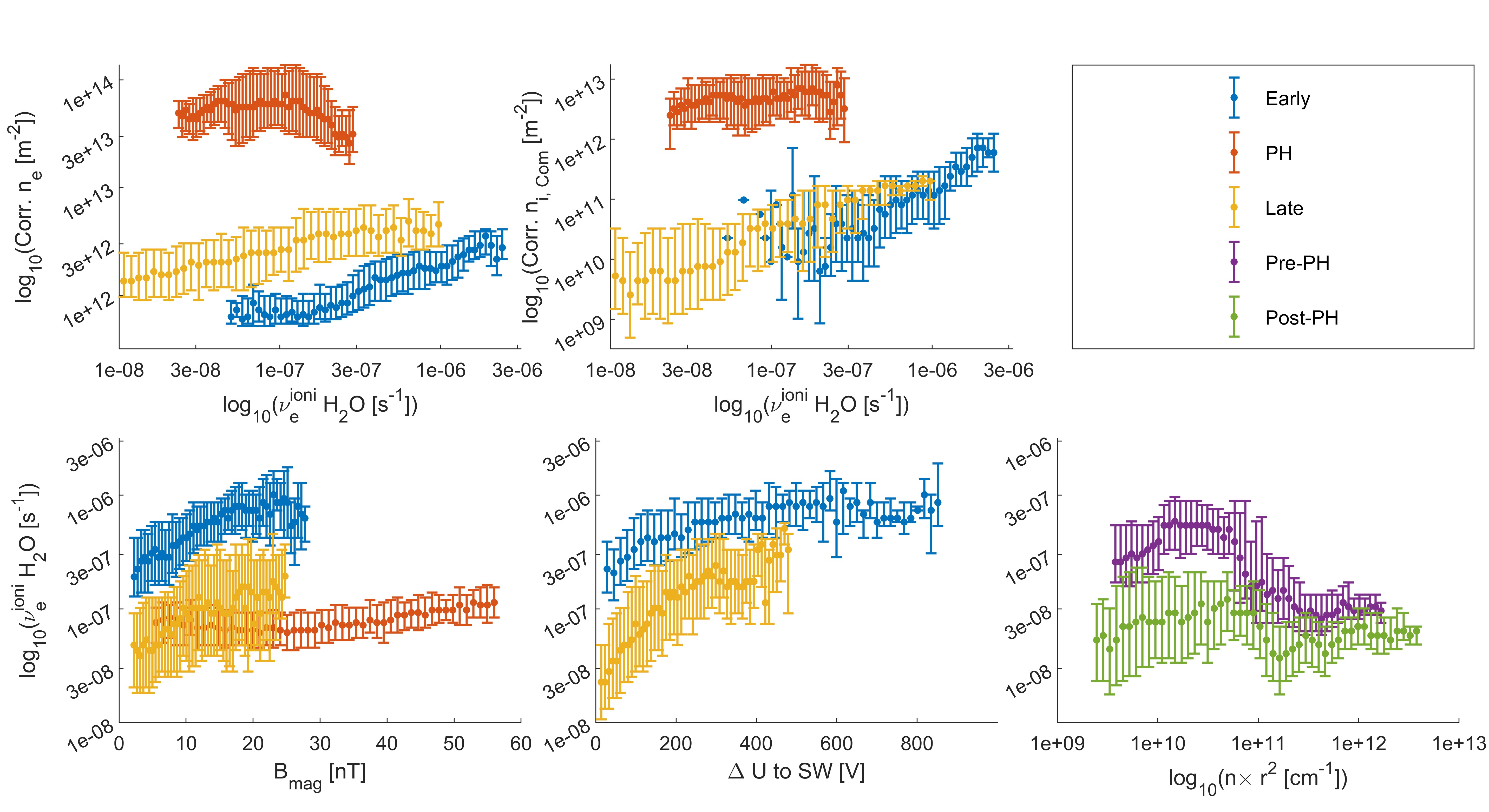}};
    \node [above left = 4.2 and 7 of EII.center, font=\bf, anchor = center] (a) {(a)};
    \node [above left = 4.2 and 1.2 of EII.center, font=\bf, anchor = center] (b) {(b)};
    \node [below left = 0.7 and 7 of EII.center, font=\bf, anchor = center] (c) {(c)};
    \node [below left = 0.7 and 1.2 of EII.center, font=\bf, anchor = center] (d) {(d)};
    \node [below right = 0.7 and 4.5 of EII.center, font=\bf, anchor = center] (e) {(e)};
    \end{tikzpicture}
    \caption{Summary of the median, 25th and 75th percentiles for Figures~\ref{fig: ioni freq vs el dens} to~\ref{fig: ioni freq vs. outgassing inb vs outb}, for each time period (see Table~\ref{tab: ioni freq time periods} and Figure~\ref{fig: CPR ioni freq fig}).}
    \label{fig: ioni freq data summary}
\end{figure*}
\comments{
Figure \ref{fig: ioni freq data summary} summarises Figures \ref{fig: ioni freq vs el dens} to \ref{fig: ioni freq vs. outgassing inb vs outb}. 
Analysis of the ionization frequencies demonstrate that, at large heliocentric distances, electron-impact ionization is the main source of ionization, as the EII frequency is positively correlated with the electron and ion densities (yellow and blue, Figures \ref{fig: ioni freq data summary}a, b). Around perihelion, the electron and ion densities show little dependence on the EII frequency (red, Figures \ref{fig: ioni freq data summary}a, b), as photoionization is more ionizing than EII during this period (green vs blue, Figure \ref{fig: CPR ioni freq fig}). 
The magnetic field strength and the Rosetta-upstream solar wind potential difference are key drivers of the EII frequency (Figures \ref{fig: ioni freq data summary}c, d), when comet 67P was far from perihelion. These indicate that the ionizing electrons originate in the solar wind and are accelerated into the inner coma by an ambipolar electric field. The EII frequency behaves non-linearly with respect to changes in the local outgassing rate (Figure \ref{fig: ioni freq data summary}e), which may reflect the changes in the ambipolar potential well. At low outgassing rates, a denser coma results in a deeper potential well and an increased EII frequency. However, at higher outgassing rates, electron-neutral collisions are more significant and lead to electron cooling and damping of the ambipolar field in the inner coma. This limits acceleration of solar wind electrons into the inner coma and leads to a reduction in the EII frequency. 
}
\section{Origin of cometary electrons with test particle simulations}\label{sec: test pl results}
When measuring electrons, it is not possible to distinguish the origin of the electron, whether from the solar wind or produced in the cometary environment through photoionization or EII. Additionally, for those electrons produced by e-impact, it is also not possible to determine the origin of the ionizing electron. With the test particle model, electrons are distinguished by their origin and, in the case of secondary electrons, by the origin of the ionizing electron. In this way, we can disentangle the most substantial electron populations within the coma and the populations which drive ionization.
\subsection{Comparison of electron sources}
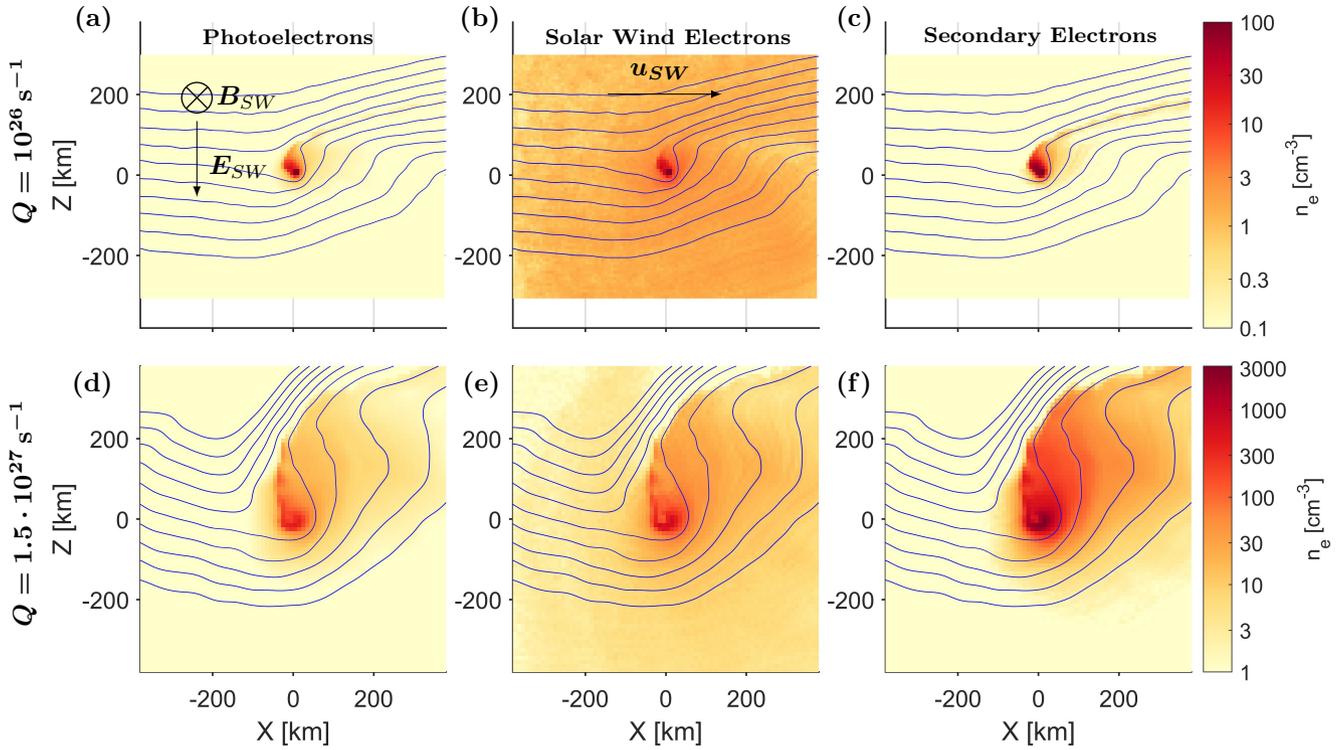
\begin{figure*}
    \centering
    \begin{tikzpicture}
       \node (EDens) at (0,0) {\includegraphics[width = 0.95\textwidth]{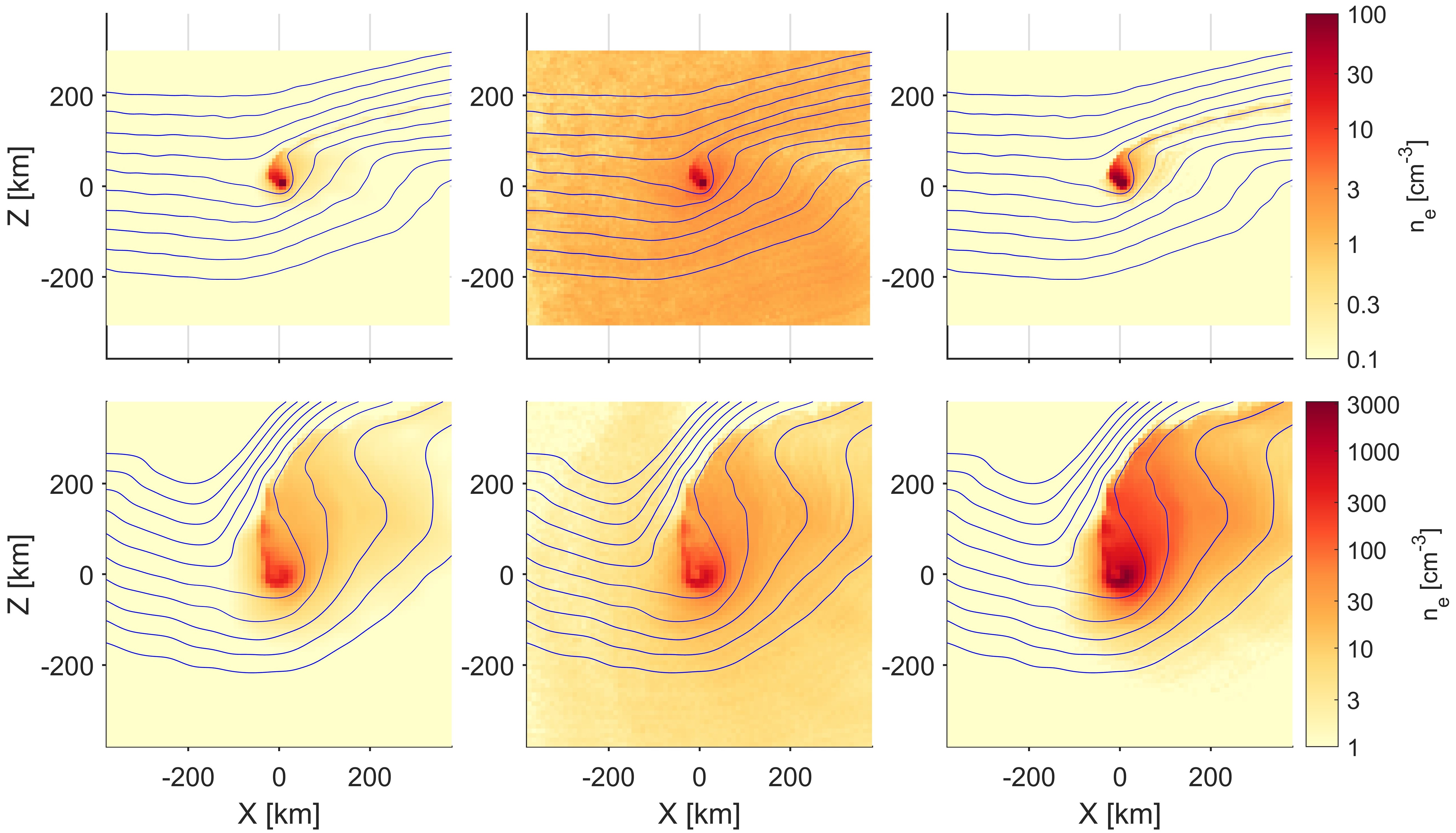}};
        \node [above left = 4.5 and 5.3 of EDens.center, font = \bf, anchor = center] (PELab) {Photoelectrons};
        \node [above left = 4.5 and 0.3 of EDens.center, font = \bf, anchor = center] (SWLab) {Solar Wind Electrons};
        \node [above right = 4.5 and 4.6 of EDens.center, font = \bf, anchor = center] (SecondaryLab) {Secondary Electrons};
        \node [above left = 9.3 and 7.5 of EDens.base, font =\large\bf] (A) {(a)};
        \node [above left = 9.3 and 2.4 of EDens.base, font =\large\bf] (B) {(b)};
        \node [above right = 9.3 and 1.8 of EDens.base, font =\large\bf] (C) {(c)};
        \node [above left = 4.5 and 7.5 of EDens.base, font =\large\bf] (D) {(d)};
        \node [above left = 4.5 and 2.4 of EDens.base, font =\large\bf] (E) {(e)};
        \node [above right = 4.5 and 1.8 of EDens.base, font =\large\bf] (F) {(f)};
        \node [above left = 9.2 and 8.5 of EDens.base, font=\large\bf, rotate = 90] (Q26) {$\bm{Q=10^{26}}$\,s$\bm{^{-1}}$};
        \node [above left = 4.7 and 8.5 of EDens.base, font=\large\bf, rotate = 90] (Q26) {$\bm{Q=1.5\cdot10^{27}}$\,s$\bm{^{-1}}$};
        \node [above left = 8.6 and 1.20 of EDens.base, anchor = center] (swArr_s) {};
        \node [right = 1.5 of swArr_s] (swArr_e) {};
        \draw [-latex] (swArr_s)--(swArr_e);
        \node [above right = 0.3 and 0.8 of swArr_s.center, font=\large, anchor = center] (vsw) {$\bm{u_{SW}}$};
        
        \node (BIn) at (-6.5,3.7) {};
        \node (EArrS) at (-6.5,3.5) {};
        \node [below = 1 of EArrS] (EArrE) {};
        \node [below right = 0.45 and 0.05 of EArrS.center, font=\large\bf] (EArrLabel) {$\bm{E}_{SW}$};
        \draw [-latex] (EArrS)--(EArrE);
        \path (BIn)  pic {vector out={line width=0.5pt, scale=0.2}} (3,0)  pic {};
        \node [right = 0.15 of BIn.center, font=\large\bf] (BLabel) {$\bm{B}_{SW}$};
\end{tikzpicture}
    \caption{Electron density of (a \& d) photoelectrons, (b \& e) solar wind electrons and (c \& f) secondary electrons  for outgassing rates of (a-c) $Q=10^{26}$~s$^{-1}$ and (d-f) $Q=1.5\times10^{27}$~s$^{-1}$. The densities are plotted in the $xz$ plane. The colour bars are over different ranges for the two outgassing cases. \comments{Streamlines are shown for electrons following the $\bm{E}\times\bm{B}$ drift velocity in the $x$-$z$ plane.}}
    \label{fig: test pl elec dens}
\end{figure*}

Figure~\ref{fig: test pl elec dens} shows the electron density calculated using the collisional test particle simulation at outgassing rates of (a-c) $Q=10^{26}$~s$^{-1}$ and (d-f) $Q=1.5 \times 10^{27}$~s$^{-1}$ \comments{(see Table~\ref{tab: test pl parameters} for other simulation parameters)}. For 67P, the outgassing rates correspond to heliocentric distances of 2.5-2.8\,au and 1.8-2\,au, respectively \citep{Laeuter2018, Biver2019}. The density of photoelectrons (a, d), solar wind electrons (b, e) and secondary electrons (c, f) are separated. \comments{Total electron densities for each outgassing rate are shown in Figure\,\ref{fig: test pl tot e dens map}, and fractional density of each population are seen in Figure\,\ref{fig: test pl ioni frac map}. }

\comments{The simulation axes are aligned to the upstream solar wind conditions with the bulk solar wind along $\hat{\bm{x}}$, the solar wind magnetic field aligned with $\hat{\bm{y}}$ and the convective electric field pointing in the $-\hat{\bm{z}}$ direction.} 

In the low outgassing case, the photoelectrons (Figure~\ref{fig: test pl elec dens}a) are confined to a small region in the inner coma, with the densest region ($n_e>10$~cm$^{-3}$) only extending 50~km from the nucleus. There is also a narrow region of high density that extends tailwards from the inner coma in the $+x$ and $+z$ direction. These electrons have been transported away from the dense innermost region by the $\bm{E}\times\bm{B}$ drift. 

The photoelectrons are primarily produced in the inner coma, where the neutral density (following $1/r^2$) is at its highest. There is no substantial absorption of the solar photons within the weakly outgassing comet ($<10\%$ for $Q<2\times10^{27}$~s$^{-1}$), so a constant photoionization frequency ($\nu_e^{h\nu} = 1.32\times10^{-7}$~s$^{-1}$, Table~\ref{tab: test pl parameters}) is used throughout the coma.

Therefore, the production of photoelectrons also scales with $1/r^2$. Produced \comments{locally} at 10~eV, the photoelectrons are effectively trapped in the coma by the ambipolar field. 

The solar wind electrons are found throughout the coma with a substantial density across the simulation domain (Figure~\ref{fig: test pl elec dens}b). At the simulation boundaries, the solar wind electron density is 1~cm$^{-3}$ and forms the dominant source of electrons at large cometocentric distances. The solar wind electron density does increase to 8~cm$^{-3}$ in the inner coma, as the electrons are funnelled towards the nucleus by the ambipolar electric field.

Close to the nucleus ($<40$~km), the solar wind and photoelectrons have comparable densities, with solar wind electrons larger by $\sim50$\%. The solar wind electrons are typically more energetic than the photoelectrons within the dense region. \comments{Consequently, it is more likely that solar wind electrons, rather than photoelectrons, will ionize neutral species, producing more secondary electrons while degrading their own energy.}

Secondary electrons form a dense population in the inner coma ($<30$~km, Figure~\ref{fig: test pl elec dens}c), with a density similar to the total of both the photoelectrons and solar wind electrons. \comments{Newly-born secondary electrons are energetic enough to themselves cause impact ionization but are trapped in the dense, inner region of the coma by the ambipolar electric field. The confinement makes the secondary electrons more likely to ionize than the primary solar wind electrons, which are unlikely to remain in the inner coma for long. With further generations, the secondary electrons become less dense and cooler as the energy is distributed between more electrons, in addition to the inelastic collisions that dissipate energy. As electrons become cold, it also becomes more likely that they will recombine with cometary ions.}

The total secondary electron density decreases rapidly with increasing cometocentric distance, and at a much faster rate than the photoelectron population. Beyond 50~km, the secondary electron density drops below the photoelectrons and beyond 100~km the solar wind electrons dominate the density. The faster decline, compared to photoelectrons, results from the fall in electron-impact ionization frequency with distance from the nucleus (see Figures~\ref{fig: test pl ioni freq map} and~\ref{fig: test pl r vs ioni freq}). 

At $Q=1.5\times10^{27}$~s$^{-1}$, the photoelectron population (Figure~\ref{fig: test pl elec dens}d) \comments{is denser and extends to larger cometocentric distances} than in the low outgassing case. The same photoionization frequency is used as in the lower outgassing case (Table~\ref{tab: test pl parameters}), so the production of photoelectrons is 15 times larger due to the increased outgassing. A similar \comments{increase} (between 10 and 20) is seen in the photoelectron density in the inner coma of the low outgassing case ($<50$~km). The region of high photoelectron density extends up to 300~km from the nucleus in the higher outgassing case, extending towards $+z$. The denser ionosphere at $Q=1.5\times10^{27}$~s$^{-1}$ produces a larger region of interaction with the solar wind plasma. 

The solar wind electrons are initiated at the boundary with the same parameters as used at $Q=10^{26}$~s$^{-1}$. However, the solar wind density is enhanced ($n_{e,SW}> 10$~cm$^{-3}$) over a larger region due to the greater extent of the ambipolar field \comments{for $Q=1.5\times10^{27}$\,s$^{-1}$}. The solar wind electrons in the inner coma closely resemble the photoelectron population in the region. The solar wind electrons, having been accelerated by the ambipolar field, undergo collisions with the neutral coma. 
They are then efficiently trapped by the ambipolar potential well, leading to further cooling, followed by transport by particle drifts.

The electrons further from the coma are not confined by the ambipolar field, although they still are accelerated and deflected by it. In the dense region of photoelectrons ($n_{e,PE}>5$~cm$^{-3}$), the solar wind electrons are a factor 3 denser than the photoelectrons (see Figures~\ref{fig: test pl elec dens}d and~\ref{fig: test pl elec dens}e), although this drops to 1 in the densest photoelectron region. 

The solar wind electrons are dense over a much larger region than the photoelectrons, but there is a depleted region extending from the inner coma towards $+z$. The low density region is associated with a peak in the ambipolar potential, which deflects electrons away from the region. 

The secondary electron cloud has a similar shape to the photoelectrons, but in the inner coma the secondary electron density reaches 1400~cm$^{-3}$. This is substantially larger than the photoelectron and solar wind electron populations, which collectively peak at 700\,cm$^{-3}$. In a non-collisional simulation at the higher outgassing rate ($Q=1.5\times10^{27}$~s$^{-1}$), the photoelectron density peaks at 900~cm$^{-3}$ while the solar wind is a minor contribution in the inner coma ($<16$\,cm$^{-3}$).
Despite the additional production (EII) and cooling (inelastic excitations) in the collisional simulation, the cometary electrons may escape more easily from the coma in the collisional case due to elastic scattering collisions. These can redistribute the parallel and perpendicular energies and allow particles to escape the potential well. Conversely, the solar wind electrons are more efficiently confined when collisions are included. They undergo inelastic collisions in the inner coma, lose energy, and become trapped by the potential well. 

The total electron density in the coma is larger in the collisional case, due to the additional substantial contribution from electron-impact ionization. The third and fourth generations of electrons are significant at $Q=1.5\times 10^{27}$\,s$^{-1}$, unlike in the lower outgassing case, as the accelerated solar wind electrons produce a population of energetic secondaries in the inner coma. The ambipolar potential well is deeper at the higher outgassing rate, meaning the secondary electrons are born at higher energies and more capable of causing ionization. Additionally, the higher outgassing rate increases the production rate of secondaries and boosts the energetic electron flux ($J(E)$, see Eq.~\ref{eq: e ioni freq}).

The impact of the additional electrons on the electric fields in the coma is assessed using the Generalised Ohm's law and assuming the magnetic field is unchanged from the collisionless case \citep{Stephenson2022TestPl}. This is a reasonable assumption, as magnetic field saturates during pile up and is resistant to changes in electron density \citep{Goetz2017}. The electron component of the Hall field ($\bm{E}_{Hall,e} =  -\bm{u}_e\times\bm{B}$) and the ambipolar field ($\bm{E}_{Ambi}=-\frac{1}{e\,n_e}\nabla p_e$) are unaffected outside the inner coma ($r>40$~km), by the additional electrons \comments{(see Figure~\ref{fig: ambipolar field})}. Closer to the nucleus, electron-neutral collisions quench the Hall and ambipolar fields, as the bulk velocity and electron temperatures are small in the region. However, the ambipolar potential well is constructed over a much larger spatial scale ($>200$~km), so the potential in the inner coma is reduced by at most 40\% \comments{(see Appendix\,\ref{sec: GOL fields})}.

In both outgassing cases, the secondary electrons are more abundant in the inner coma than photoelectrons. Solar wind electrons also have a substantial contribution to the total density. Electron-impact ionization is the major source of cometary electrons within a weakly outgassing comet. \comments{As the outgassing increases, it is likely that photoionization will become dominant, as the cooling of the ionizing electrons becomes more efficient. This was observed at 67P during the Rosetta mission around perihelion (see Figure~\ref{fig: CPR ioni freq fig}a), but the high outgassing cases cannot be addressed with the test particle model.  }

\subsection{Source of ionizing electrons}
\subsubsection{EII frequency from different electron populations}
\begin{figure*}
    \centering
    \begin{tikzpicture}
        \node (EIoni) at (0,0) {\includegraphics[width = 0.95\textwidth]{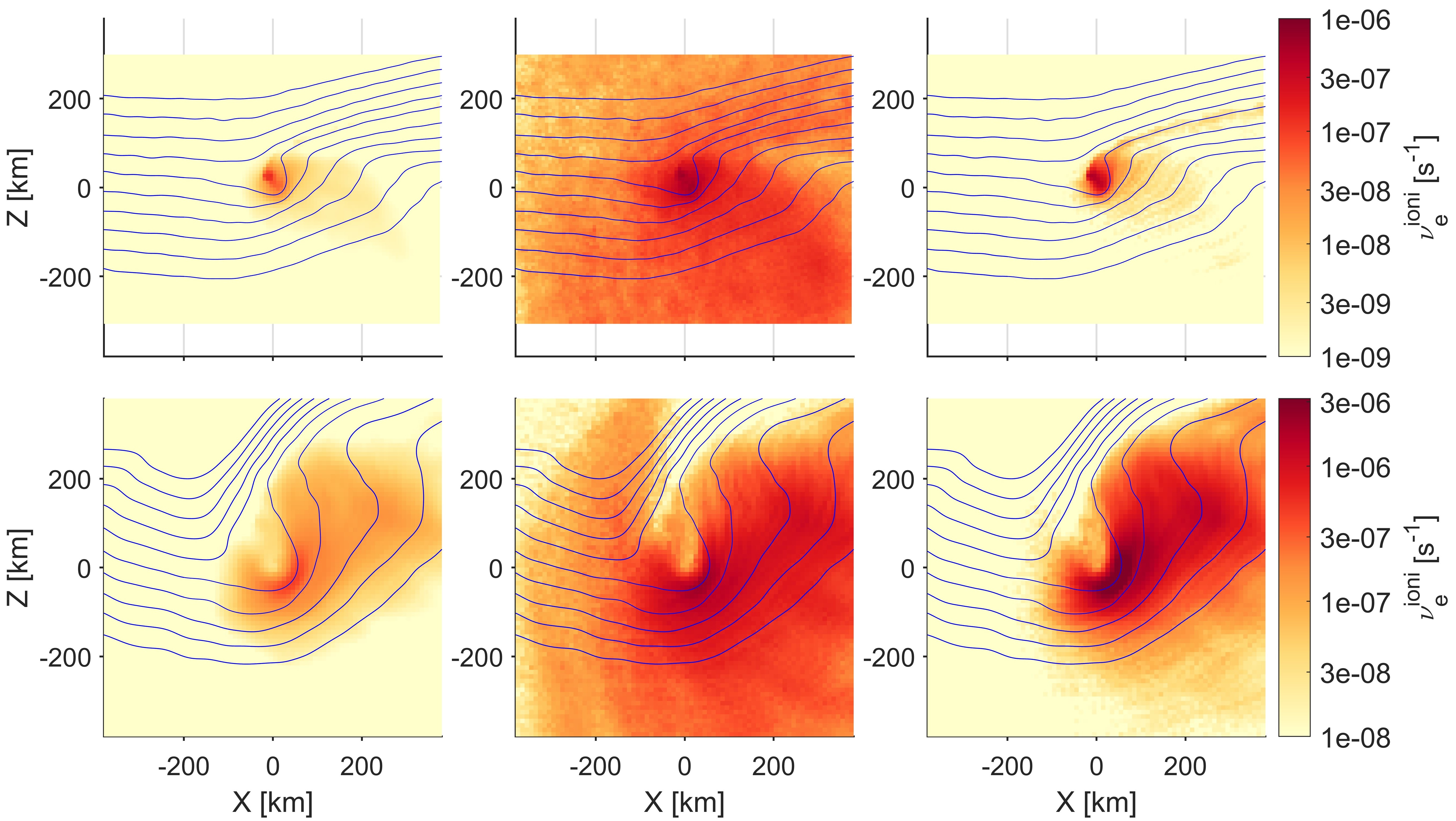}};
        \node [above left = 4.5 and 5.3 of EIoni.center, font = \bf, anchor = center] (PELab) {Photoelectrons};
        \node [above left = 4.5 and 0.3 of EIoni.center, font = \bf, anchor = center] (SWLab) {Solar Wind Electrons};
        \node [above right = 4.5 and 4.6 of EIoni.center, font = \bf, anchor = center] (SecondaryLab) {Secondary Electrons};
        \node [above left = 9.3 and 7.5 of EIoni.base, font =\large\bf] (A) {(a)};
        \node [above left = 9.3 and 2.3 of EIoni.base, font =\large\bf] (B) {(b)};
        \node [above right = 9.3 and 1.8 of EIoni.base, font =\large\bf] (C) {(c)};
        \node [above left = 4.5 and 7.5 of EIoni.base, font =\large\bf] (D) {(d)};
        \node [above left = 4.5 and 2.3 of EIoni.base, font =\large\bf] (E) {(e)};
        \node [above right = 4.5 and 1.8 of EIoni.base, font =\large\bf] (F) {(f)};
        \node [above left = 9.2 and 8.5 of EIoni.base, font=\large\bf, rotate = 90] (Q26) {$\bm{Q=10^{26}}$\,s$\bm{^{-1}}$};
        \node [above left = 4.7 and 8.5 of EIoni.base, font=\large\bf, rotate = 90] (Q26) {$\bm{Q=1.5\cdot10^{27}}$\,s$\bm{^{-1}}$};
        \node [above left = 8.3 and 1.20 of EDens.base, anchor = center] (swArr_s) {};
        \node [right = 1.5 of swArr_s] (swArr_e) {};
        \draw [-latex] (swArr_s)--(swArr_e);
        \node [above right = 0.3 and 0.8 of swArr_s.center, font=\large, anchor = center] (vsw) {$\bm{u_{SW}}$};
        
        \node (BIn) at (-6.5,3.7) {};
        \node (EArrS) at (-6.5,3.5) {};
        \node [below = 1 of EArrS] (EArrE) {};
        \node [below right = 0.45 and 0.05 of EArrS.center, font=\large\bf] (EArrLabel) {$\bm{E}_{SW}$};
        \draw [-latex] (EArrS)--(EArrE);
        \path (BIn)  pic {vector out={line width=0.5pt, scale=0.2}} (3,0)  pic {};
        \node [right = 0.15 of BIn.center, font=\large\bf] (BLabel) {$\bm{B}_{SW}$};
\end{tikzpicture}
    \caption{Electron-impact ionization frequency from collisions by (a,d) photoelectrons, (b,e) solar wind electrons and (c,f) secondary electrons for outgassing rates of (a-c) $Q=10^{26}$~s$^{-1}$ and (d-f) $Q=1.5\times10^{27}$~s$^{-1}$. \comments{Streamlines are shown for electrons following the $\bm{E}\times\bm{B}$ drift velocity in the $x$-$z$ plane.}}
    \label{fig: test pl ioni freq map}
\end{figure*}
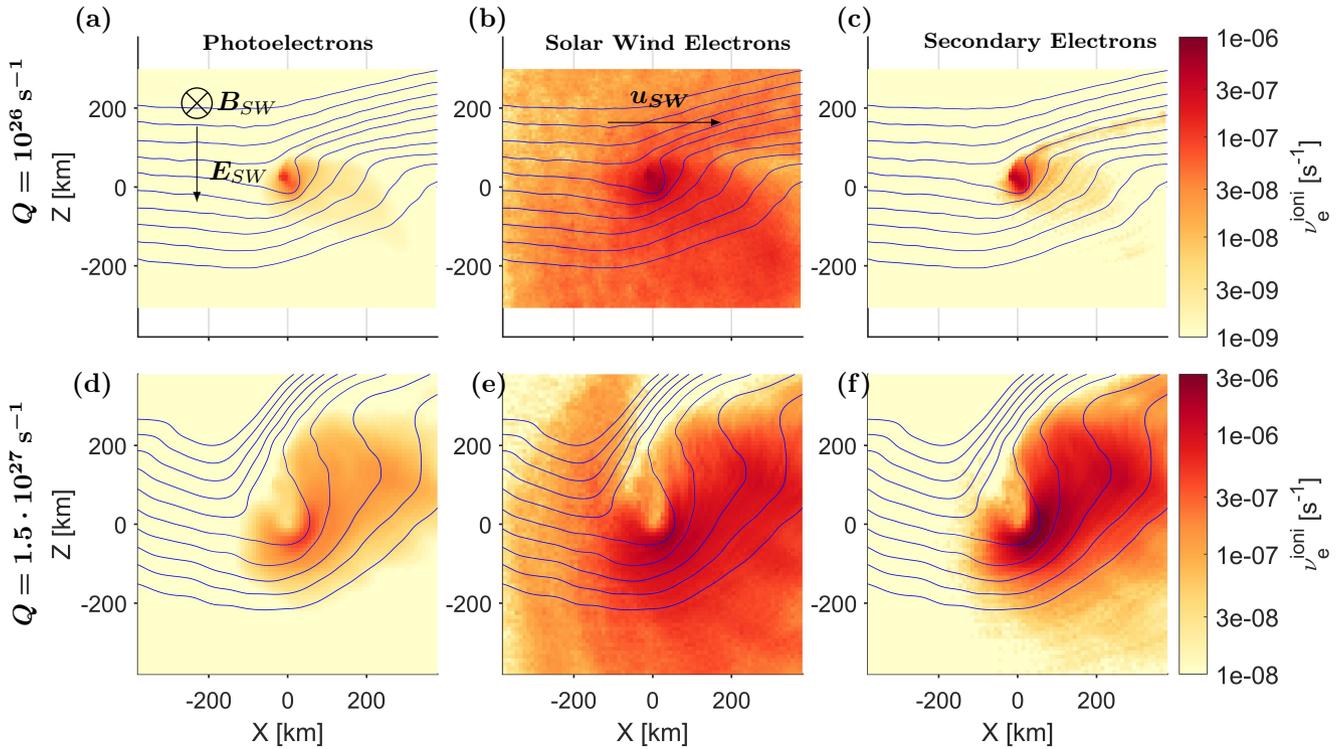

Having determined that the bulk of cometary electrons are produced by electron-impact ionization at medium-to-large heliocentric distances, we now examine the origin of the ionizing electrons. Figure~\ref{fig: test pl ioni freq map} shows the electron-impact ionization frequency from each of the electron populations, calculated from the collisional test particle model using Eq.~\ref{eq: eimpact ioni freq edf}.

In the lowest outgassing case, the EII frequency from photoelectrons peaks at $6\times 10^{-8}$~s$^{-1}$ in the inner coma (see Figure~\ref{fig: test pl ioni freq map}a). This decreases quickly with cometocentric distance to $10^{-8}$~s$^{-1}$ at 50~km. The number of photoelectrons reduces further from the nucleus (see Figure~\ref{fig: test pl elec dens}a), and the typical electron energy also falls in a shallower region of the potential well. The EII frequency from photoelectrons is smaller than from photoionization ($\nu_e^{h\nu} = 1.32 \times 10^{-7}$~s$^{-1}$) throughout the coma, which is typical for an optically thin coma. 

The solar wind electrons are much more ionizing than photoelectrons throughout the coma (Figure~\ref{fig: test pl ioni freq map}b), with a peak at $6\times10^{-7}$~s$^{-1}$. The solar wind electrons are ionizing over a large region of the coma, with the increased frequency close to the nucleus driven by the acceleration in the ambipolar potential well. The EII frequency from solar wind electrons exceeds the photoionization frequency up to 200~km from the nucleus, and dominates the contribution to the EII frequency within the coma (compared with photoelectrons).

Secondary electrons generate additional electrons within the coma through electron-impact collisions. This process is only substantial over a small region in the inner coma (see Figure~\ref{fig: test pl ioni freq map}c), as secondary electrons are well confined by the ambipolar field. Very close to the nucleus ($<70$~km), the EII frequency from secondary electrons exceeds that from photoelectrons. The secondary electrons, produced by collisions of solar wind electrons, are much more energetic than the newly born photoelectrons and therefore more likely to cause ionization. 

At $Q=1.5\times10^{27}$~s$^{-1}$, the EII frequency from photoelectron collisions (Figure~\ref{fig: test pl ioni freq map}d) is substantial over a large region extending 100~km upstream of the nucleus and up to 380~km downstream. The largest EII frequencies are not found in the innermost coma, despite the large electron density in the region (see Figure~\ref{fig: test pl elec dens}d-f). The peak EII frequencies in all three populations are found at $(50,0, -50)$~km (Figure~\ref{fig: test pl ioni freq map}). Here, the ambipolar potential well is deep ($>150$~V) and the neutral coma is not dense enough to efficiently cool the electrons in the region.

As seen at lower outgassing, solar wind electrons are more ionizing than photoelectrons throughout the simulation domain. However, the secondary electrons exceed the EII frequency of the solar wind electrons in the inner coma. The secondary electrons, particularly those produced by collisions of solar wind electrons in the inner coma, are very energetic and can drive electron-impact ionizations. The secondaries are also produced within the ambipolar potential well, so are more efficiently trapped than the solar wind electrons. Therefore, they are more likely to collide with the neutral gas and produce further generations of electrons. The further generations of electrons are less ionizing as the input energy is distributed between more electrons and some energy is dissipated \comments{through supplying the threshold energy for} inelastic excitations. The secondary electrons that drive ionization are almost entirely descendents of the accelerated solar wind population, rather than the lower energy photoelectrons. 


\subsubsection{EII frequency vs magnetic field strength}\label{sec: test pl eii vs B}
\begin{figure*}
    \centering
    \begin{tikzpicture}
    \node (BMag) at (0,0) {\includegraphics[width =\textwidth]{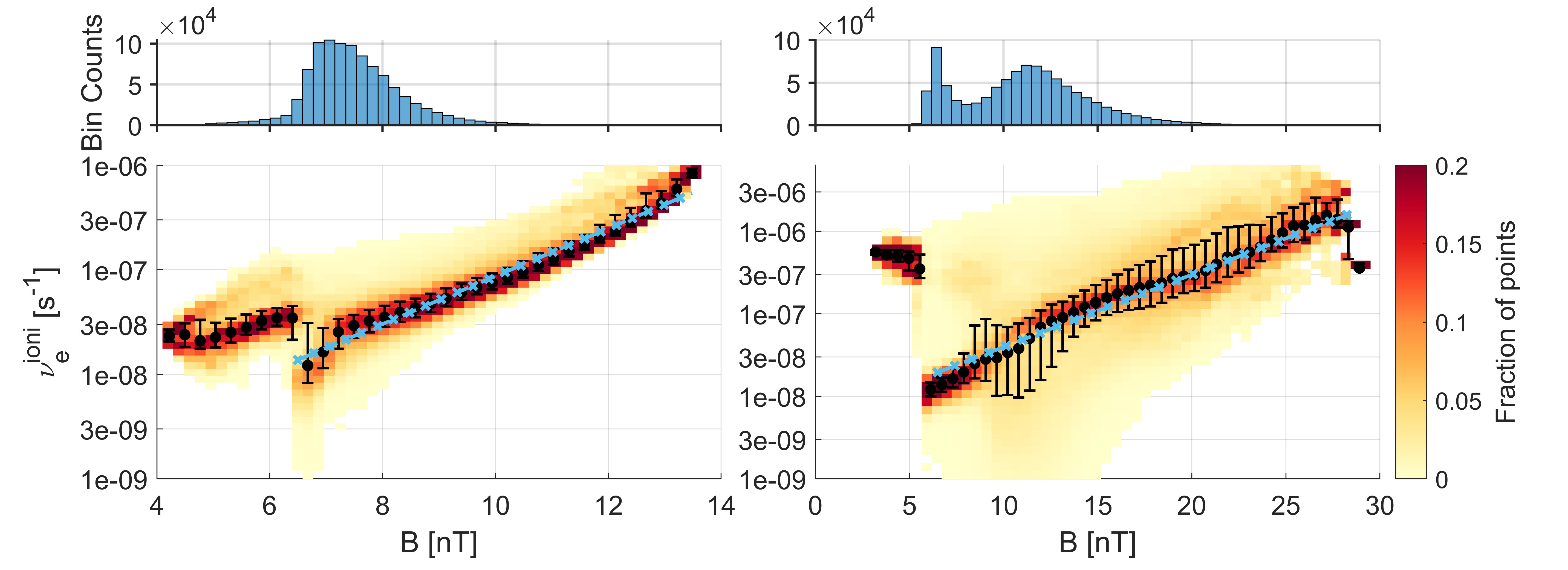}};

    \node [above left = 6.5 and 6.3 of BMag.base, font =\large\bf] (A) {(a)};
    \node [above right = 6.5 and 0.5 of BMag.base, font =\large\bf] (B) {(b)};
    \node [above left = 3.6 and 3.0 of BMag.base, align=left, font =\bf] (R2a) {$\log_{10}(\frac{y}{\mathrm{s}^{-1}}) = 0.229\frac{x}{\mathrm{nT}} -9.349$\\$R^2=0.9788$};
        \node [above right = 1.0 and 0.8 of BMag.base, align=left, font =\bf] (R2c) {$\log_{10}(\frac{y}{\mathrm{s}^{-1}}) = 0.0877\frac{x}{\mathrm{nT}} -8.2795$\\$R^2=0.9442$};

    \end{tikzpicture}
    
    \caption{Electron-impact ionization frequency against magnetic field strength throughout the coma for outgassing rates of (a) $Q=10^{26}$~s$^{-1}$ and (b) $Q=1.5\times10^{27}$~s$^{-1}$. Median (black circles) and quartile values (error bars) are plotted for each. \comments{Linear fits are given for both outgassing rates (blue crosses).}}
    \label{fig: test pl B vs ioni freq}
\end{figure*}
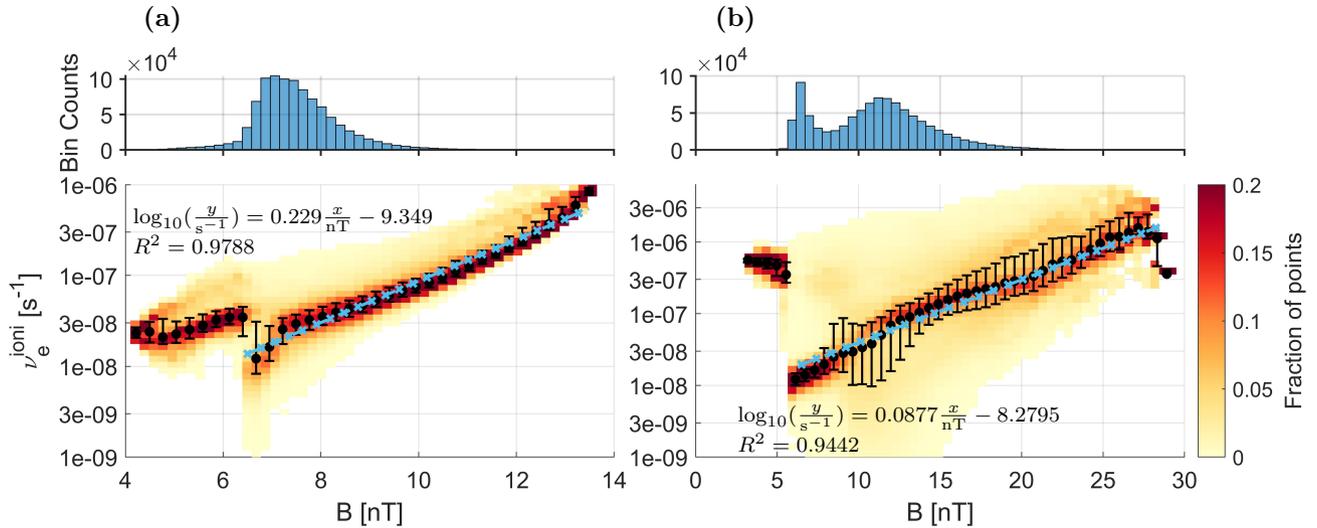

The analysis applied to the data from Rosetta in Section~\ref{sec: EII Rosetta} can also be applied to the outputs of the test particle simulation. The values calculated for each cell of the test particle simulation are organized into a 2D histogram in Figures~\ref{fig: test pl B vs ioni freq} and~\ref{fig: test pl r vs ioni freq}, with the median and the quartiles for each bin \edits{of the magnetic field strength} also shown. The results from the test particle model represent a single 3D snapshot under a single set of conditions, whereas the Rosetta data provided a single measurement point that moved through an evolving coma. 

Figure~\ref{fig: test pl B vs ioni freq} shows the relation between the EII frequency and the magnetic field strength in the test particle simulations. Under both outgassing rates, the EII frequency increased with magnetic field strength. At the lower outgassing rate, the EII frequency increases from $3.2\times10^{-8}$~s$^{-1}$ at 8~nT to $10^{-6}$~s$^{-1}$ at 13~nT (Figure~\ref{fig: test pl B vs ioni freq}a). 

At $Q=1.5\times10^{27}$~s$^{-1}$, the EII frequency increases from $10^{-8}$~s$^{-1}$ at 6~nT to $1.6\times10^{-6}$~s$^{-1}$ at 25~nT (Figure~\ref{fig: test pl B vs ioni freq}b). \comments{The} EII frequency decreases at very large magnetic field strengths ($B>25$~nT, Figure~\ref{fig: test pl B vs ioni freq}b). These cells are found close to the nucleus, where the electron-neutral collisions efficiently degrade the high energy tail of the solar wind electrons, quenching the ionization processes.

\comments{At both outgassing rates,} there is an apparent discontinuity in the EII frequency between 6 and 6.5~nT, \comments{which reflects that very few regions in the simulation domain have magnetic fields below 6.5\,nT (see top of Figure~\ref{fig: test pl B vs ioni freq}a). This is more a reflection of the sampled domain rather than any physical process. Field strengths of 6.5\,nT} correspond to the region of \edits{upstream} solar wind \edits{which has been slightly compressed near the upstream boundary}. The electrons are unaccelerated solar wind electrons ($T_{e,SW}=10$~eV), which are not energetic enough to cause substantial ionization. 
The smallest magnetic field strengths \comments{($<6$\,nT)} are found $\sim 300$~km downstream of the nucleus between  $y=\pm30$~km. 

\comments{At $Q=1.5\times10^{27}$\,s$^{-1}$, magnetic field strengths below 6\,nT are confined to a small region} found $\sim 300$~km downstream of the nucleus between  $y=\pm30$~km. The lowest EII frequencies at $10^{-9}$~s$^{-1}$ are found at (-50, 0, 200)~km in the region depleted of electrons (see Figure~\ref{fig: test pl elec dens}d-f). This region is a maximum in the parallel electric potential at 50~V above the undisturbed solar wind. The potential barrier decelerates and deflects electrons, reducing the probability of ionization collisions occurring in the region.  

The positive correlation with magnetic field strength agrees qualitatively with the results from the Rosetta mission (see Section~\ref{sec: eii vs B field})\comments{, although the gradients of the correlations are much larger in the fits to the test particle model (0.23 and 0.088 vs. 0.024 and 0.028).} The ionizing electrons are accelerated solar wind electrons, which are funnelled into the inner coma. The electrons are then trapped and pile up in a dense region where they can produce many secondary electrons. 

\subsubsection{EII frequency vs cometocentric distance}\label{sec: test pl eii vs r com}
\begin{figure*}
    \centering

    \begin{tikzpicture}

        \node (BMag) at (0,0) {\includegraphics[width =\textwidth]{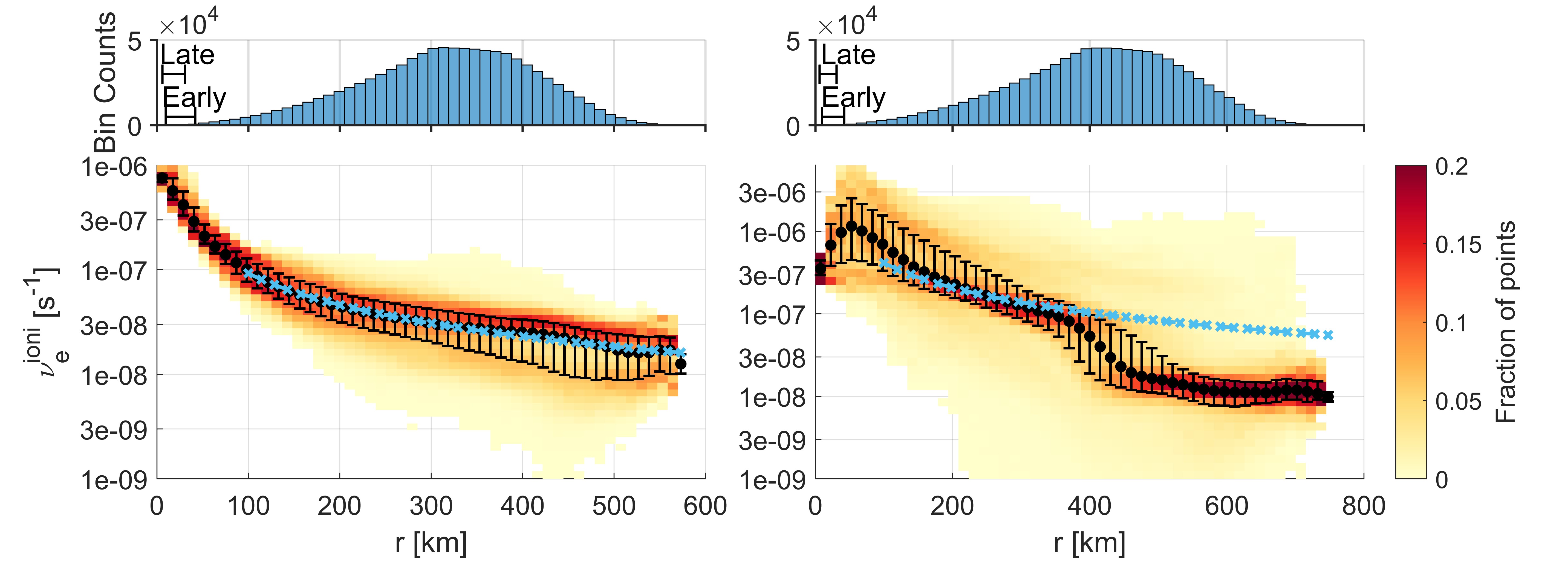}};
    
        \node [above left = 5.7 and 8.0 of BMag.base, font =\large\bf] (A) {(a)};
        \node [above left = 5.7 and 0.1 of BMag.base, font =\large\bf] (B) {(b)};
        \node [above left = 4.0 and 3.0 of BMag.base, align=left, font =\bf] (R2a) {$\frac{y}{\mathrm{s}^{-1}} = 9.36\times10^{-6}\mathrm{km}/x$\\$R^2=0.980$};
        \node [above right = 4.0 and 2.8 of BMag.base, align=left, font =\bf] (R2c) {$\frac{y}{\mathrm{s}^{-1}} = 4.162\times10^{-5}\mathrm{km}/x$\\$R^2=0.736$};

    \end{tikzpicture}
    \caption{Total electron-impact ionization frequency against cometocentric distance throughout the coma from test particle simulations at outgassing rates of (a) $Q=10^{26}$~s$^{-1}$ and (b) $Q=1.5\times10^{27}$~s$^{-1}$. Median (black circles) and quartile values (error bars) are plotted for each bin of \edits{cometocentric distance, $r$. }\comments{The range of cometocentric distances probed by Rosetta im the early and late periods (see Table~\ref{tab: ioni freq time periods}) are show in the top plots for each period. Fits of a $1/r$ profile are given for both outgassing rates (blue crosses) for $r>100$\,km. At $Q=1.5\times10^{27}$\,s$^{-1}$, the function is fit between 100 and 400\,km. The spatial resolution of the fields (see Table~\ref{tab: test pl parameters}) and lack of feedback of collisional processes on the fields limits the validity of fits below $r=100$\,km.}}
    \label{fig: test pl r vs ioni freq}
\end{figure*}
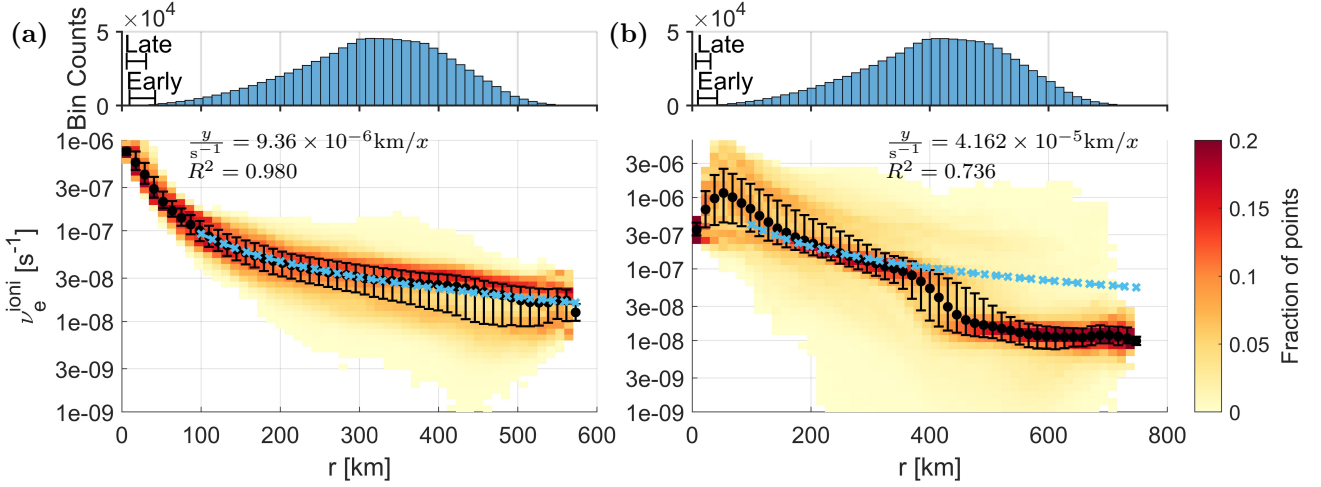

Figure~\ref{fig: test pl r vs ioni freq} shows the variation of the EII frequency with cometocentric distance throughout the coma. Under both outgassing conditions, the EII frequency increases with decreasing distance to the nucleus. 

In the low outgassing case, the EII frequency is fairly constant at large distances from the coma (Figure~\ref{fig: test pl r vs ioni freq}a), as these points are in the undisturbed solar wind. Closer to the nucleus, the electrons are accelerated by the ambipolar potential well and the EII frequency increases. The ionization frequency is dominated by  collisions of solar wind electrons or their secondary electrons. The EII frequency increases from $10^{-7}$~s$^{-1}$ to $10^{-6}$~s$^{-1}$ between $r=100$~km and the nucleus. The cooling processes in the inner coma do cause some degradation at $Q=10^{26}$~s$^{-1}$, but only a small fraction of the total electrons become cold. Consequently, the EII frequency continues to increase all the way to the surface. There is some uncertainty in the behaviour of the electrons in the inner coma, as the field resolution (7.66~km, Table~\ref{tab: test pl parameters}) is larger than the comet nucleus \citep[$1.7$~km,][]{Jorda2016}. 

At $Q=1.5\times10^{27}$~s$^{-1}$, the behaviour is very similar to the case at $Q=10^{26}$~s$^{-1}$ away from the inner coma. Far from the nucleus ($r>400$~km) in the (somewhat) pristine solar wind, the EII frequency is steady at $10^{-8}$~s$^{-1}$. All three electron populations are more ionizing closer to the nucleus (50-350~km), with the total EII frequency increasing by a factor 10. Even closer to the nucleus ($r<50$~km), the EII frequency decreases as the neutral gas is dense enough to substantially cool the high energy electrons. 

\comments{The median EII frequencies at the low outgassing rate are very close to a $\nu_e^{\text{ioni}}\propto1/r$ profile (blue dashed line, Figure~\ref{fig: test pl r vs ioni freq}) throughout the wider coma. Similar behaviour is seen between 100 and 400 km at $Q=1.5\times10^{27}$\,s$^{-1}$, but the collisional cooling in the inner coma causes deviation from this trend. At small cometocentric distances ($r<50$\,km), the EII frequency increases with cometocentric distance at $Q=1.5\times10^{27}$\,s$^{-1}$, with a gradient of $\Delta(\log_{10}\nu_e^{\text{ioni}})/\Delta r=0.0119$.} 


The behaviour of the EII frequency in the inner coma is subject to several caveats, particularly in the higher outgassing case. Firstly, the resolution of the fields is coarse (7.66~km and 10~km, Table~\ref{tab: test pl parameters}) compared to the size of the nucleus, which may have impacts on the inner coma. Additionally, there is no feedback from collisional processes onto the fields in the test particle simulations. Cooling of electrons in the inner coma reduces the electron pressure gradient and consequently the ambipolar field in the inner coma \citep{Stephenson2022TestPl}. \comments{The quenched ambipolar field in the inner coma results in a potential well that is flatter and reduced in depth by 20\% close to the nucleus (grey region, Figure~\ref{fig: ambipolar well}), compared to when collisions are neglected. However, the ambipolar field is still efficient at trapping of cometary electrons near the nucleus and accelerating solar wind electrons into the inner coma, as the potential well is built up over a large range of cometocentric distances, including regions without substantial electron cooling.} 

\comments{The flattening of the potential well in the inner coma would result in a region of constant ionization frequency, as energetic solar wind electrons would not be funnelled closer to the nucleus or undergo further acceleration. In Figure~\ref{fig: test pl r vs ioni freq}, this would lead to a plateau in the ionization frequency below $\approx50$\,km, and a deviation away from the $1/r$ trend, which occurs without any collisional feedback.}

\section{Ion acceleration and electron impact ionization away from perihelion}\label{sec: discussion}
\begingroup
\renewcommand*{\arraystretch}{1.5}
\begin{table}
    \centering
    \caption{Set of test cases to examine ion acceleration in the coma. Rosetta has been set at a cometocentric distance of 30~km. }
    \label{tab: ion acc cases}
    \begin{tabular}{c c c}
         Case & Ion bulk velocity & Electron-impact freq \\
         \hline
         1 & Constant at 0.6~km\,s$^{-1}$ & Constant at $3 \times 10^{-7}$~s$^{-1}$ \\
         2 & Constant at 5~km\,s$^{-1}$ & Constant at $3 \times 10^{-7}$~s$^{-1}$ \\
         \thead{3\\{ }\\{ }} & \thead{Linear increase. \\0.6~km\,s$^{-1}$ at surface. \\ 5~km\,s$^{-1}$ at Rosetta.} & \thead{Constant at $3 \times 10^{-7}$~s$^{-1}$ \\{ } \\{ } }\\
         \thead{4\\{ }\\{ }} & \thead{Linear increase.\\ 0.6~km\,s$^{-1}$ at surface.\\ 5~km\,s$^{-1}$ at Rosetta.} & \thead{$\propto 1/r$.\\ $3 \times 10^{-7}$~s$^{-1}$ at Rosetta \\ { }}\\
    \end{tabular}
    
\end{table}
\endgroup

\begin{figure}
    \centering
    \begin{tikzpicture}
        \node (exb) at (0,0) {\includegraphics[width = 0.5 \textwidth]{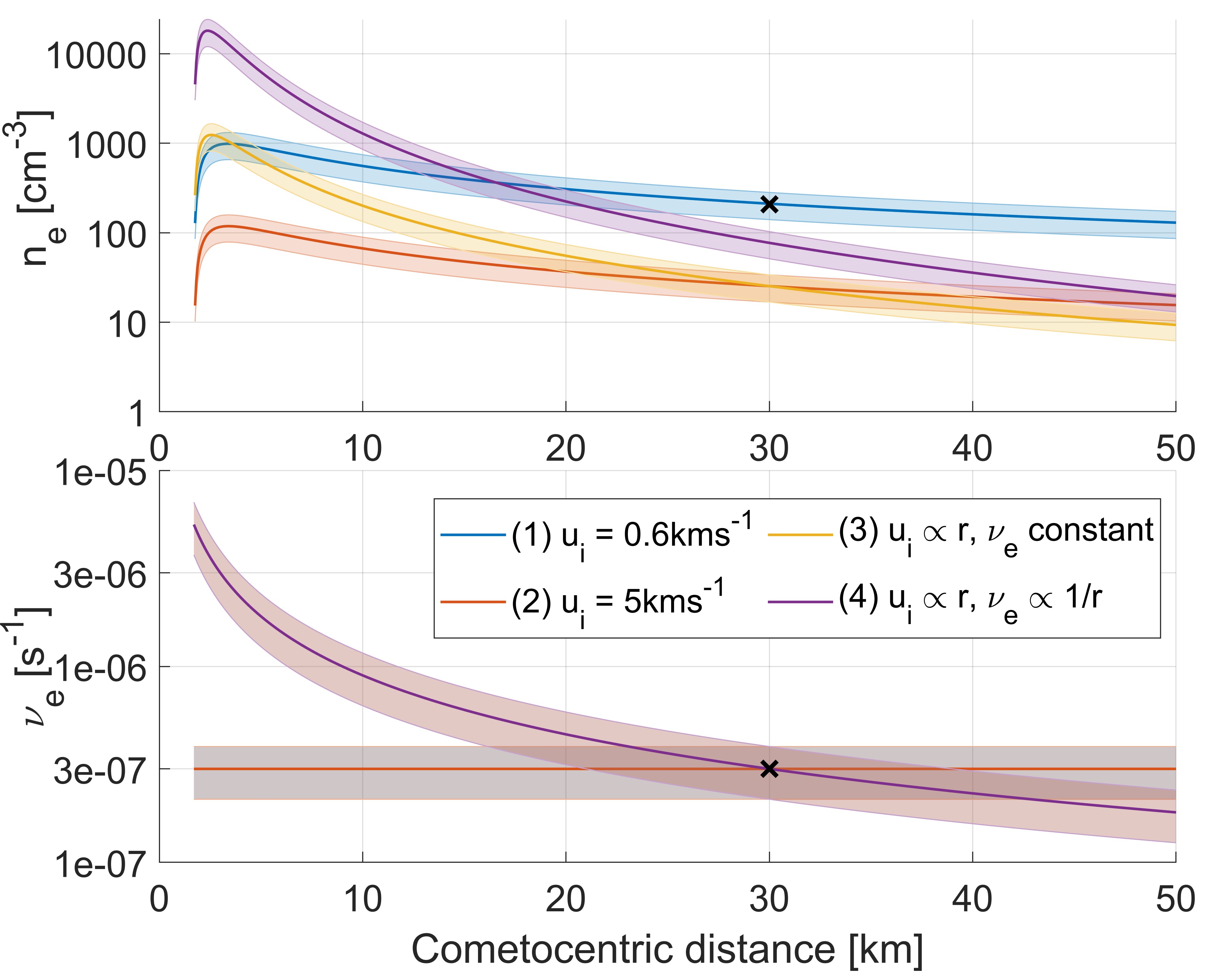}};
        \node [above left = 2.9 and 4.1 of exb.center, font = \bf] (A) {(a)};
        \node [above left = 3.1 and 4.1 of exb.base, font = \bf] (B) {(b)};

    \end{tikzpicture}
    
    \caption{(a) Electron density and (b) electron-impact ionization frequency as a function of cometocentric distance for the four test cases (1) - (4) outlined in Table~\ref{tab: ion acc cases}. The black crosses give the electron density and ionization frequency measured at Rosetta (at 30~km). The ionization frequencies for cases (1) - (3) are identical in (b). \edits{The error bars allow for 10\% error in the outgassing rate and outflow velocity, as well as 30\% error on the electron impact ionization frequency.}}
    \label{fig: ion acc cases}
\end{figure}
\begin{figure}
    \centering
    \includegraphics[width = 0.5\textwidth]{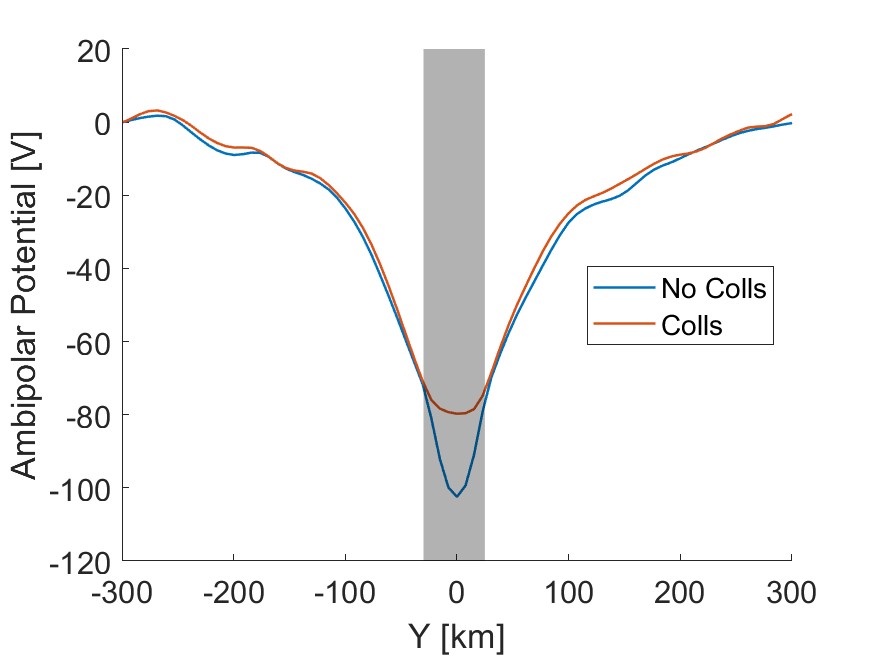}
    \caption{\edits{Ambipolar potential well along $y$ at $x,z=7.66$~km for a collisionless (blue) and collisional (red) electron test particle simulation at $Q=10^{26}$\,s$^{-1}$ (see Table~\ref{tab: test pl parameters}). The collisional region of the coma is highlighted in grey and is associated with a flattened potential well.}} 
    \label{fig: ambipolar well}
\end{figure}

Multi-instrument models have accurately reproduced the measured electron density (and ion density through quasineutrality) at Rosetta, when far from perihelion \citep{Galand2016, Heritier2017Vert, Heritier2018source}. These are based on the continuity equation for the total ion density, which is given by \citep{Galand2016}: 
\begin{equation}\label{eq: ion cont eq}
    \frac{1}{r^2}\frac{\mathrm{d}}{\mathrm{d}r}\bigg(n_i(r) r^2 u_i(r)\bigg) = P_i(r) - L_i(r),
\end{equation}
where $n_i$ is the total ion density, $P_i$ is the production rate of ions through ionization and $L_i$ is the chemical loss rate through electron-ion recombination. Far from perihelion, transport dominated over recombination at 67P, so the last term can be neglected \citep{Galand2016}. \edits{In Eq.~\ref{eq: ion cont eq}, we have assumed a spherically symmetric cometary plasma so $u_i$ is only the radial ion velocity.}

The multi-instrument approaches assume ions flow radially outwards at the neutral gas velocity and that the ionization frequency is constant throughout the coma (in the optically thin regime).
\edits{The neutral coma and electron-ionization frequency are derived from measurements at Rosetta taken at similar times.}

The production $P_i$ of cometary ions and electrons are dominated by electron-impact ionization at large heliocentric distances (see Section~\ref{sec: EII Rosetta}). \edits{Therefore, the electron density measured at Rosetta $r_{Ros}$ is subject to the outgassing rate and \comments{to} two main, competing drivers: (1) the electron impact ionization frequency at $r<r_{Ros}$ \comments{(through $P_i = (\nu_{h\nu}^{ioni}+\nu_e^{\text{ioni}}) n_n$, where $n_n = Q/(4\pi u_{gas}r^2)$ is the neutral density)} and (2) the ion velocity, $u_i(r)$ \comments{(contained in the left hand side of Eq. \ref{eq: ion cont eq})} at Rosetta. The outgassing rate and EII frequency at Rosetta varied by orders of magnitude within several hours, so the multi-instrument modelled densities must be driven by almost-concurrent measurements at Rosetta and cannot be used to explain the electron densities over long periods with a single set of inputs.

Given that the electron density magnitude and variation is well captured with the multi-instrument model \citep{Galand2016, Heritier2017Vert, Heritier2018source}, an increase from the modelled ion velocity would have to be compensated by increasing the modelled EII frequency in the coma. However, the EII frequency is also constrained by measurements at Rosetta, so would have to be varied between Rosetta and the nucleus' surface.}

Away from perihelion, cometary ion radial bulk and effective total speeds have been measured in excess of 10~km\;s$^{-1}$ in the inner coma by RPC/LAP \citep{Johansson2021plasDens} and RPC/ICA \citep{Nilsson2020}, which have substantial implications on the EII frequency throughout the coma. 
\comments{The key uncertainty in estimates of the ion bulk velocity is the impact of the spacecraft potential, which acclerates and deflects ions near the spacecraft, with low energy cometary ions being most strongly deflected \citep{Bergman2020VSC}. Low energy water ions moving through a typical spacecraft potential of $V_{S/C}=-15$\,V could increase in speed by up to 12.5\,km\,s$^{-1}$.}

For ions to achieve the large measured radial outflow velocities, they must undergo substantial acceleration in the inner coma, as they are born at the neutral flow speed \citep[$0.4-1$~km\,s$^{-1}$;][]{Marshall2017, Biver2019}. The radial ion velocity measurements are substantially larger than estimated by \cite{Vigren2017a} at $Q=2\times10^{27}$~s$^{-1}$. The 1D model with a constant ambipolar field resulted in an ion velocity of 2.5\,km\,s$^{-1}$ at 100\,km.

\edits{For consistency to be maintained between the multi-instrument models of electron density and the high estimates of ion bulk velocity, higher EII frequencies are required close to the nucleus. The effective bulk speeds $\sim$10~km\;s$^{-1}$ could have non-radial components, which would moderate the high EII frequencies needed to maintain the observed plasma density.}

\edits{There \comments{has} also been a number of measurements of the ion bulk velocities close to perihelion, particularly in the vicinity of the diamagnetic cavity \citep{Bergman2021cavSpeeds, Bergman2021CavityFlow}.These also find ions with large bulk speed ($>5$\,km\,s$^{-1}$), as well as a population of ions streaming towards the comet. The presence of the diamagnetic cavity substantially alters the plasma environment and the coma is much more collisional than at large heliocentric distances. Therefore, this discussion only applies to the case at large heliocentric distances and does not comment on the conditions or measurements around perihelion.}

Figure~\ref{fig: ion acc cases} illustrates the ion density profile at a point in time derived under the prior assumptions of the multi-instrument \comments{model} approach (blue, case 1), using an outgassing of $Q=10^{26}$~s$^{-1}$. The conditions for each test case are outlined in Table~\ref{tab: ion acc cases}. The electron density and EII frequency are both measured at Rosetta, which we have chosen to be at $r=30$~km (black crosses, Figures~\ref{fig: ion acc cases}a and b). \edits{This was a typical cometocentric distance of Rosetta when away from perihelion.}

We have neglected photoionization throughout the coma, as EII was found to be the dominant source of electrons in the coma far from perihelion (see Figure~\ref{fig: CPR ioni freq fig}), \comments{especially during the early phase of the mission}. If the ions undergo linear acceleration and the electron-impact ionization frequency varies as $\nu^{\text{ioni}}_{e}(r)\propto \frac{1}{r^n}$, the plasma density \comments{from Eq.~\ref{eq: ion cont eq}} is given by:

\begin{equation}\label{eq: plasma dens EII r^n}
\begin{split}
n_i(r) = \frac{Q \nu^{\text{ioni}}_{e}(r_{Ros})r_{Ros}}{4\pi u_{gas}\times u_{i}(r)\times r^2} \times \quad\quad\quad\quad\quad\quad\quad\quad\quad\\ \quad\quad\quad\quad\begin{cases}
\log(\frac{r}{r_{67P}}) & \text{if } n = 1\\
\frac{1}{n-1}\big(\frac{r_{Ros}}{r}\big)^{n-1}\bigg[\big(\frac{r}{r_{67P}}\big)^{n-1} - 1\bigg] & \text{otherwise};
\end{cases}
\end{split}
\end{equation}
\comments{using the boundary condition $n_i=0$ at $r=r_{67P}$.} 

\comments{The relation $\nu^{\text{ioni}}_{e}(r)\propto r^{-1}$ is observed in the test particle simulations (though this conclusion has limitations discussed hereafter), particularly at $Q=10^{26}$\,s$^{-1}$ (see Figure~\ref{fig: test pl r vs ioni freq}a). It has previously been used in studies of FUV emissions, when all electrons were assumed to be energetic enough to excite \citep{Feldman2015}. This is indeed the dependence expected in the case where the EII frequency follows the total electron density.}

\comments{From in-situ observations,} the variation of EII frequency with cometocentric distance at comet 67P \comments{remains} uncertain at large heliocentric distances. The data from the Rosetta mission do not show a clear correlation to or consistent behaviour with cometocentric distance (see Section~\ref{sec: eii vs rcom}).

The test particle simulations suggest that the EII frequency could increase by a factor of 10 within 100~km of the nucleus. However, these results do not include any feedback of collisional processes onto the electric and magnetic fields, which may damp the electron-impact ionization in the inner coma (see Section~\ref{sec: test pl eii vs r com}), resulting in a more constant ionization frequency close to the nucleus.

Cases 2 (red) and 3 (yellow; Figure~\ref{fig: ion acc cases}a) show the ion density profiles with higher ion velocities, but no change in the EII frequency \edits{(same as Case 1, blue)}. The radial ion velocity is set to 5~km\;s$^{-1}$ at Rosetta, an average radial velocity according to the RPC/LAP and RPC/ICA measurements away from perihelion. In both scenarios, the ion density cannot be reproduced at Rosetta, regardless of whether the ions are born at 5~km\;s$^{-1}$ (red) or are accelerated from $u_{gas} = 0.6$\,km\,s$^{-1}$ in the coma (yellow). This is inconsistent with the constraints of the multi-instrument models.

\edits{In Case 4, we have increased the ionization frequency (purple, Figure~\ref{fig: ion acc cases}b) close to the nucleus to compensate for the high ion velocities, using a $\nu_e^{\text{ioni}}\propto1/r$ profile.} This profile assumes the EII frequency varies similarly to the bulk plasma density, as has previously been used in FUV emission models \citep{Feldman2015}. This relation with ionization frequency restores the electron density to within a factor of 3 of those observed at Rosetta.

However, the energetic electron population (which drives EII) is greatly boosted near the nucleus in Case 4, compared to the previously assumed case (yellow, Figure~\ref{fig: ion acc cases}b). The energetic electrons are increased by a factor of 10 in the inner coma, and would need to be further increased to fully reproduce the plasma density measured at Rosetta. Using $\nu_e^{\text{ioni}}\propto1/r^{1.89}$ would fully resolve the discrepancy in the electron density (purple vs. black cross, Figure~\ref{fig: ion acc cases}a). This profile would increase the  energetic electron population in the inner coma even further. 

\edits{While this would reconcile the high ion velocity and plasma density measurements, the EII frequency throughout the coma is not a free parameter. Multi-instrument models of FUV emissions from the coma have shown that emissions are driven by electron impact and are consistent with a constant EII frequency along the line of sight \citep{Galand2020, Stephenson2021FUV}, as was concluded by \cite{Chaufray2017}.} FUV emissions are driven by the same energetic electron population as the ionizing electrons, so the emission frequency is expected to show the same spatial variation as the EII frequency. The FUV emissions are also more sensitive to the EII frequency in the inner coma than the electron density, which is accumulated over the radial column. Each shell of width $\mathrm{d}r$ contributes $\nu_e^{em}(r) \mathrm{d}r / r^2$ to the total brightness and $\nu_e^{\text{ioni}}(r) \mathrm{d}r$ to the electron density, where $\nu_e^{em}(r)$ is the emission frequency. 

Consequently, an increase by a factor 10-15 in the energetic electron population in the inner coma would result in an increase in FUV emissions by an order of magnitude compared to the case of constant electron flux. While there are uncertainties in the multi-instrument FUV studies, an increase in the energetic electron population by a factor of 10 would \comments{produce} a clear discrepancy between the modelled and observed emission brightnesses.  

Additionally, the direct measurements of the energetic electrons by RPC/IES do not show any clear variation with cometocentric distance, when at large heliocentric distance (see Figure~\ref{fig: ioni freq vs R com hist}). This is reinforced by the RPC/IES measurements at the end of mission, which did not vary significantly down to the surface \citep{Heritier2017Vert}.

In contrast, the test particle simulations do indicate that the ionization frequency should increase closer to the nucleus (see Figure~\ref{fig: test pl r vs ioni freq}). However, the EII frequency increases by only a factor 3 within 30~km of the nucleus. This is much less than the required increase to reconcile the ion flow speed and electron density measurements.
\begin{figure}
\begin{tikzpicture}

\node (diag) at (0,0) {\includegraphics[width = 0.5\textwidth]{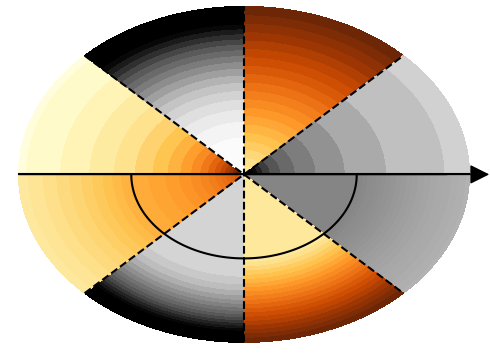}};
\node [above = 3.5 of diag.center, font=\large\bf] (noColl) {No Collisional Cooling};
\node [below = 3.5 of diag.center, font=\large\bf] (Coll) {Collisional Cooling};
\node [above left = 1.1 and 3.7 of diag.center, font=\large\bf] (ne) {$\bm{n_e}$};
\node [below left = 1.1 and 3.7 of diag.center, font=\large\bf] (ne) {$\bm{n_e}$};
\node [above right = 1.1 and 3.8 of diag.center, font=\large\bf] (ne) {$\nu_e^{\text{ioni}}$};
\node [below right = 1.1 and 3.8 of diag.center, font=\large\bf] (ne) {$\nu_e^{\text{ioni}}$};
\node [above right = 3.1 and 1.1 of diag.center, font=\large\bf] (ne) {$\bm{V_{Pot}}$};
\node [below right = 3.1 and 1.1 of diag.center, font=\large\bf] (ne) {$\bm{V_{Pot}}$};
\node [above left = 3.1 and 1.1 of diag.center, font=\large\bf] (ne) {$\bm{u_{i}}$};
\node [below left = 3.1 and 1.1 of diag.center, font=\large\bf] (ne) {$\bm{u_{i}}$};
\node [right = 4.4 of diag.center, font=\large\bf] (ne) {$\bm{r}$};

\end{tikzpicture}
\caption{Illustration of how electron density $n_e$, ion bulk speed, $u_i$, ambipolar potential and EII frequency may vary with cometocentric distance in a coma (top half) without collisional cooling and (bottom half) with collisional cooling. The black line in the bottom half bounds the collisional region of the coma, inside which the potential, EII frequency and ion bulk speed are constant. Inside the collisional region, the ion speed, ambipolar potential, and EII frequency are constant. In the non collisional case, the ions accelerate throughout the coma, from the comet surface, and the ambipolar potential decreases all the way to the surface. This also results in a peak in the EII frequency very close to the nucleus.}
\label{fig: discussion diagram}
\end{figure}
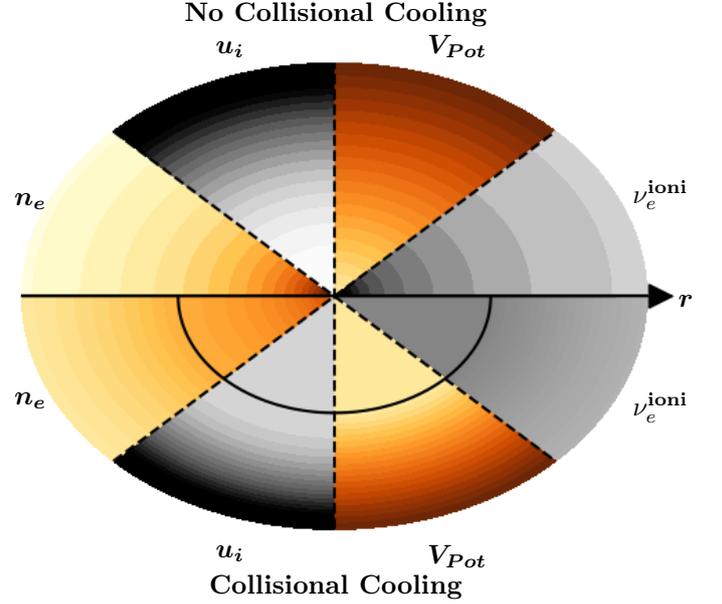
\edits{The test particle simulations also lack any feedback from electron collisions on the fields in the simulation. Using the outputs of the test particle simulations, we can estimate the feedback of collisional process on the ambipolar field using the Generalised Ohm's Law \citep{Stephenson2022TestPl}:
\begin{equation}\label{eq: ambipolar field}
    \bm{E}_{Ambi} = -\frac{1}{e n_e}\nabla(n_e k_B T_e).
\end{equation}
Figure~\ref{fig: ambipolar well} shows the ambipolar potential well along the magnetic field line that passes through (10, 0, 10)~km at $Q=10^{26}$~s$^{-1}$, for a collisionless (blue) and collisional (red) electron simulation. In the collisionless case, the potential well would lead to an increasing ionization frequency down to the surface as solar wind electrons undergo acceleration into the potential well. This would be consistent with Case 4 and the ion speed measurements, although the increase in ionization frequency would \comments{still} not be of the required magnitude  (see Figure~\ref{fig: test pl r vs ioni freq}a).

When collisions are included the electron temperatures are reduced in a collisional region in the inner coma \citep{Stephenson2022TestPl}. This reduces the ambipolar field strength and flattens the potential well near the nucleus. In the grey highlighted region, electrons would undergo little acceleration and the electron impact ionization frequency would be approximately constant. This is consistent with the approach in Case 1 and used for the multi-instrument observations. The consistent cold electron observations at Rosetta, even at large heliocentric distances, indicate that the spacecraft was often in this region of lower temperatures and the flattened ambipolar potential \citep{Eriksson2017, Engelhardt2018, Gilet2020, Wattieaux2020}. Therefore, the case of constant EII frequency between Rosetta and the surface (as used in Case 1) is the more reasonable assumption when the electric fields are accounted for. Beyond the collisional region (grey Figure~\ref{fig: ambipolar well}), acceleration of the cometary ions is expected. The 80~\comments{V} potential difference translates to 28~km\,s$^{-1}$ for water ions, although this acceleration, which occurs over 300\,km (potential well scale) is small compared to the acceleration by pick up ($E_{conv} = -2.4\times10^{-3}$\,V\,m$^{-1}$ results in 720\,V over 300~km). } 

While there may be some variation in the EII frequency with cometocentric distance, a variation of factor 10 within 30~km of the nucleus is inconsistent with the particle spectral measurements and the FUV emissions \comments{on the one hand}, and \comments{with the} results of the test particle model \comments{on the other hand}. \edits{When allowing the EII frequency to increase by a factor of 3 close to the nucleus (more than any discrepancy seen in the multi-instrument FUV studies), the maximum radial ion velocity allowed is twice the outflow gas velocity, significantly smaller than the estimated ion bulk speeds \comments{from Rosetta observations}.} Direct measurements of the energetic electrons do not suggest such substantial enhancement in the inner coma.

\comments{Figure~\ref{fig: discussion diagram} illustrates how the electron density, ion bulk speed, ambipolar potential and EII frequency may vary in a weakly collisional coma (bottom half) compared to the case without collisions (top half), with the collisional region bounded by the black line in the bottom half. Within the collisional region ion bulk speed, the potential and EII frequency are approximately constant with cometocentric distance, consistent with the multi-instrument models. Outside the collisional region, the ambipolar potential increases, leading to ion acceleration and reduction of the EII frequency (possibly with a $1/r$ profile as seen in Figure\,\ref{fig: test pl r vs ioni freq})}.

As the EII frequency does not substantially vary between Rosetta and the nucleus at large heliocentric distances, it still remains unclear how the large radial ion bulk speeds can be reconciled with the direct measurements and multi-instrument models of the plasma density and FUV brightnesses. 

\section{Conclusion}\label{sec: conclusion}
We have calculated a new dataset of the electron-impact ionization frequency throughout the Rosetta mission, using data from RPC/IES. Electron-impact ionization is the dominant source of electrons in the coma at large heliocentric distances (Figure~\ref{fig: CPR ioni freq fig}) and varies by up to three orders of magnitude over short periods. Far from perihelion, the electron density at Rosetta is strongly dependent on the EII frequency (Figure~\ref{fig: ioni freq vs el dens}). This relationship is also seen in respect to the spacecraft potential and the cometary ion density (Figure~\ref{fig: EII vs VSC}). 

Around perihelion, the EII frequency is much less variable (one order of magnitude) and is less ionizing than photoionization (see Figure~\ref{fig: CPR ioni freq fig}). Hence, it is not surprising that the electron and ion populations are not strongly dependent on the EII frequency near perihelion (Figures~\ref{fig: ioni freq vs el dens} and~\ref{fig: ioni freq vs Low energy ion Pop}).

Away from perihelion, the EII frequency is positively correlated with the solar wind potential difference (estimated from RPC/ICA measurements; see Figure~\ref{fig: EII vs potwell}), confirming that the ionizing electrons originate in the solar wind and are accelerated in the cometary environment. The solar wind electrons are accelerated by an ambipolar potential well. 

The EII frequency is well correlated with magnetic field strength (Figure~\ref{fig: ioni freq vs B Mag}), indicating that EII is more ionizing in regions where the solar wind slows down and piles up. 
The outgassing rate (Figure~\ref{fig: ioni freq vs. outgassing inb vs outb}) is also a key driver of the EII frequency, although non-linearly. At low outgassing rates, the EII frequency increases with outgassing \comments{rate}, before plateauing, and subsequently \comments{decreasing}. This may reflect the variation in the ambipolar potential well with outgassing. At low outgassing, there are few electron-neutral collisions in the coma so increasing outgassing leads to a deeper potential well, \comments{as the spatial scale of the cometary ionosphere increases.}
At intermediate rates, collisions start to become important and cold electrons form in the inner coma. This feedback reduces the ambipolar field in the inner coma and limits the depth of the potential well. As the outgassing increases further, the electron cooling increases and extends farther from the nucleus and the ambipolar potential well is damped. 

The origin of electrons within the coma is simulated using a 3D collisional test particle model of electrons at a comet (see Section~\ref{sec: test pl results}). At outgassing rates of $Q=10^{26}$~s$^{-1}$ and $Q=1.5\times10^{27}$~s$^{-1}$, electron-impact ionization was found to be the major source of electrons within the coma (Figure~\ref{fig: test pl elec dens}) and the ionizing electrons are solar wind electrons, or cometary electrons that have been produced by solar wind electron-impacts. The test particle model also shows a positive correlation between magnetic field strength, as was seen in the Rosetta data. 

There is not a clear variation in the EII frequency with cometocentric distance away from perihelion, using the RPC/IES data. 
In the test particle simulations, the EII frequency does increase closer to the nucleus, as the electrons are accelerated into the inner coma by the ambipolar potential well. However, there is no feedback of collisions onto the fields in the test particle model, which damps the ambipolar field in the inner coma and limits the EII frequency. \edits{When the feedback of collisions on the ambipolar field is considered (see Figure~\ref{fig: ambipolar well}), the potential well has a much flatter shape in the collisional region of the coma. This would result in a roughly constant ionization frequency in the collisional region, where cold electrons are present and electron temperatures are low. Cold electrons were observed throughout the Rosetta mission, so there should be little difference in the ambipolar potential between Rosetta and 67P.} 
\comments{For more thorough examination of this feedback between collisional processes and the fields in the cometary environment, a collisional PiC model would be necessary. This approach would also allow application to higher outgassing rates, where the test particle approach is limited by the lack of feedback between fields and collisions.}
We consider recent measurements of the ion flow speeds at comet 67P \edits{far from perihelion}, which indicate ions undergo acceleration from the neutral outflow speed to over 5~km\;s$^{-1}$ between production and Rosetta. This acceleration would have to be compensated by increased electron-impact ionization within the inner coma, to restore the electron density to the observed levels. However, the required factor 10-15 increase in EII frequency close to the nucleus is inconsistent with the FUV emissions from the coma. It is also well beyond the increase in EII frequency in the test particle model (factor 3 within 30~km), which can be treated as an upper bound. 

The discrepancy between the ion flow speed measurements and the multi-instrument studies \edits{applied to large heliocentric distances} \comments{remains} an outstanding question, but cannot reasonably be explained by an arbitrary enhancement in the electron-impact ionization close to the nucleus.

\section*{Acknowledgements}
We gratefully acknowledge the work of the many engineers, technicians and scientists involved in the Rosetta mission. In particular, we acknowldege the work of the ROSINA team, with PI K. Altwegg, and the RPC instrument teams, including the PIs: A. Eriksson of RPC/LAP, J. Burch of RPC/IES, KH. Glassmeier of RPC/MAG and C. Carr of RPC/PIU. Without their contributions, ROSINA and RPC would not have produced such outstanding results.
Rosetta is an ESA mission with contributions from its member states and NASA.

Work at Imperial College has been supported by the Science and Technology Facilities Council (STFC) of the UK under studentships ST/S505432/1 and ST/W507519/1 and under grants ST/5000364/1 and ST/W001071/1. Work at the University of Bern was funded by the Canton of Bern and the Swiss National Science Foundation (200020\_207312). Work at LPC2E and Lagrange laboratories is co-funded by CNES. Work at Umeå University was supported by the Swedish National Space Agency (SNSA) grant 108/18. J. D. gratefully acknowledge support from NASA’s Rosetta Data Analysis Program, Grant No. 80NSSC19K1305, and the NASA High-End Computing (HEC) Program through the NASA Advanced Supercomputing (NAS) Division at Ames Research Center.
\section*{Data Availability}
The data underlying this article are available on the PSA at \url{https://psa.esa.int}. The electron-impact ionization frequencies are available at \url{10.5281/zenodo.8048344}. 
\bibliography{my_collection}

\begin{appendices}
\section{Variation of electron impact ionization frequency during the Rosetta mission}
In this section, we show the correlation of the electron-impact ionization frequency (EII) with additional properties of the cometary environment to those discussed in Section~\ref{sec: EII Rosetta}.
\subsection{EII vs. heliocentric distance}\label{sec: EII vs rh}
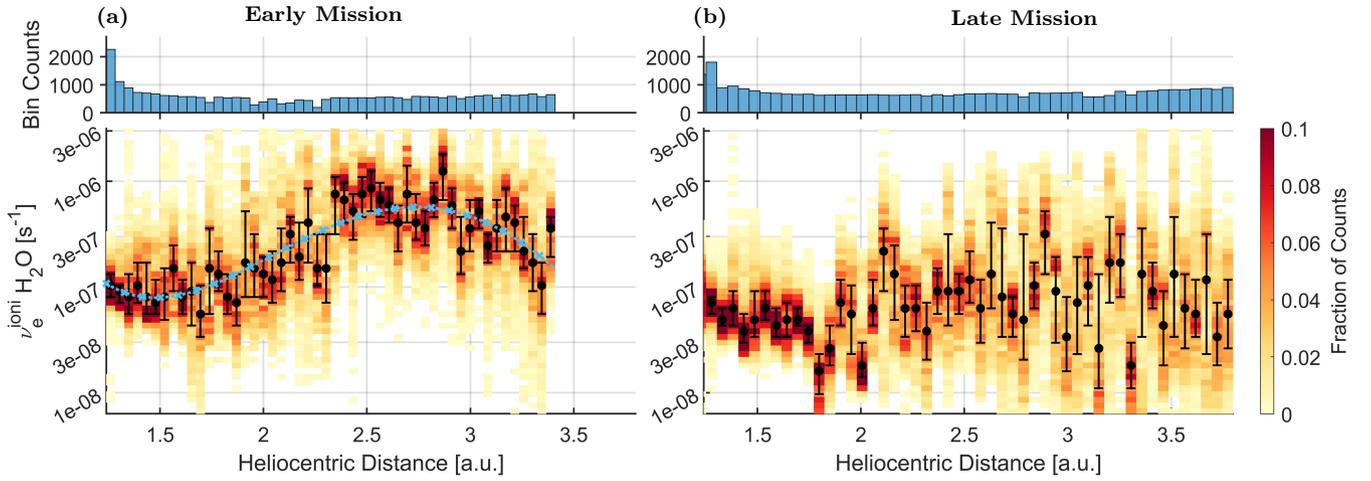
\begin{figure*}
\begin{tikzpicture}

\node (USW) at (0,0) {\includegraphics[width = \textwidth]{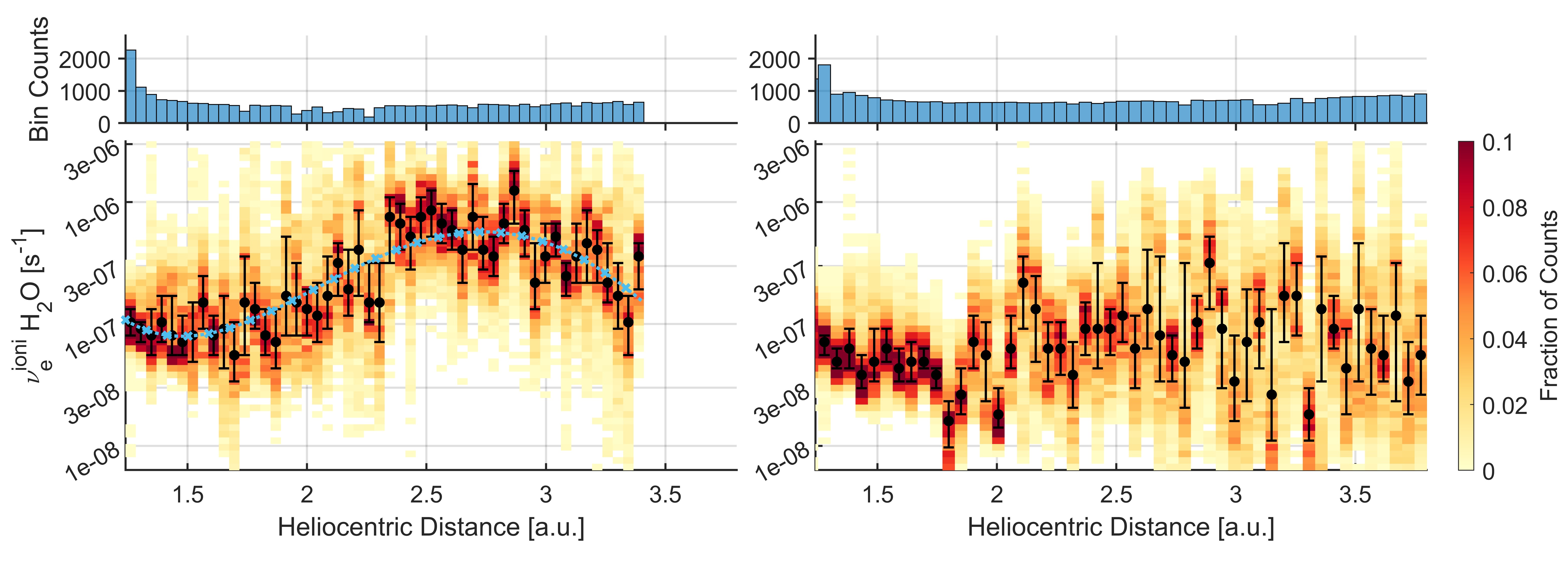}};
 \node [above left = 3.0 and 3.5 of USW.center, font = \bf] (USWLab) {Early Mission};
    \node [above right = 3.0 and 3.5 of USW.center, font = \bf] (OutBLab) {Late Mission};
    \node [above left = 6.3 and 7.1 of USW.base, font =\bf] (A) {(a)};
    \node [above right = 6.3 and 0.1 of USW.base, font =\bf] (B) {(b)};
\end{tikzpicture}
\caption{Electron-impact ionization frequency vs. the heliocentric distance of comet 67P (a) pre-perihelion and (b) post-perihelion. Median (black circles) and quartile values (error bars) are plotted for each bin along the $x$-axis. The top panels show the number of counts for each bin of heliocentric distance. The fit to the EII frequency with heliocentric distance is shown in blue crosses in the left panel (see Table~\ref{tab: ioni freq fits}).}
\label{fig: EII vs rh}
\end{figure*}

Firstly, Figure~\ref{fig: EII vs rh} shows the relationship between heliocentric distance and EII frequency, in addition to the fit to the EII frequency outlined in Table~\ref{tab: ioni freq fits}. In the early mission period, the EII frequency varies non-linearly with heliocentric distance, reflecting the same behaviour as observed in the relationship with local outgassing rate (see Figure~\ref{fig: ioni freq vs. outgassing inb vs outb}). However, post-perihelion the correlation with heliocentric distance becomes much weaker and shows significant variance around any trend. This may result from the more asymmetric outgassing rate post perihelion, with strong \ce{CO2} outgassing from the southern hemisphere \citep{Laeuter2018}.

\subsection{EII vs. Spacecraft potential}\label{sec: EII vs VSC}
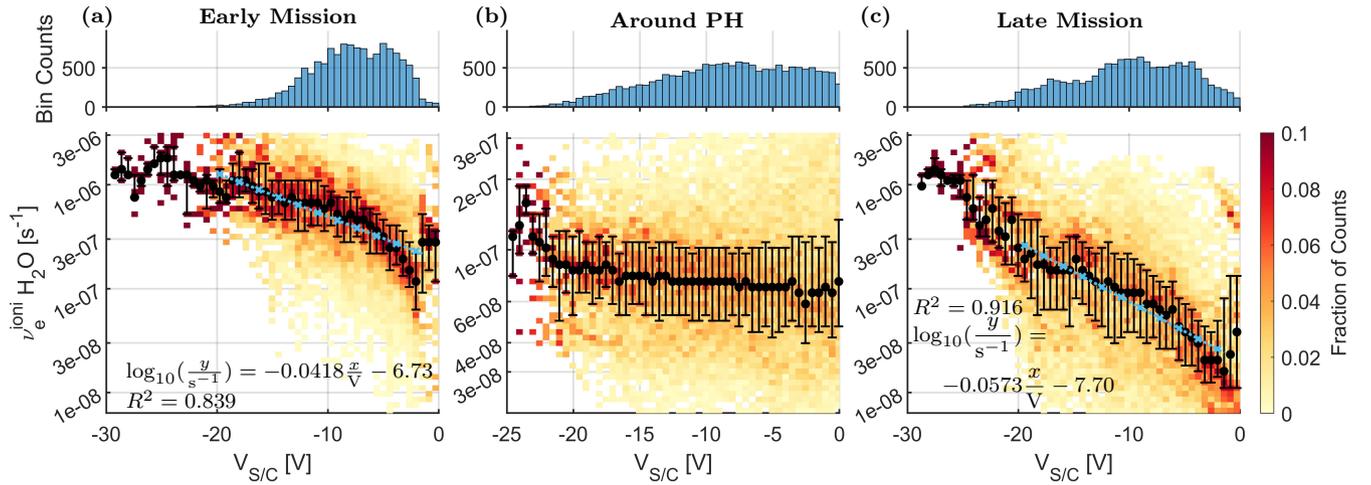
\begin{figure*}
\begin{tikzpicture}

\node (eIoni) at (0,0) {\includegraphics[width = \textwidth]{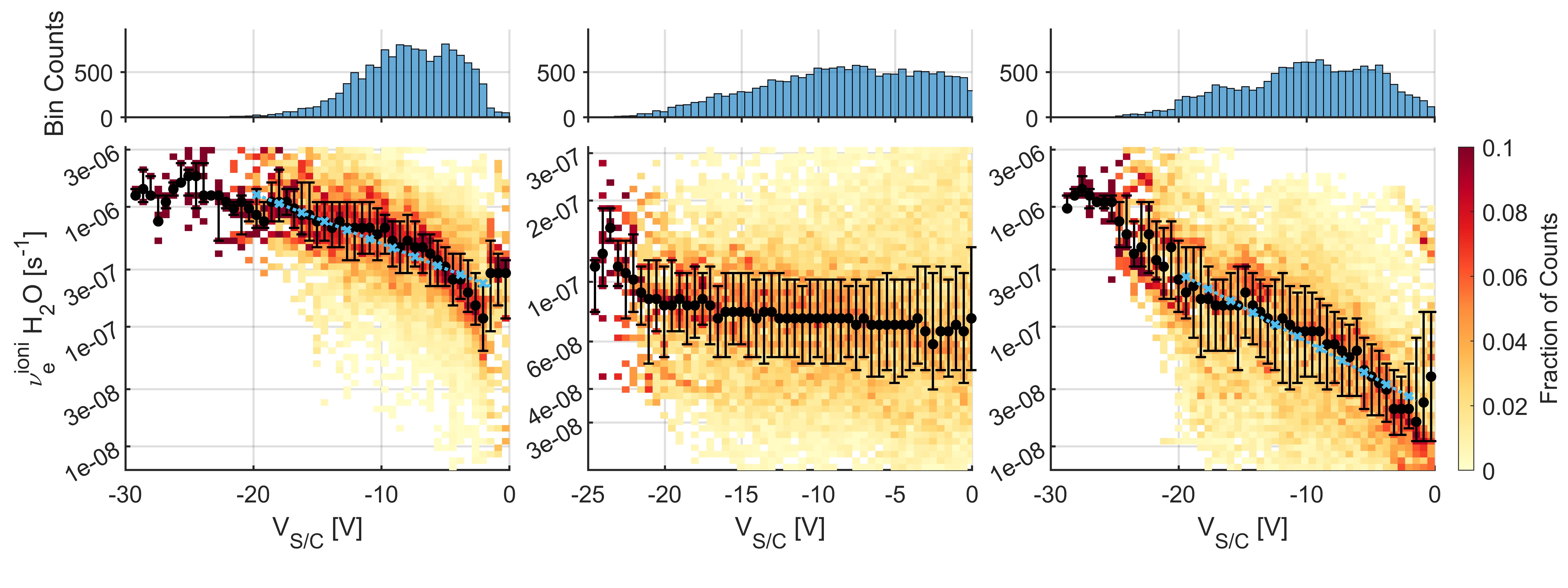}};
\node [above left = 6.3 and 4.1 of eIoni.base, font = \bf] (EarlyLab) {Early Mission};
        \node [above = 6.3 of eIoni.base, font = \bf] (PHLab) {Around PH};
        \node [above right = 6.3 and 4.1 of eIoni.base, font = \bf] (LateLab) {Late Mission};
        \node [above left = 6.3 and 7.3 of eIoni.base, font =\bf] (A) {(a)};
        \node [above left = 6.3 and 2.1 of eIoni.base, font =\bf] (B) {(b)};
        \node [above right =6.3 and 2.3 of eIoni.base, font =\bf] (C) {(c)};
        \node [above left = 1.3 and 3.1 of eIoni.base, align=left, font =\bf] (R2a) {$\log_{10}(\frac{y}{\mathrm{s}^{-1}}) = -0.0418\frac{x}{\mathrm{V}} -6.73$\\$R^2=0.839$};
        \node [above right = 1.3 and 3.0 of eIoni.base, align=left, font =\bf] (R2c) {$R^2=0.916$\\$\begin{aligned}
        \log_{10}(\frac{y}{\mathrm{s}^{-1}}) &=\\ -0.0573&\frac{x}{\mathrm{V}} -7.70\end{aligned}$};
\end{tikzpicture}
\caption{Electron-impact ionization frequency vs. the spacecraft potential measured by RPC/LAP (a) early in the mission, (b) around perihelion and (c) late in the mission. Median (black circles) and quartile values (error bars) are plotted for each bin along the $x$-axis. The top panels show the number of intervals for each bin along $x$. \comments{Linear fits are given for the early and late periods (blue crosses).}}
\label{fig: EII vs VSC}
\end{figure*}

The EII frequency is negatively correlated with the spacecraft potential (Figure~\ref{fig: EII vs VSC}), as measured by RPC/LAP, during both the early and late mission periods (see Table~\ref{tab: ioni freq time periods}). Around perihelion, the EII frequency is much less dependent on the spacecraft potential and is largely constant at $\nu_e^{\text{ioni}}\approx8\times10^{-8}$\,s$^{-1}$.

The spacecraft potential is a function of the electron temperature and density \citep{Odelstad2015b, Johansson2021plasDens}, so the correlations in Figure~\ref{fig: EII vs VSC} can be interpreted in the same way as Figure~\ref{fig: ioni freq vs el dens}. Denser, hotter plasmas result in higher EII frequencies and more negative spacecraft potentials. This behaviour is seen at large heliocentric distances, as electron impact ionization was the major source of electrons within the coma during these periods. 
\subsection{EII vs magnetic field angles}\label{sec: EII vs mag angles}
\begin{figure*}
\begin{tikzpicture}

\node (eIoni) at (0,0) {\includegraphics[width = \textwidth]{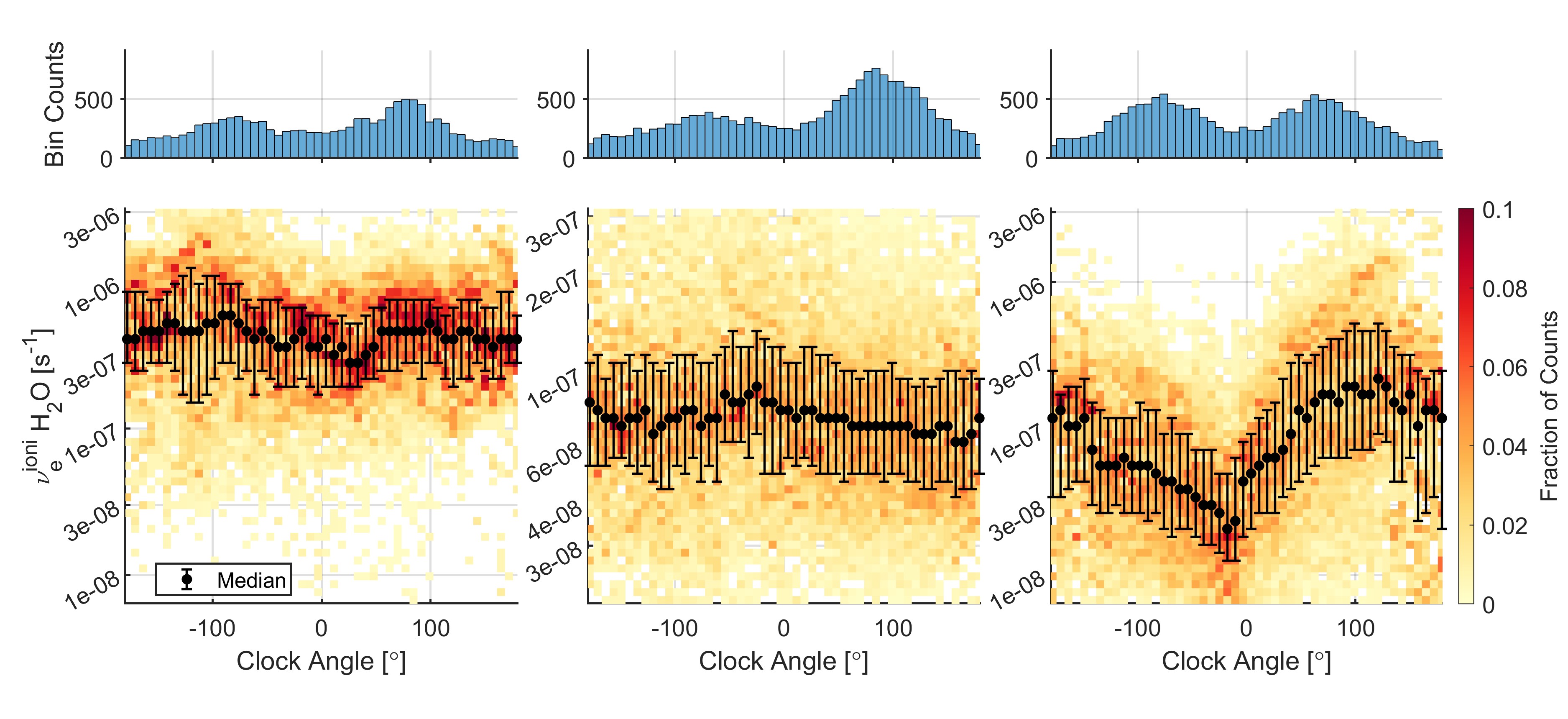}};
 \node [above left = 7.5 and 4.1 of eIoni.base, font = \bf] (EarlyLab) {Early Mission};
\node [above = 7.5 of eIoni.base, font = \bf] (PHLab) {Around PH};
\node [above right = 7.5 and 4.1 of eIoni.base, font = \bf] (LateLab) {Late Mission};
\node [above left = 7.5 and 7.3 of eIoni.base, font =\bf] (A) {(a)};
\node [above left = 7.5 and 2.1 of eIoni.base, font =\bf] (B) {(b)};
\node [above right =7.5 and 2.3 of eIoni.base, font =\bf] (C) {(c)};
\end{tikzpicture}
\caption{Electron-impact ionization frequency vs. the clock angle of the magentic field (see Eq.~\ref{eq: clock angle}) (a) early in the mission, (b) around perihelion and (c) late in the mission. Median (black circles) and quartile values (error bars) are plotted for each bin along the $x$-axis. The top panels show the number of intervals for each bin along $x$.}
\label{fig: EII vs clock angle}
\end{figure*}

\begin{figure*}
\begin{tikzpicture}

\node (eIoni) at (0,0) {\includegraphics[width = \textwidth]{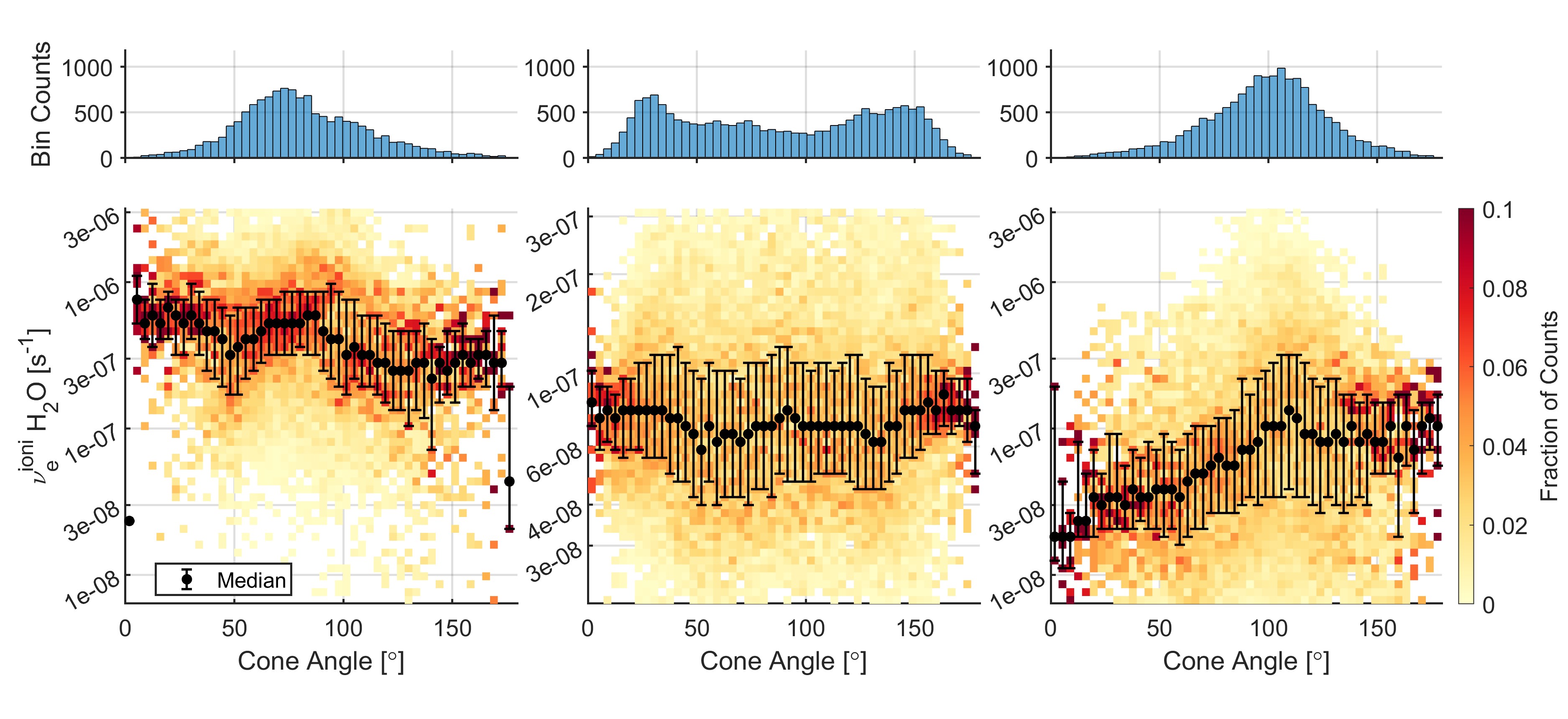}};
 \node [above left = 7.5 and 4.1 of eIoni.base, font = \bf] (EarlyLab) {Early Mission};
\node [above = 7.5 of eIoni.base, font = \bf] (PHLab) {Around PH};
\node [above right = 7.5 and 4.1 of eIoni.base, font = \bf] (LateLab) {Late Mission};
\node [above left = 7.5 and 7.3 of eIoni.base, font =\bf] (A) {(a)};
\node [above left = 7.5 and 2.1 of eIoni.base, font =\bf] (B) {(b)};
\node [above right =7.5 and 2.3 of eIoni.base, font =\bf] (C) {(c)};
\end{tikzpicture}
\caption{Electron-impact ionization frequency vs. the cone angle of the magnetic field (see Eq.~\ref{eq: cone angle}) (a) early in the mission, (b) around perihelion and (c) late in the mission. Median (black circles) and quartile values (error bars) are plotted for each bin along the $x$-axis. The top panels show the number of intervals for each bin along $x$.}
\label{fig: EII vs Cone angle}
\end{figure*}

In Section~\ref{sec: eii vs B field} and Section~\ref{sec: test pl eii vs B}, we have demonstrated the dependence of the EII frequency on the magnetic field strength in the coma. In this section, we examine the correlation betweent the EII frequency and the orientation of the magnetic field. We consider the clock and cone angles which are given by:
\begin{align}
\theta_{\text{Clock}} &= \arctan(B_y/B_z) \label{eq: clock angle}\\
\theta_{\text{Cone}} &=\arctan(\sqrt{B^2_y+B^2_z}/B_x)\label{eq: cone angle}
\end{align}
Figures~\ref{fig: EII vs clock angle} and~\ref{fig: EII vs Cone angle} show the dependence of the EII frequency on the clock and cone angles during the early, perihelion and late mission periods (see Table~\ref{tab: ioni freq time periods}). The EII frequency does not depend substantially on either the clock or cone angles in the early and perihelion periods. In the late mission period, there is a non-linear relationship with the clock angle with a minimum near 0$^\circ$. Additionally, the EII frequency increases with cone angle during this time (see Figure~\ref{fig: EII vs Cone angle}c).

\begin{figure*}
\begin{tikzpicture}

\node (USW) at (0,0) {\includegraphics[width = \textwidth]{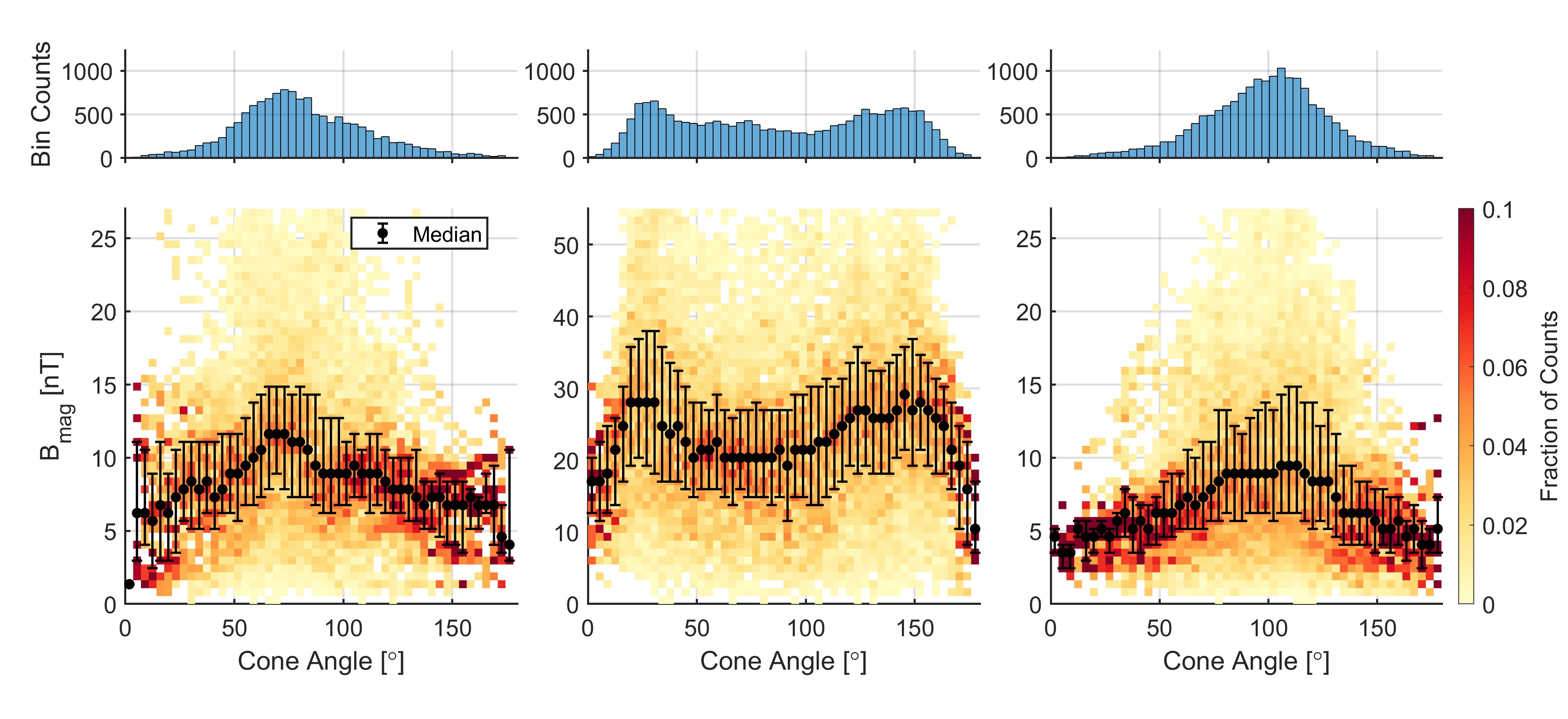}};
 \node [above left = 7.5 and 4.1 of eIoni.base, font = \bf] (EarlyLab) {Early Mission};
\node [above = 7.5 of eIoni.base, font = \bf] (PHLab) {Around PH};
\node [above right = 7.5 and 4.1 of eIoni.base, font = \bf] (LateLab) {Late Mission};
\node [above left = 7.5 and 7.3 of eIoni.base, font =\bf] (A) {(a)};
\node [above left = 7.5 and 2.1 of eIoni.base, font =\bf] (B) {(b)};
\node [above right =7.5 and 2.3 of eIoni.base, font =\bf] (C) {(c)};
\end{tikzpicture}
\caption{The cone angle of the magentic field (see Eq.~\ref{eq: cone angle}) vs. the magnetic field strength (a) early in the mission, (b) around perihelion and (c) late in the mission. Median (black circles) and quartile values (error bars) are plotted for each bin along the $x$-axis. The top panels show the number of intervals for each bin along $x$.}
\label{fig: B_mag vs Cone angle}
\end{figure*}

The non-linear behaviour with clock angle is likely an effect of the sampling with magnetic field strength, as the field strength shows similar behaviour with clock angle as is exhbited by the EII frequency. The positive correlation with cone angle could be interpreted as increased draping of magnetic field lines (at high cone angles) being associated with higher EII frequencies. However, the cone angle is also correlated with the magnetic field strength (see Figure~\ref{fig: B_mag vs Cone angle}), which is shown to be a driver of the EII frequency in Section~\ref{sec: eii vs B field}. The positive correlation with EII frequency between $\theta_{\text{Cone}}=[0,100]^\circ$ and the subsequent decrease to $\theta_{\text{Cone}}=125^\circ$ can be attributed to the varying magnetic field strength. Beyond $\theta_{\text{Cone}}=150^\circ$, the EII frequency increases while the median magnetic field strength decreases, in contrast to expectations. However, the number of datapoints is small at large cone angles and the variability in the median and quartile EII frequencies is large.

\begin{figure*}
\begin{tikzpicture}

\node (eIoni) at (0,0) {\includegraphics[width = \textwidth]{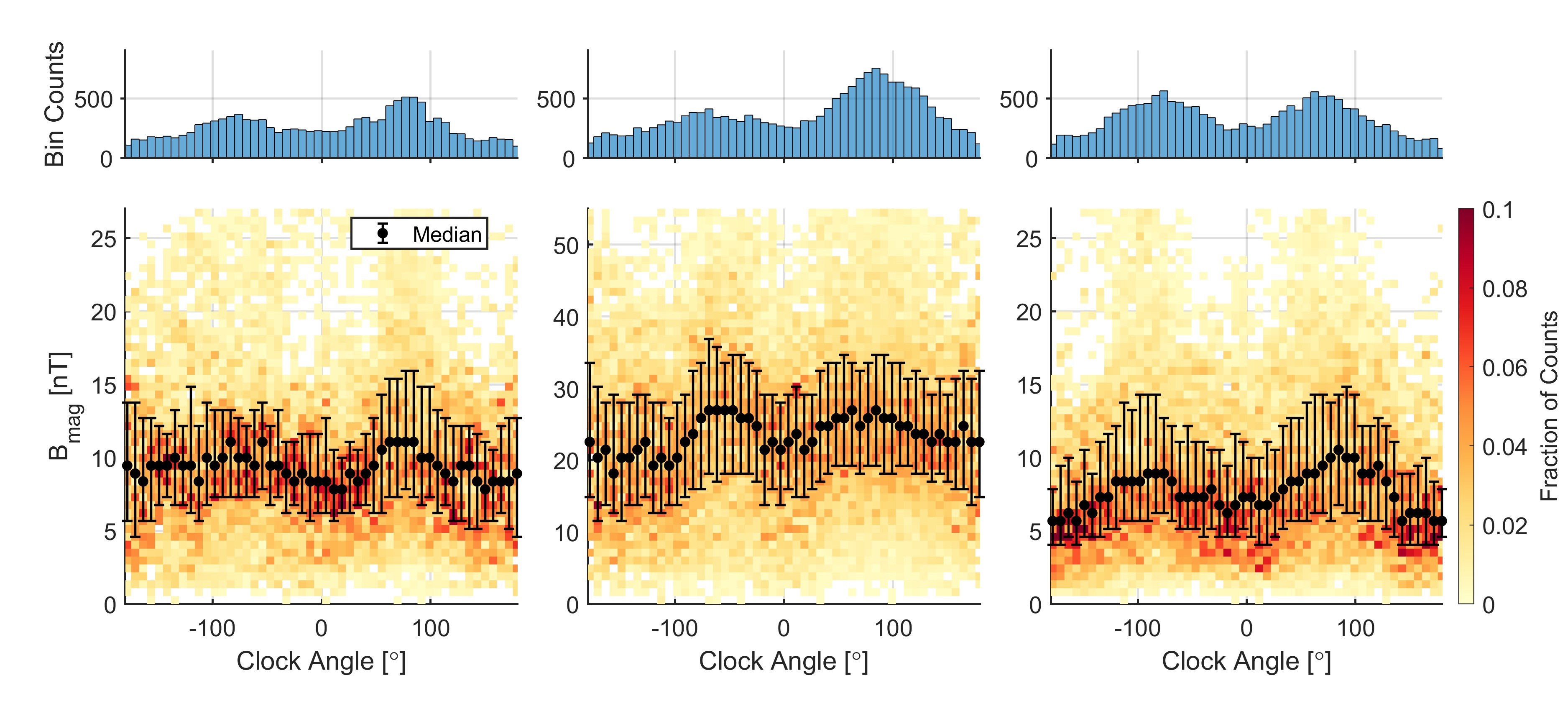}};
  \node [above left = 7.5 and 4.1 of eIoni.base, font = \bf] (EarlyLab) {Early Mission};
\node [above = 7.5 of eIoni.base, font = \bf] (PHLab) {Around PH};
\node [above right = 7.5 and 4.1 of eIoni.base, font = \bf] (LateLab) {Late Mission};
\node [above left = 7.5 and 7.3 of eIoni.base, font =\bf] (A) {(a)};
\node [above left = 7.5 and 2.1 of eIoni.base, font =\bf] (B) {(b)};
\node [above right =7.5 and 2.3 of eIoni.base, font =\bf] (C) {(c)};
\end{tikzpicture}
\caption{The clock angle of the magnetic field (see Eq.~\ref{eq: clock angle})  vs. the magnetic field strength (a) early in the mission, (b) around perihelion and (c) late in the mission. Median (black circles) and quartile values (error bars) are plotted for each bin along the $x$-axis. The top panels show the number of intervals for each bin along $x$.}
\label{fig: B vs clock angle}
\end{figure*}

\section{Total electron density and population fractions throughout the coma}
\begin{figure*}
    \centering
    \begin{tikzpicture}
        \node (EIoni) at (0,0) {\includegraphics[width = 0.8\textwidth]{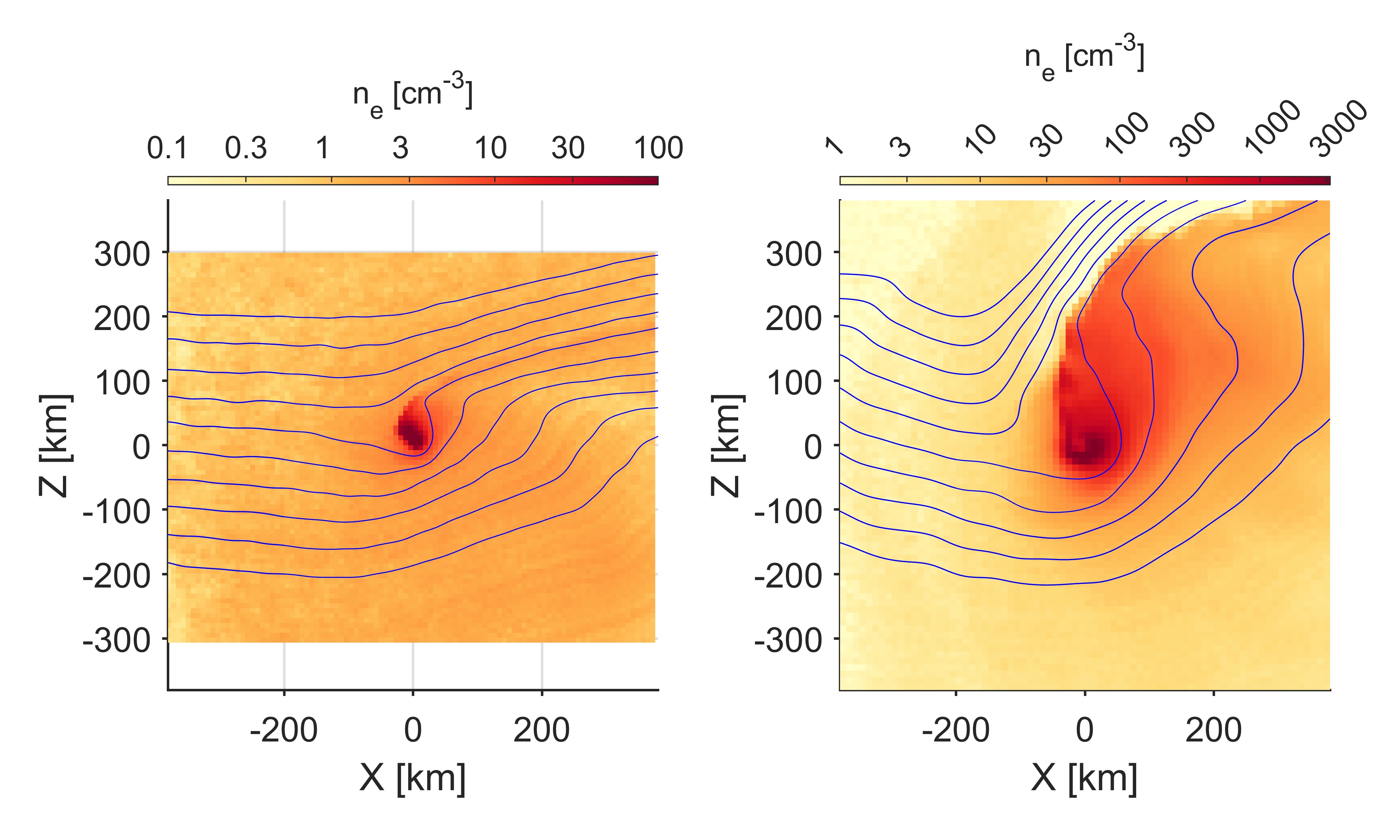}};

        \node [above left = 6.3 and 6.0 of EIoni.base, font =\large\bf] (A) {(a)};
        \node [above right = 6.3 and 0.5 of EIoni.base, font =\large\bf] (B) {(b)};

        \node [above left = 6.0 and 3.60 of EDens.base, anchor = center] (swArr_s) {};
        \node [right = 1.5 of swArr_s] (swArr_e) {};
        \draw [-latex] (swArr_s)--(swArr_e);
        \node [above right = 0.3 and 0.8 of swArr_s.center, font=\large, anchor = center] (vsw) {$\bm{u_{SW}}$};
        
        \node (BIn) at (-4.7,1.2) {};
        \node (EArrS) at (-4.7,1.0) {};
        \node [below = 1 of EArrS] (EArrE) {};
        \node [below right = 0.45 and 0.05 of EArrS.center, font=\large\bf] (EArrLabel) {$\bm{E}_{SW}$};
        \draw [-latex] (EArrS)--(EArrE);
        \path (BIn)  pic {vector out={line width=0.5pt, scale=0.2}} (3,0)  pic {};
        \node [right = 0.15 of BIn.center, font=\large\bf] (BLabel) {$\bm{B}_{SW}$};
\end{tikzpicture}
    \caption{Total electron density obtained for simulations at outgassing rates of (a) $Q=10^{26}$~s$^{-1}$ and (b) $Q=1.5\times10^{27}$~s$^{-1}$ in the $x$-$z$ plane. Streamlines are shown for electrons following the $\bm{E}\times\bm{B}$ drift velocity in the $x$-$z$ plane.}
    \label{fig: test pl tot e dens map}
\end{figure*}

\begin{figure*}
    \centering
    \begin{tikzpicture}
        \node (EIoni) at (0,0) {\includegraphics[width = 0.95\textwidth]{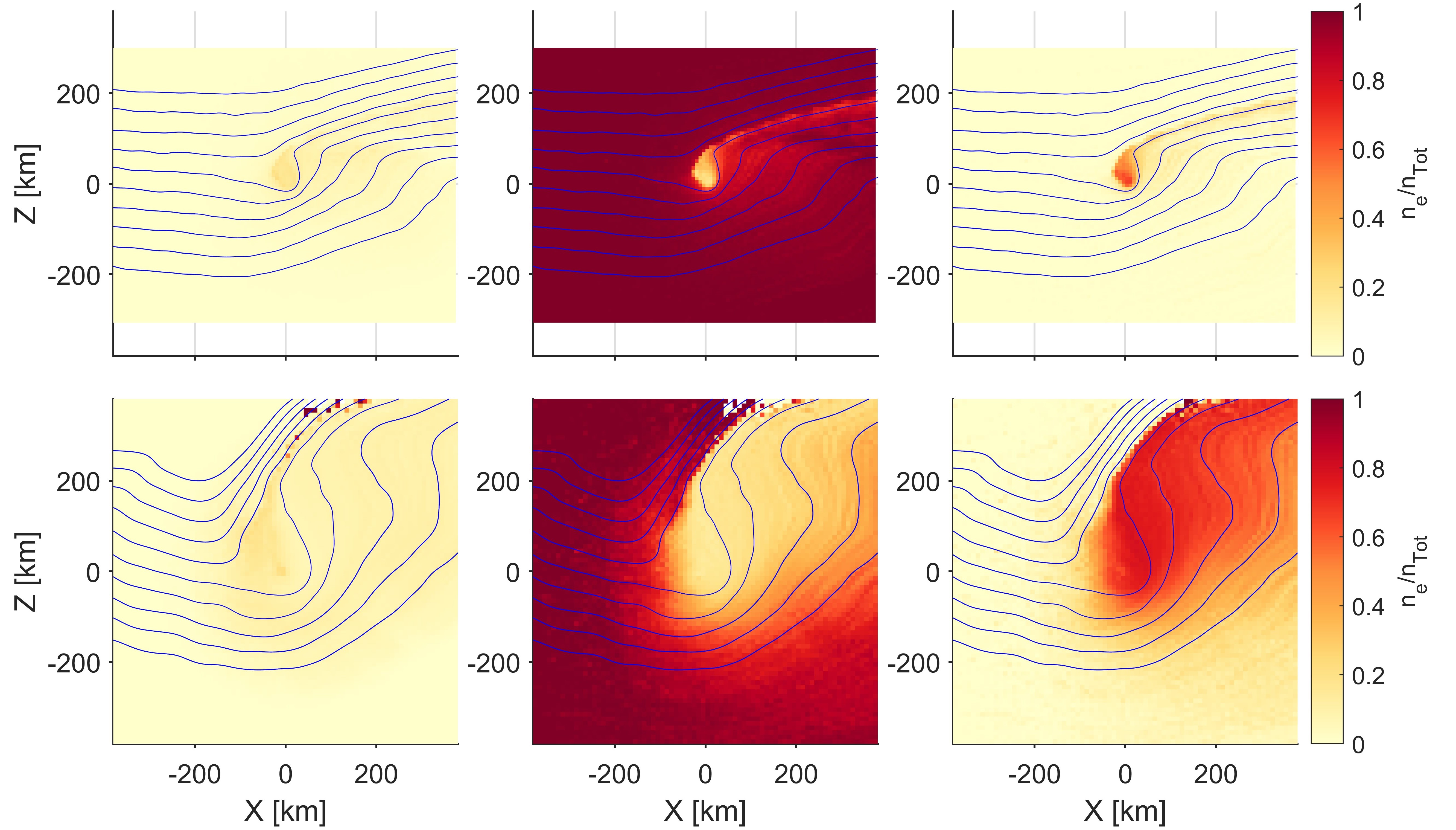}};
        \node [above left = 4.5 and 5.3 of EIoni.center, font = \bf, anchor = center] (PELab) {Photoelectrons};
        \node [above left = 4.5 and 0.3 of EIoni.center, font = \bf, anchor = center] (SWLab) {Solar Wind Electrons};
        \node [above right = 4.5 and 4.6 of EIoni.center, font = \bf, anchor = center] (SecondaryLab) {Secondary Electrons};
        \node [above left = 9.3 and 7.5 of EIoni.base, font =\large\bf] (A) {(a)};
        \node [above left = 9.3 and 2.3 of EIoni.base, font =\large\bf] (B) {(b)};
        \node [above right = 9.3 and 1.8 of EIoni.base, font =\large\bf] (C) {(c)};
        \node [above left = 4.5 and 7.5 of EIoni.base, font =\large\bf] (D) {(d)};
        \node [above left = 4.5 and 2.3 of EIoni.base, font =\large\bf] (E) {(e)};
        \node [above right = 4.5 and 1.8 of EIoni.base, font =\large\bf] (F) {(f)};
        \node [above left = 9.2 and 8.5 of EIoni.base, font=\large\bf, rotate = 90] (Q26) {$\bm{Q=10^{26}}$\,s$\bm{^{-1}}$};
        \node [above left = 4.7 and 8.5 of EIoni.base, font=\large\bf, rotate = 90] (Q26) {$\bm{Q=1.5\cdot10^{27}}$\,s$\bm{^{-1}}$};
        \node [above left = 8.5 and 5.3 of EDens.base, anchor = center] (swArr_s) {};
        \node [right = 1.5 of swArr_s] (swArr_e) {};
        \draw [-latex] (swArr_s)--(swArr_e);
        \node [above right = 0.3 and 0.8 of swArr_s.center, font=\large, anchor = center] (vsw) {$\bm{u_{SW}}$};
        
        \node (BIn) at (-6.5,3.7) {};
        \node (EArrS) at (-6.5,3.5) {};
        \node [below = 1 of EArrS] (EArrE) {};
        \node [below right = 0.45 and 0.05 of EArrS.center, font=\large\bf] (EArrLabel) {$\bm{E}_{SW}$};
        \draw [-latex] (EArrS)--(EArrE);
        \path (BIn)  pic {vector out={line width=0.5pt, scale=0.2}} (3,0)  pic {};
        \node [right = 0.15 of BIn.center, font=\large\bf] (BLabel) {$\bm{B}_{SW}$};
\end{tikzpicture}
    \caption{Fraction of total electron density made up by (a,d) photoelectrons, (b,e) solar wind electrons and (c,f) secondary electrons for outgassing rates of (a-c) $Q=10^{26}$~s$^{-1}$ and (d-f) $Q=1.5\times10^{27}$~s$^{-1}$ in the $x$-$z$ plane. Streamlines are shown for electrons following the $\bm{E}\times\bm{B}$ drift velocity in the $x$-$z$ plane.}
    \label{fig: test pl ioni frac map}
\end{figure*}

Figure~\ref{fig: test pl tot e dens map} shows the total electron density from the test particle simulation for the two outgassing rates (see Table~\ref{tab: test pl parameters}). The fractional contribution from each electron population (photoelectron, solar wind and secondary electrons) is separated in Figure~\ref{fig: test pl ioni frac map}. Away from the inner coma, the electron density is dominated by the solar wind population. Closer to the nucleus, the photoelectron and secondary electrons make up an increasingly large fraction of the total density. For both outgassing rates, electron impact ionization is stronger than photoionization close to the nucleus, resulting in large fractions of secondary electrons ($>0.5$).

\section{Ambipolar and Hall fields in the cometary coma}\label{sec: GOL fields}
\begin{figure*}
    \centering
    \begin{tikzpicture}
        \node (EIoni) at (0,0) {\includegraphics[width = 0.95\textwidth]{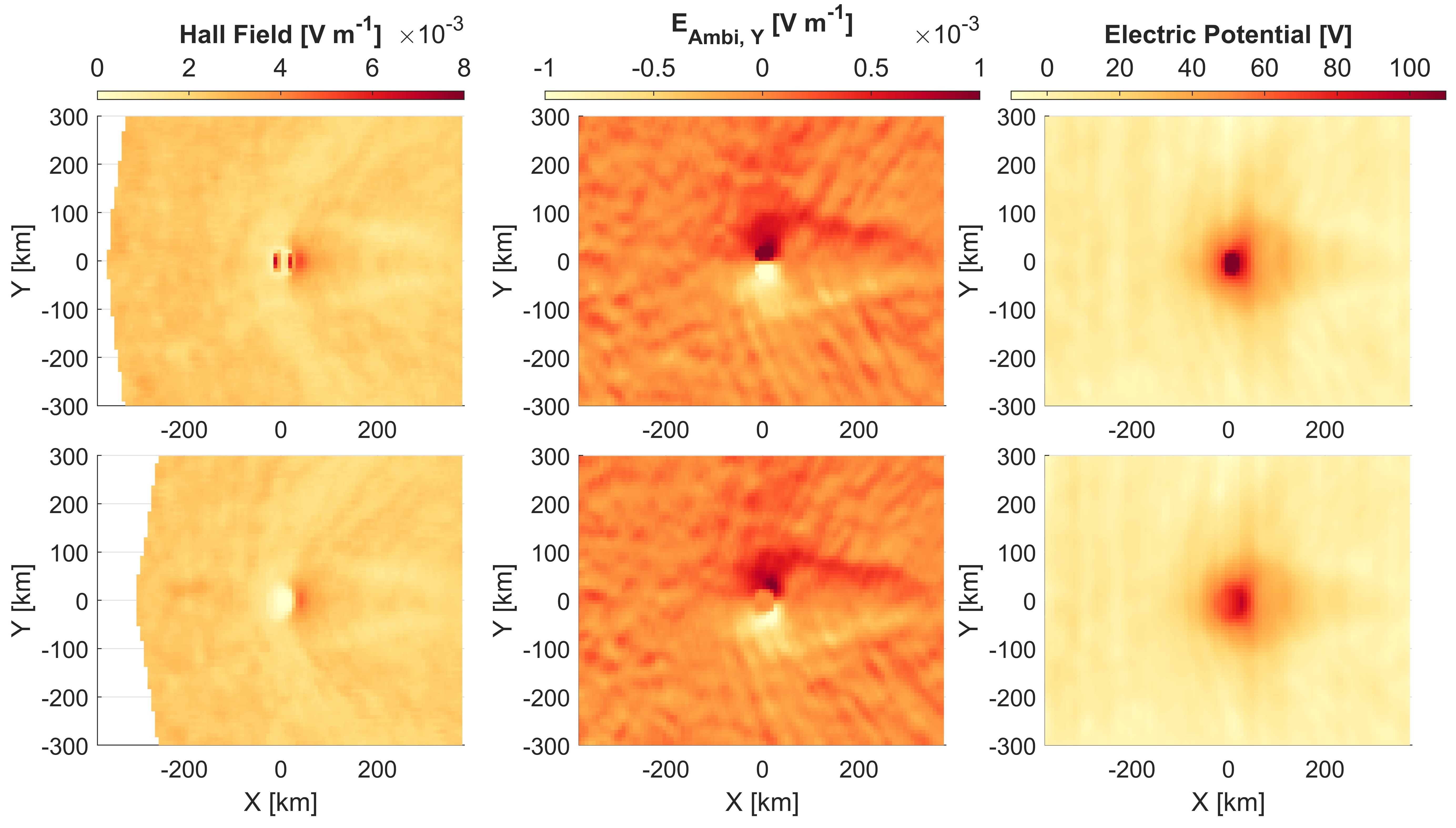}};
        \node [above left = 7.8 and 8.0 of EIoni.base, font =\large\bf] (A) {(a)};
        \node [above left = 7.8 and 2.3 of EIoni.base, font =\large\bf] (B) {(b)};
        \node [above right = 7.8 and 2.4 of EIoni.base, font =\large\bf] (C) {(c)};
        \node [above left = 4.1 and 8.0 of EIoni.base, font =\large\bf] (D) {(d)};
        \node [above left = 4.1 and 2.3 of EIoni.base, font =\large\bf] (E) {(e)};
        \node [above right = 4.1 and 2.4 of EIoni.base, font =\large\bf] (F) {(f)};
        \node [above left = 7.6 and 8.5 of EIoni.base, font=\large\bf, rotate = 90] (Q26) {Collisionless};
        \node [above left = 3.7 and 8.5 of EIoni.base, font=\large\bf, rotate = 90] (Q26) {Collisional};
        \node [above left = 7.9 and 5.40 of EDens.base, anchor = center] (swArr_s) {};
        \node [right = 1.5 of swArr_s] (swArr_e) {};
        \draw [-latex] (swArr_s)--(swArr_e);
        \node [above right = 0.3 and 0.8 of swArr_s.center, font=\large, anchor = center] (vsw) {$\bm{u_{SW}}$};
        
        \node (BIn) at (-6.5,3.2) {};
        \node (EArrS) at (-6.5,3.0) {};
        \node [below = 1 of EArrS] (EArrE) {};
        \node [below right = 0.45 and 0.05 of EArrS.center, font=\large\bf] (EArrLabel) {$\bm{E}_{SW}$};
        \draw [-latex] (EArrS)--(EArrE);
        \path (BIn)  pic {vector out={line width=0.5pt, scale=0.2}} (3,0)  pic {};
        \node [right = 0.15 of BIn.center, font=\large\bf] (BLabel) {$\bm{B}_{SW}$};
\end{tikzpicture}
    \caption{(a,d) Electron component of the Hall field (see Eq.~\ref{eq: Hall field}), (b,e) the ambipolar electric field (Eq.~\ref{eq: ambipolar field}) and (c,f) ambipolar potential well from (a-c) collisionless and (d-f) collisional test particle simulations at $Q=10^{26}$\,s$^{-1}$. }
    \label{fig: ambipolar field}
\end{figure*}

Figure~\ref{fig: ambipolar field} shows the hall and ambipolar fields from non-collisional and collisional test particle simulations of a comet with outgassing $Q=10^{26}$\,s$^{-1}$. The Hall field only includes the electron component, as the ions are not modelled in the particle simulations.
\begin{align}\label{eq: Hall field}
        \bm{E}_{Hall,e}& =  -\bm{u}_e\times\bm{B}
\end{align}
 While this does not capture the total Hall field, the difference between the collisional and non-collisional cases shows the role of collisions in mitigating the Hall field in the inner coma.

\end{appendices}
\end{document}

%% file: preamble.tex
\usepackage[T1]{fontenc}
\usepackage[english]{babel}

\usepackage{amsmath,subcaption, caption}
\usepackage{multirow}
\usepackage{siunitx}
\usepackage{mhchem}
\usepackage{multicol}
\usepackage{graphicx, xcolor, wrapfig}
\usepackage{esint,bm}
\usepackage{setspace}
\usepackage{amssymb}
\usepackage{blindtext}
\usepackage{tcolorbox}
\usepackage[toc,page]{appendix}
\usepackage{tikz, mathtools}
\usepackage{multirow}
\usepackage{outlines}
\usepackage{relsize}

\tikzset{fontscale/.style = {font=\relsize{#1}}
    }

\usetikzlibrary{shapes.geometric, arrows,arrows.meta,positioning,quotes,automata,decorations.text}
\usepackage{natbib}
\setcitestyle{authoryear-comp,open={(},close={)}}

\newcommand{\edits}[1]{{\color{black}#1}}
\newcommand{\comments}[1]{{\color{black}#1}}
\usepackage{tabularx}
\usepackage{makecell}

\usetikzlibrary{plotmarks}

\usepackage{pdflscape}
\pgfdeclarelayer{bg}    
\pgfsetlayers{bg,main}  

\tikzset{
   pics/.cd,
   vector out/.style={
      code={
         \draw[#1] (0,0)  circle (1) (45:1) -- (225:1) (135:1) -- (315:1);
      }
   }
}

\tikzset{
   pics/.cd,
   vector in/.style={
      code={
        \draw[#1] (0,0)  circle (1);
        \fill[#1] (0,0)  circle (.1);
      }
   }
}

%% file: main.bbl
\newcommand{\noop}[1]{}
\begin{thebibliography}{}
\makeatletter
\relax
\def\mn@urlcharsother{\let\do\@makeother \do\$\do\&\do\#\do\^\do\_\do\%\do\~}
\def\mn@doi{\begingroup\mn@urlcharsother \@ifnextchar [ {\mn@doi@}
  {\mn@doi@[]}}
\def\mn@doi@[#1]#2{\def\@tempa{#1}\ifx\@tempa\@empty \href
  {http://dx.doi.org/#2} {doi:#2}\else \href {http://dx.doi.org/#2} {#1}\fi
  \endgroup}
\def\mn@eprint#1#2{\mn@eprint@#1:#2::\@nil}
\def\mn@eprint@arXiv#1{\href {http://arxiv.org/abs/#1} {{\tt arXiv:#1}}}
\def\mn@eprint@dblp#1{\href {http://dblp.uni-trier.de/rec/bibtex/#1.xml}
  {dblp:#1}}
\def\mn@eprint@#1:#2:#3:#4\@nil{\def\@tempa {#1}\def\@tempb {#2}\def\@tempc
  {#3}\ifx \@tempc \@empty \let \@tempc \@tempb \let \@tempb \@tempa \fi \ifx
  \@tempb \@empty \def\@tempb {arXiv}\fi \@ifundefined
  {mn@eprint@\@tempb}{\@tempb:\@tempc}{\expandafter \expandafter \csname
  mn@eprint@\@tempb\endcsname \expandafter{\@tempc}}}

\bibitem[\protect\citeauthoryear{Balsiger et~al.,}{Balsiger
  et~al.}{2007}]{Balsiger2007}
Balsiger H.,  et~al., 2007, \mn@doi [Space Science Reviews]
  {10.1007/s11214-006-8335-3}, 128, 745

\bibitem[\protect\citeauthoryear{Behar, Nilsson, Alho, Goetz  \&
  Tsurutani}{Behar et~al.}{2017}]{Behar2017}
Behar E.,  Nilsson H.,  Alho M.,  Goetz C.,   Tsurutani B.,  2017, \mn@doi
  [Monthly Notices of the Royal Astronomical Society] {10.1093/mnras/stx1871},
  469, S396

\bibitem[\protect\citeauthoryear{Bergman, Stenberg~Wieser, Wieser, Johansson
  \& Eriksson}{Bergman et~al.}{2020}]{Bergman2020VSC}
Bergman S.,  Stenberg~Wieser G.,  Wieser M.,  Johansson F.~L.,   Eriksson A.,
  2020, \mn@doi [Journal of Geophysical Research: Space Physics]
  {https://doi.org/10.1029/2019JA027478}, 125, e2019JA027478

\bibitem[\protect\citeauthoryear{Bergman et~al.,}{Bergman
  et~al.}{2021a}]{Bergman2021cavSpeeds}
Bergman S.,  et~al., 2021a, \mn@doi [Monthly Notices of the Royal Astronomical
  Society] {10.1093/mnras/stab584}, 503, 2733

\bibitem[\protect\citeauthoryear{Bergman, Stenberg Wieser, Wieser, Nilsson,
  Vigren, Beth, Masunaga  \& Eriksson}{Bergman
  et~al.}{2021b}]{Bergman2021CavityFlow}
Bergman S.,  Stenberg Wieser G.,  Wieser M.,  Nilsson H.,  Vigren E.,  Beth
  A.,  Masunaga K.,   Eriksson A.,  2021b, \mn@doi [Monthly Notices of the
  Royal Astronomical Society] {10.1093/mnras/stab2470}, 507, 4900

\bibitem[\protect\citeauthoryear{Beth, Galand  \& Heritier}{Beth
  et~al.}{2019}]{Beth2019}
Beth A.,  Galand M.,   Heritier K.~L.,  2019, \mn@doi [A\&A]
  {10.1051/0004-6361/201833517}, 630, A47

\bibitem[\protect\citeauthoryear{Beth, Galand, Wedlund  \& Eriksson}{Beth
  et~al.}{2022}]{beth2022cometary}
Beth A.,  Galand M.,  Wedlund C.~S.,   Eriksson A.,  2022, arXiv preprint
  arXiv:2211.03868

\bibitem[\protect\citeauthoryear{Biver et~al.,}{Biver et~al.}{2019}]{Biver2019}
Biver N.,  et~al., 2019, \mn@doi [A\&A] {10.1051/0004-6361/201834960}, 630, A19

\bibitem[\protect\citeauthoryear{Broiles et~al.,}{Broiles
  et~al.}{2016a}]{Broiles2016Kappa}
Broiles T.~W.,  et~al., 2016a, \mn@doi [Journal of Geophysical Research: Space
  Physics] {10.1002/2016JA022972}, 121, 7407

\bibitem[\protect\citeauthoryear{Broiles et~al.,}{Broiles
  et~al.}{2016b}]{Broiles2016Supratherm}
Broiles T.~W.,  et~al., 2016b, \mn@doi [Monthly Notices of the Royal
  Astronomical Society] {10.1093/mnras/stw2942}, 462, S312

\bibitem[\protect\citeauthoryear{Burch, Goldstein, Cravens, Gibson, Lundin,
  Pollock, Winningham  \& Young}{Burch et~al.}{2007}]{Burch2007}
Burch J.~L.,  Goldstein R.,  Cravens T.~E.,  Gibson W.~C.,  Lundin R.~N.,
  Pollock C.~J.,  Winningham J.~D.,   Young D.~T.,  2007, \mn@doi [Space
  Science Reviews] {10.1007/s11214-006-9002-4}, 128, 697

\bibitem[\protect\citeauthoryear{Carr et~al.,}{Carr et~al.}{2007}]{Carr2007}
Carr C.,  et~al., 2007, \mn@doi [\ssr] {10.1007/s11214-006-9136-4}, 128, 629

\bibitem[\protect\citeauthoryear{Chaufray et~al.,}{Chaufray
  et~al.}{2017}]{Chaufray2017}
Chaufray J.-Y.,  et~al., 2017, \mn@doi [Monthly Notices of the Royal
  Astronomical Society] {10.1093/mnras/stx1895}, 469, S416

\bibitem[\protect\citeauthoryear{Cho, Park, Tanaka  \& Buckman}{Cho
  et~al.}{2004}]{cho2004measurements}
Cho H.,  Park Y.~S.,  Tanaka H.,   Buckman S.~J.,  2004, \mn@doi [Journal of
  Physics B: Atomic, Molecular and Optical Physics]
  {10.1088/0953-4075/37/3/008}, 37, 625

\bibitem[\protect\citeauthoryear{Clark et~al.,}{Clark et~al.}{2015}]{Clark2015}
Clark G.,  et~al., 2015, \mn@doi [Astronomy {\&} Astrophysics]
  {10.1051/0004-6361/201526351}, 583, A24

\bibitem[\protect\citeauthoryear{Cravens, Kozyra, Nagy, Gombosi  \&
  Kurtz}{Cravens et~al.}{1987}]{Cravens1987ImpactIoni}
Cravens T.~E.,  Kozyra J.~U.,  Nagy A.~F.,  Gombosi T.~I.,   Kurtz M.,  1987,
  \mn@doi [Journal of Geophysical Research] {10.1029/JA092iA07p07341}, 92, 7341

\bibitem[\protect\citeauthoryear{Deca, Divin, Henri, Eriksson, Markidis,
  Olshevsky  \& Hor{\'{a}}nyi}{Deca et~al.}{2017}]{Deca2017}
Deca J.,  Divin A.,  Henri P.,  Eriksson A.,  Markidis S.,  Olshevsky V.,
  Hor{\'{a}}nyi M.,  2017, \mn@doi [Physical Review Letters]
  {10.1103/PhysRevLett.118.205101}, 118, 205101

\bibitem[\protect\citeauthoryear{Deca, Henri, Divin, Eriksson, Galand, Beth,
  Ostaszewski  \& Hor\'anyi}{Deca et~al.}{2019}]{Deca2019}
Deca J.,  Henri P.,  Divin A.,  Eriksson A.,  Galand M.,  Beth A.,  Ostaszewski
  K.,   Hor\'anyi M.,  2019, \mn@doi [Phys. Rev. Lett.]
  {10.1103/PhysRevLett.123.055101}, 123, 055101

\bibitem[\protect\citeauthoryear{Divin, Deca, Eriksson, Henri, Lapenta,
  Olshevsky  \& Markidis}{Divin et~al.}{2020}]{Divin2020}
Divin A.,  Deca J.,  Eriksson A.,  Henri P.,  Lapenta G.,  Olshevsky V.,
  Markidis S.,  2020, \mn@doi [The Astrophysical Journal Letters]
  {10.3847/2041-8213/ab6662}, 889, L33

\bibitem[\protect\citeauthoryear{Edberg et~al.,}{Edberg
  et~al.}{2016a}]{Edberg2016CIR}
Edberg N. J.~T.,  et~al., 2016a, \mn@doi [Journal of Geophysical Research A:
  Space Physics] {10.1002/2015JA022147}, 121, 949

\bibitem[\protect\citeauthoryear{Edberg et~al.,}{Edberg
  et~al.}{2016b}]{Edberg2016CME}
Edberg N.~J.,  et~al., 2016b, \mn@doi [Monthly Notices of the Royal
  Astronomical Society] {10.1093/mnras/stw2112}, 462, S45

\bibitem[\protect\citeauthoryear{Engelhardt, Eriksson, Vigren, Vali{\`{e}}res,
  Rubin, Gilet  \& Henri}{Engelhardt et~al.}{2018}]{Engelhardt2018}
Engelhardt I. A.~D.,  Eriksson A.~I.,  Vigren E.,  Vali{\`{e}}res X.,  Rubin
  M.,  Gilet N.,   Henri P.,  2018, \mn@doi [Astronomy \& Astrophysics]
  {10.1051/0004-6361/201833251}, 15

\bibitem[\protect\citeauthoryear{Eriksson et~al.,}{Eriksson
  et~al.}{2007}]{Eriksson2007}
Eriksson A.~I.,  et~al., 2007, \mn@doi [Space Science Reviews]
  {10.1007/s11214-006-9003-3}, 128, 729

\bibitem[\protect\citeauthoryear{Eriksson et~al.,}{Eriksson
  et~al.}{2017}]{Eriksson2017}
Eriksson A.~I.,  et~al., 2017, \mn@doi [Astronomy {\&} Astrophysics]
  {10.1051/0004-6361/201630159}, 605

\bibitem[\protect\citeauthoryear{Faure, Gorfinkiel  \& Tennyson}{Faure
  et~al.}{2004}]{faure2004low}
Faure A.,  Gorfinkiel J.~D.,   Tennyson J.,  2004, \mn@doi [Journal of Physics
  B: Atomic, Molecular and Optical Physics] {10.1088/0953-4075/37/4/007}, 37,
  801

\bibitem[\protect\citeauthoryear{Feldman et~al.,}{Feldman
  et~al.}{2015}]{Feldman2015}
Feldman P.~D.,  et~al., 2015, \mn@doi [Astronomy {\&} Astrophysics]
  {10.1051/0004-6361/201525925}, 583, A8

\bibitem[\protect\citeauthoryear{Fougere et~al.,}{Fougere
  et~al.}{2016}]{Fougere2016}
Fougere N.,  et~al., 2016, \mn@doi [Monthly Notices of the Royal Astronomical
  Society] {10.1093/mnras/stw2388}, 462, S156

\bibitem[\protect\citeauthoryear{Galand et~al.,}{Galand
  et~al.}{2016}]{Galand2016}
Galand M.,  et~al., 2016, \mn@doi [Monthly Notices of the Royal Astronomical
  Society] {10.1093/mnras/stw2891}, 462, S331

\bibitem[\protect\citeauthoryear{Galand et~al.,}{Galand
  et~al.}{2020}]{Galand2020}
Galand M.,  et~al., 2020, \mn@doi [Nature Astronomy]
  {10.1038/s41550-020-1171-7}

\bibitem[\protect\citeauthoryear{Gasc et~al.,}{Gasc
  et~al.}{2017}]{Gasc2017MNRAS}
Gasc S.,  et~al., 2017, \mn@doi [Monthly Notices of the Royal Astronomical
  Society] {10.1093/mnras/stx1412}, 469, S108

\bibitem[\protect\citeauthoryear{Gilet et~al.,}{Gilet et~al.}{2020}]{Gilet2020}
Gilet N.,  et~al., 2020, \mn@doi [A\&A] {10.1051/0004-6361/201937056}, 640,
  A110

\bibitem[\protect\citeauthoryear{Glassmeier et~al.,}{Glassmeier
  et~al.}{2007}]{Glassmeier2007MAG}
Glassmeier K.-H.,  et~al., 2007, \mn@doi [Space Science Reviews]
  {10.1007/s11214-006-9114-x}, 128, 649

\bibitem[\protect\citeauthoryear{Goetz et~al.,}{Goetz
  et~al.}{2016}]{Goetz2016CavStruct}
Goetz C.,  et~al., 2016, \mn@doi [Monthly Notices of the Royal Astronomical
  Society] {10.1093/mnras/stw3148}, 462, S459

\bibitem[\protect\citeauthoryear{Goetz, Volwerk, Richter  \& Glassmeier}{Goetz
  et~al.}{2017}]{Goetz2017}
Goetz C.,  Volwerk M.,  Richter I.,   Glassmeier K.-H.,  2017, \mn@doi [Monthly
  Notices of the Royal Astronomical Society] {10.1093/mnras/stx1570}, 469, S268

\bibitem[\protect\citeauthoryear{Hajra et~al.,}{Hajra
  et~al.}{2018}]{Hajra2018CIRs}
Hajra R.,  et~al., 2018, \mn@doi [Monthly Notices of the Royal Astronomical
  Society] {10.1093/mnras/sty2166}, 480, 4544

\bibitem[\protect\citeauthoryear{Hansen et~al.,}{Hansen
  et~al.}{2016}]{Hansen2016}
Hansen K.~C.,  et~al., 2016, \mn@doi [Monthly Notices of the Royal Astronomical
  Society] {10.1093/mnras/stw2413}, 462, stw2413

\bibitem[\protect\citeauthoryear{Haser}{Haser}{1957}]{Haser1957}
Haser L.,  1957, Bulletin de la Classe des Sciences de l'Acad{\'{e}}mie Royale
  de Belgique, vol. 43, p. 740-750 (1957)., 43, 740

\bibitem[\protect\citeauthoryear{Heritier et~al.,}{Heritier
  et~al.}{2017a}]{Heritier2017Vert}
Heritier K.~L.,  et~al., 2017a, \mn@doi [Monthly Notices of the Royal
  Astronomical Society] {10.1093/mnras/stx1459}, 469, S118

\bibitem[\protect\citeauthoryear{Heritier et~al.,}{Heritier
  et~al.}{2017b}]{Heritier2017IonComp}
Heritier K.~L.,  et~al., 2017b, \mn@doi [Monthly Notices of the Royal
  Astronomical Society] {10.1093/mnras/stx1912}, 469, S427

\bibitem[\protect\citeauthoryear{Heritier et~al.,}{Heritier
  et~al.}{2018}]{Heritier2018source}
Heritier K.~L.,  et~al., 2018, \mn@doi [\aap] {10.1051/0004-6361/201832881},
  618, A77

\bibitem[\protect\citeauthoryear{Itikawa \& Mason}{Itikawa \&
  Mason}{2005}]{Itikawa2005}
Itikawa Y.,  Mason N.,  2005, \mn@doi [Journal of Physical and Chemical
  Reference Data] {10.1063/1.1799251}, 34, 1

\bibitem[\protect\citeauthoryear{Johansson et~al.,}{Johansson
  et~al.}{2021}]{Johansson2021plasDens}
Johansson F.~L.,  et~al., 2021, \mn@doi [A\&A] {10.1051/0004-6361/202039959},
  653, A128

\bibitem[\protect\citeauthoryear{Jorda et~al.,}{Jorda et~al.}{2016}]{Jorda2016}
Jorda L.,  et~al., 2016, \mn@doi [Icarus] {10.1016/J.ICARUS.2016.05.002}, 277,
  257

\bibitem[\protect\citeauthoryear{Lean, Warren, Mariska  \& Bishop}{Lean
  et~al.}{2003}]{Lean2003}
Lean J.~L.,  Warren H.~P.,  Mariska J.~T.,   Bishop J.,  2003, \mn@doi [Journal
  of Geophysical Research: Space Physics]
  {https://doi.org/10.1029/2001JA009238}, 108

\bibitem[\protect\citeauthoryear{Lindsay \& Mangan}{Lindsay \&
  Mangan}{2003}]{LindsayMangan2003}
Lindsay B.~G.,  Mangan M.~A.,  2003, Interactions of Photons and Electrons with
  Molecules {\textperiodcentered} 5.1 Ionization: Datasheet from
  Landolt-B{\"o}rnstein - Group I Elementary Particles, Nuclei and Atoms
  {\textperiodcentered} Volume 17C: ``Interactions of Photons and Electrons
  with Molecules'' in SpringerMaterials
  (https://doi.org/10.1007/10874891{\_}2), \mn@doi{10.1007/10874891_2}, \url
  {https://materials.springer.com/lb/docs/sm_lbs_978-3-540-45843-2_2}

\bibitem[\protect\citeauthoryear{Läuter, Kramer, Rubin  \& Altwegg}{Läuter
  et~al.}{2018}]{Laeuter2018}
Läuter M.,  Kramer T.,  Rubin M.,   Altwegg K.,  2018, \mn@doi [Monthly
  Notices of the Royal Astronomical Society] {10.1093/mnras/sty3103}, 483, 852

\bibitem[\protect\citeauthoryear{Marshall et~al.,}{Marshall
  et~al.}{2017}]{Marshall2017}
Marshall D.~W.,  et~al., 2017, \mn@doi [Astronomy {\&} Astrophysics]
  {10.1051/0004-6361/201730502}, 603, A87

\bibitem[\protect\citeauthoryear{Myllys et~al.,}{Myllys
  et~al.}{2019}]{Myllys2019}
Myllys M.,  et~al., 2019, \mn@doi [Astronomy {\&} Astrophysics]
  {10.1051/0004-6361/201834964}

\bibitem[\protect\citeauthoryear{Myllys et~al.,}{Myllys
  et~al.}{2021}]{Myllys2021}
Myllys M.,  et~al., 2021, \mn@doi [A\&A] {10.1051/0004-6361/201936633}, 652,
  A73

\bibitem[\protect\citeauthoryear{Nilsson et~al.,}{Nilsson
  et~al.}{2007}]{Nilsson2007}
Nilsson H.,  et~al., 2007, \mn@doi [Space Science Reviews]
  {10.1007/s11214-006-9031-z}, 128, 671

\bibitem[\protect\citeauthoryear{Nilsson et~al.,}{Nilsson
  et~al.}{2017}]{Nilsson2017}
Nilsson H.,  et~al., 2017, \mn@doi [Monthly Notices of the Royal Astronomical
  Society] {10.1093/mnras/stx1491}, 469, S252

\bibitem[\protect\citeauthoryear{Nilsson et~al.,}{Nilsson
  et~al.}{2020}]{Nilsson2020}
Nilsson H.,  et~al., 2020, \mn@doi [Monthly Notices of the Royal Astronomical
  Society] {10.1093/mnras/staa2613}, 498, 5263

\bibitem[\protect\citeauthoryear{Nilsson et~al.,}{Nilsson
  et~al.}{2022}]{Nilsson2022}
Nilsson H.,  et~al., 2022, \mn@doi [A\&A] {10.1051/0004-6361/202142867}, 659,
  A18

\bibitem[\protect\citeauthoryear{Odelstad et~al.,}{Odelstad
  et~al.}{2015}]{Odelstad2015b}
Odelstad E.,  et~al., 2015, \mn@doi [Geophysical Research Letters]
  {10.1002/2015GL066599}, 42, 10,126

\bibitem[\protect\citeauthoryear{Odelstad et~al.,}{Odelstad
  et~al.}{2018}]{Odelstad2018}
Odelstad E.,  et~al., 2018, \mn@doi [Journal of Geophysical Research: Space
  Physics] {10.1029/2018JA025542}, 123, 5870

\bibitem[\protect\citeauthoryear{{Simon Wedlund} et~al.,}{{Simon Wedlund}
  et~al.}{2019}]{SimonWedlund2019b}
{Simon Wedlund} C.,  et~al., 2019, \mn@doi [Astronomy {\&} Astrophysics]
  {10.1051/0004-6361/201834881}

\bibitem[\protect\citeauthoryear{Sishtla, Divin, Deca, Olshevsky  \&
  Markidis}{Sishtla et~al.}{2019}]{Sishtla2019trapping}
Sishtla C.~P.,  Divin A.,  Deca J.,  Olshevsky V.,   Markidis S.,  2019,
  \mn@doi [Physics of Plasmas] {10.1063/1.5115456}, 26, 102904

\bibitem[\protect\citeauthoryear{Song et~al.,}{Song et~al.}{2021}]{Song2021}
Song M.-Y.,  et~al., 2021, \mn@doi [Journal of Physical and Chemical Reference
  Data] {10.1063/5.0035315}, 50

\bibitem[\protect\citeauthoryear{Stephenson et~al.,}{Stephenson
  et~al.}{2021}]{Stephenson2021FUV}
Stephenson P.,  et~al., 2021, \mn@doi [A\&A] {10.1051/0004-6361/202039155},
  647, A119

\bibitem[\protect\citeauthoryear{Stephenson, Galand, Deca, Henri  \&
  Carnielli}{Stephenson et~al.}{2022}]{Stephenson2022TestPl}
Stephenson P.,  Galand M.,  Deca J.,  Henri P.,   Carnielli G.,  2022, \mn@doi
  [Monthly Notices of the Royal Astronomical Society] {10.1093/mnras/stac055},
  511, 4090

\bibitem[\protect\citeauthoryear{Straub, Lindsay, Smith  \& Stebbings}{Straub
  et~al.}{1998}]{Straub1994}
Straub H.~C.,  Lindsay B.~G.,  Smith K.~A.,   Stebbings R.~F.,  1998, \mn@doi
  [The Journal of Chemical Physics] {10.1063/1.475367}, 108, 109

\bibitem[\protect\citeauthoryear{Trotignon et~al.,}{Trotignon
  et~al.}{2007}]{Trotignon2007}
Trotignon J.~G.,  et~al., 2007, \mn@doi [Space Science Reviews]
  {10.1007/s11214-006-9005-1}, 128, 713

\bibitem[\protect\citeauthoryear{Vigren \& Eriksson}{Vigren \&
  Eriksson}{2017}]{Vigren2017a}
Vigren E.,  Eriksson A.~I.,  2017, \mn@doi [The Astronomical Journal]
  {10.3847/1538-3881/aa6006}, 153, 150

\bibitem[\protect\citeauthoryear{Vigren \& Galand}{Vigren \&
  Galand}{2013}]{Vigren2013}
Vigren E.,  Galand M.,  2013, \mn@doi [The Astrophysical Journal]
  {10.1088/0004-637X/772/1/33}, 772, 33

\bibitem[\protect\citeauthoryear{Vigren et~al.,}{Vigren
  et~al.}{2017}]{Vigren2017}
Vigren E.,  et~al., 2017, \mn@doi [Monthly Notices of the Royal Astronomical
  Society] {10.1093/mnras/stx1472}, 469, S142

\bibitem[\protect\citeauthoryear{Wattieaux, Henri, Gilet, Valli\`eres  \&
  Deca}{Wattieaux et~al.}{2020}]{Wattieaux2020}
Wattieaux G.,  Henri P.,  Gilet N.,  Valli\`eres X.,   Deca J.,  2020, \mn@doi
  [A\&A] {10.1051/0004-6361/202037571}, 638, A124

\bibitem[\protect\citeauthoryear{Woods \& Rottman}{Woods \&
  Rottman}{2002}]{Woods2002}
Woods T.~N.,  Rottman G.~J.,  2002, Solar Ultraviolet Variability Over Time
  Periods of Aeronomic Interest.
American Geophysical Union (AGU), pp 221--233,
  \mn@doi{https://doi.org/10.1029/130GM14}

\bibitem[\protect\citeauthoryear{Woods et~al.,}{Woods et~al.}{2005}]{Woods2005}
Woods T.~N.,  et~al., 2005, \mn@doi [Journal of Geophysical Research]
  {10.1029/2004JA010765}, 110, A01312

\bibitem[\protect\citeauthoryear{Woods et~al.,}{Woods et~al.}{2012}]{Woods2012}
Woods T.~N.,  et~al., 2012, \mn@doi [Solar Physics]
  {10.1007/s11207-009-9487-6}, 275, 115

\makeatother
\end{thebibliography}
